\documentclass[12pt,english]{report}

\usepackage{natbib}
\usepackage{url}
\makeatletter
\def\url@leostyle{%
    \def\UrlFont{\sf}}{\def\UrlFont{\small\ttfamily}}
\makeatother
\usepackage{geometry}
\usepackage{setspace}

\usepackage[stable]{footmisc}

\usepackage{fancyhdr}

\usepackage[font=small,labelfont=bf,justification=raggedright]{caption}

\usepackage{authblk}

\usepackage{appendix}

\usepackage{tocvsec2}

\input{Qcircuit}
\renewcommand{\Qcircuit}[1][0em]{\xymatrix @*=<#1>}

\usepackage{graphicx}
\usepackage{babel}
\usepackage{mathrsfs}
\usepackage{amsmath}
\usepackage{amsfonts}
\usepackage{gensymb}
\usepackage{units}

\usepackage{listings}
\usepackage{tikz}
\usetikzlibrary{arrows,automata}


\numberwithin{equation}{chapter}

\raggedright
\begin{document}

\pagestyle{plain}
\pagenumbering{roman}

\thispagestyle{empty}

\vspace*{\fill}
\begin{center}
\textsc{On the Physical Explanation for Quantum Computational
  Speedup}


\par
\vspace{12pt}
(Thesis format: Monograph)

\par
\vspace{0.5in}
by

\par
\vspace{0.5in}
Michael E. \underline{Cuffaro}

\par
\vspace{0.5in}
Graduate Program in Philosophy

\par
\vspace{0.5in}
A thesis submitted in partial fulfilment

of the requirements for the degree of

Doctor of Philosophy

\par
\vspace{0.5in}
The School of Graduate and Postdoctoral Studies

The University of Western Ontario

London, Ontario, Canada

\par
\vspace{0.5in}
\textcopyright \mbox{ } Michael E. Cuffaro, 2013
\end{center}
\vspace*{\fill}

%
%
%
%
%
%
%
%
%
%
%
%
%
%
%
%
%
%
%
%

\newpage

\onehalfspacing

\addcontentsline{toc}{chapter}{Abstract}
\begin{center}\textbf{\textsc{Abstract}}\end{center}
\begin{center}
On the Physical Explanation for Quantum Computational Speedup

Michael E. Cuffaro
\end{center}

The aim of this dissertation is to clarify the debate over the
explanation of \emph{quantum speedup} and to submit, for the reader's
consideration, a tentative resolution to it. In particular, I argue,
in this dissertation, that the physical explanation for quantum
speedup is precisely the fact that the phenomenon of \emph{quantum
entanglement} enables a quantum computer to fully exploit the
representational capacity of Hilbert space. This is impossible for
classical systems, joint states of which must always be representable
as product states.

I begin the dissertation by considering, in Chapter \ref{ch:mwi}, the
most popular of the candidate physical explanations for quantum
speedup: the \emph{many worlds explanation of quantum computation}. I
argue that, although it is inspired by the neo-Everettian
interpretation of quantum mechanics, unlike the latter it does not
have the conceptual resources required to overcome objections such as
the so-called `preferred basis objection'. I further argue that the
many worlds explanation, at best, can serve as a good description of
the physical process which takes place in
so-called \emph{network-based} computation, but that it is
incompatible with other models of computation such as \emph{cluster
state} quantum computing. I next consider, in Chapter \ref{ch:nec}, a
common component of most other candidate explanations of quantum
speedup: \emph{quantum entanglement}. I investigate whether
entanglement can be said to be a necessary component of any
explanation for quantum speedup, and I consider two major purported
counter-examples to this claim. I argue that neither of these, in
fact, show that entanglement is unnecessary for speedup, and that, on
the contrary, we should conclude that it is. In Chapters \ref{ch:suff}
and \ref{ch:sig} I then ask whether entanglement can be said to be
sufficient as well. In Chapter \ref{ch:suff} I argue that despite a
result that seems to indicate the contrary, entanglement, considered
as a resource, can be seen as sufficient to enable quantum
speedup. Finally, in Chapter \ref{ch:sig} I argue that entanglement is
sufficient \emph{to explain} quantum speedup as well.

\mbox{} \\


\newpage

\addcontentsline{toc}{chapter}{Dedication}
\noindent \begin{Large}\emph{To my nephews and nieces}\end{Large}

\newpage

\addcontentsline{toc}{chapter}{Acknowledgements}
\begin{center}\textbf{\textsc{Acknowledgements}}\end{center}

I sincerely thank my supervisor, Professor Wayne Myrvold, for his
incisive criticisms of my earlier drafts of this dissertation,
for his kind and helpful comments and advice, and for his patience and
encouragement as I worked my way through the more intricate details of
the argument. I also thank Professors Robert DiSalle, Chris Smeenk,
William Harper, and William Demopoulos for their kindness,
encouragement, and support. And though he was neither directly nor
indirectly involved in this work, I thank Professor John Bell, for
inspiring in me the desire to produce the best work that I am capable
of.

I am also indebted to Erik Curiel, Emerson Doyle, Lucas Dunlap,
Nicolas Fillion, Dylan Gault, Sona Ghosh, Robert Moir, Ryan Samaroo,
Morgan Tait, and Jos Uffink for our discussions on this and other
topics in the philosophy of physics. I am indebted to Molly Kao, both
for discussion and most of all for her generosity of spirit,
invaluable to me through all of the trials and tribulations that go
along with the writing of a work of this length and scope. Finally, I
am indebted to the rest of the faculty, students, and staff of the
Philosophy department at The University of Western Ontario for making
my experience here one that I will always remember with fondness.

\singlespacing

\newpage

\newpage

\tableofcontents


\newpage


\pagenumbering{arabic}
\onehalfspacing

\fancypagestyle{plain}{
  \fancyhf{} 
  \rhead{\thepage}
}

\pagestyle{fancy}
\fancyhf{}
\rhead{\thepage}

\chapter{Overview}
\label{ch:over}

\settocdepth{subsection}

\section{Introduction}
\label{over:s:intro}

Of the many and varied applications of quantum information theory,
perhaps the most fascinating is the sub-field of quantum
computation. In this sub-field, computational algorithms are designed
which utilise the resources available in quantum systems to compute
solutions to computational problems with, in some cases, exponentially
fewer resources than any known classical algorithm. But while the fact
of quantum computational speedup is almost beyond doubt,\footnote{Just
  as with other important problems in computational complexity
  theory, such as the \textbf{P} = \textbf{NP} problem, there is
  currently no proof, though it is very strongly suspected to be
  true, that the class of problems efficiently solvable by a quantum
  computer is larger than the class of problems efficiently solvable
  by a classical computer (cf. Appendix \ref{ch:cc}).} the source of
quantum speedup is still a matter of debate. Candidate explanations of
quantum speedup range from the purported ability of quantum computers
to perform multiple function evaluations simultaneously
\citep[][]{deutsch1997,duwell2004,hewittHorsman2009} to the purported
ability of a quantum computer to compute a global property of a
function by performing fewer, not more, computations
\citep[e.g.][]{steane2003,bub2010} than classical computers.

The aim of this dissertation is to clarify this debate and to submit,
for the reader's consideration, a tentative resolution to it. In the
following pages I will argue that the explanation for quantum speedup
is precisely the following. The phenomenon of \emph{quantum
  entanglement} enables a quantum computer to fully exploit the
representational capacity of Hilbert space. This is impossible for
classical systems, joint states of which must always be representable
as product states. Since the number of distinct product states of
$n$-fold $d$-dimensional systems is exponentially fewer than the total
number of states representable in the corresponding Hilbert space, a
classical computer will, in general, require exponentially more steps
than a quantum computer to solve a computational problem that requires
one to take full advantage of this representational capacity.

\section{Synopsis of this dissertation}

\subsection{Chapter summaries}

\subsubsection{Chapter \ref{ch:mwi}}

Chapter \ref{ch:mwi} examines what is arguably the most well-known of
the candidate explanations for quantum speedup: the so-called
\emph{many worlds explanation of quantum computation}. This
explanation of quantum computation draws its inspiration from the
many-worlds interpretation of quantum mechanics. According to this
explanation, when a quantum computer effects a transition such
as:
\begin{eqnarray}
\label{over:eqn:parallel}
\sum_{x=0}^{2^n-1} |
x \rangle | 0 \rangle \rightarrow \sum_{x=0}^{2^n-1} | x \rangle |
f(x) \rangle,
\end{eqnarray}
it literally performs, simultaneously and in different physical
worlds or universes, local function evaluations on all of the possible
values of $x$.

The many worlds explanation is, on the one hand, very attractive as an
explanation of quantum speedup. If one takes the transition
\eqref{over:eqn:parallel} at face value, i.e., as exhibiting the fact
that the quantum computer is actually physically performing, somehow,
multiple function evaluations of different values of $x$, then the
many worlds explanation directly answers the question of \emph{where}
this parallel processing is occurring (i.e., in distinct physical
universes) in a way in which other explanations do not. Thus it is,
plausibly, the most intuitive explanation of quantum speedup.

As I argue in this chapter, however, the many worlds explanation,
unlike the many worlds interpretation of quantum mechanics from which
it is inspired, cannot avail itself of many of the arguments which
appeal to decoherence as a criterion for distinguishing worlds in
order to address the so-called preferred basis objection. The
criterion for world decomposition that is adopted (as a substitute for
decoherence) by advocates of the many worlds explanation, meanwhile,
cannot fulfil this role except in an ad hoc way.

A second, perhaps more significant, problem for the many worlds
explanation is the relatively recent development of an alternative
model of quantum computation: the cluster state model. The standard
network model (also known as the `circuit' model) and the cluster
state model are computationally equivalent in the sense that one can
be used to efficiently simulate the other; but while an explanation of
the network model in terms of many worlds seems (prima facie, at
least) intuitive and plausible, this is far from being true for the
case of cluster state computation. Indeed, as I will argue, the many
worlds explanation of quantum computing is, in an important sense,
incompatible with the cluster state model.

Based on these considerations I conclude that we must reject the many
worlds explanation.

\subsubsection{Chapter \ref{ch:nec}}

Given that we must reject the popular many worlds explanation, the
question arises as to whether any of the other candidate explanations
for quantum speedup are correct. When one examines these apparently
disparate explanations, however, one finds that each of them (and the
many worlds explanation as well, in fact) include a central role for
the phenomenon of quantum entanglement. Given this, the question then
arises as to whether entanglement can be said to be a necessary
element of any candidate explanation for quantum speedup.

On the one hand, a positive answer to this question is supported by
the well known theoretical result \citep[]{jozsa2003} that when one
restricts oneself to computation over pure states, one requires a
quantum computer to be in an entangled state in order to achieve a
quantum speedup over classical computation. On the other hand it is
not clear that the same holds true for mixed states. In particular, it
seems as though it is possible to achieve a modest (sub-exponential)
speedup over classical computation using certain mixed states which
are, by definition, unentangled. Additionally, it seems as though it
is possible to achieve a substantial (i.e., exponential) speedup over
classical computation using certain mixed states that contain only a
vanishingly small amount of entanglement. In light of these results,
it is tempting to conclude that one need not appeal to entanglement
after all in order to explain quantum speedup.

Despite these purported counter-examples, I argue in this chapter that
such a conclusion is premature. In the first type of counterexample,
where sub-exponential speedup has been demonstrated with unentangled
mixed states, it can be argued, and I do argue, that when one
considers the initially mixed state of the computer as representing a
space of possible pure state preparations for the system, it is
evident that the speedup obtainable from this system stems from the
fact that the quantum computer evolves some of these possible pure
state preparations to entangled states. As for the second type of
counter-example, where exponential speedup is achieved with only a
vanishingly small amount of entanglement (thus bringing into doubt its
efficacy and thus its necessity for enabling quantum speedup), I argue
that when one considers the pure state representation of the initial
state of such a system, in which the system's correlations with the
environment are included as part of the overall description of the
system, then the role that entanglement plays in the speedup displayed
by the system is both clarified and indeed confirmed by recent
research on the physical characteristics of such systems. Since pure
states, as I also argue, represent a more fundamental representation
of quantum systems than mixed states, one should conclude that
entanglement is necessary for the speedup exhibited by such systems.

\subsubsection{Chapter \ref{ch:suff}}

If it is concluded that entanglement is a necessary component in any
explanation of quantum speedup, then the natural next question to ask
is whether it is also sufficient. In this chapter I begin to answer
this question by first asking whether entanglement can be said to be a
sufficient physical resource for enabling quantum speedup.

The answer to this question is commonly held to be no. According to
the Gottesman-Knill theorem \citep[464]{nielsenChuang2000}, any
quantum algorithm or protocol which exclusively utilises the elements
of a certain restricted set of quantum operations can be efficiently
simulated by classical means. Yet, since some of the algorithms and
protocols falling into this category involve entangled states, it is
usually concluded that entanglement cannot, therefore, be sufficient
to enable quantum speedup.

In this short chapter I argue that this conclusion is misleading. As I
explain, the quantum operations to which the Gottesman-Knill theorem
applies are precisely those which will never yield a violation of the
Bell inequalities, for they all involve rotations of the Bloch sphere
representation of the state space for a single qubit given in
multiples of $\pi/2$. It is well known, however, that the correlations
present in entangled quantum systems whose subsystems always take on
orientations with respect to one another that are multiples of $\pi/2$
are reproducible by a classical hidden variables theory. Thus it
should be no surprise that entangled quantum states which only undergo
operations in the Gottesman-Knill group of operations are efficiently
simulable by a classical computer.

What the Gottesman-Knill theorem shows us, I argue, is that one must
\emph{use} an entangled quantum state to its full potential in order
to achieve a quantum speedup; if one only utilises the portion of the
system's state space efficiently accessible by a classical system, no
speedup will be achieved, even when the system is
entangled. Nevertheless, there is a meaningful sense in which an
entangled quantum state is sufficient for quantum speedup: an
entangled quantum state \emph{provides sufficient physical resources
  to enable} quantum speedup, whether or not one elects to use these
resources to their full potential.

\subsubsection{Chapter \ref{ch:sig}}

In this chapter I address the questions of whether and in what sense
entanglement is sufficient \emph{to explain} quantum computational
speedup. I begin by distilling the argumentation of the previous
chapters into the tentative explanation for quantum speedup that I
gave above; i.e., that since the state spaces available to classical
systems are exponentially smaller than those available to quantum
systems, one requires, in general, exponentially more resources to
simulate a quantum system by classical means. I argue that this
explanation can be taken as explanatory in the following sense: just
as the essential physical characteristics of classical computational
systems can be taken, in computability theory and in computational
complexity theory, to be explanations of their computational
capabilities\textemdash of \emph{how it is possible that} such systems
are able to compute particular classes of problems using a specified
number of resources\textemdash, so can the essential physical
characteristics of \emph{quantum} computational systems be so
taken. These essential characteristics are, just as for classical
systems, the properties of the states and state transitions available
to quantum systems.

In the remainder of the chapter I argue that this candidate
explanation for quantum speedup is compatible with accounts of
physical explanation that require explanations to be causal in
nature. In particular, I consider a challenge to the view that
entanglement itself can be given a causal physical explanation: an
argument, due to \citet[]{stachel1997}, that entanglement should not
be characterised as essentially involving physical interactions, but
rather as arising from a more abstract set of requirements. I argue
that these abstract requirements themselves can be accounted for in
terms of physical interactions, and that the notion of physical
interaction involved in the description of entangled quantum systems
can therefore be made compatible with a suitably intuitive notion of
causation.

\subsection{Common chapter elements}

Chapters \ref{ch:mwi}, \ref{ch:nec}, and \ref{ch:suff} include a
``Preliminaries'' and a ``Next steps'' section, which follow upon the
chapter introduction and chapter conclusion, respectively. The
``Preliminaries'' section contains some of the technical details that
are required in order to comprehend the argumentation of the
chapter. They are placed in this section for ease of reference, as
they will often be referred to in subsequent chapters. Readers already
familiar with these technical details may skim\textemdash but not
skip\textemdash this section. The purpose of the ``Next steps''
section is to link the content of the current chapter to the subject
matter and argumentation that are to be pursued in the next.

The reader will also occasionally be referred to the appendices. These
contain more detailed discussions of various technical topics which
are useful for comprehending the overall argument of the dissertation,
but inessential to its exposition.

\section{Basic terminology and notational conventions}
\label{intro:s:term}

\paragraph{Qubit.}
A qubit is the basic unit of quantum information, analogous to a
classical bit. It can be physically realised by any two-level quantum
mechanical system. Like a bit, it can be ``on'' or ``off'', but unlike
a bit it can also be in a superposition of these values.

\paragraph{Computational basis.}
The computational, or classical, basis for a single qubit is the basis
$\{| 0 \rangle, | 1 \rangle\},$ which can be used to represent the
classical bit states $\{\uparrow,\downarrow\}$, where $| 0 \rangle =
\left (\begin{smallmatrix} 1 \\ 0 \end{smallmatrix}\right ),$ and $| 1
\rangle = \left (\begin{smallmatrix} 0 \\ 1 \end{smallmatrix}\right
).$

\paragraph{+,- basis.}
An alternative basis for representing qubits is the basis $\{| +
\rangle, | - \rangle\},$ where $| + \rangle = \frac{1}{\sqrt{2}}\left
(\begin{smallmatrix} 1 \\ 1 \end{smallmatrix}\right ) = \frac{| 0
  \rangle + | 1 \rangle}{\sqrt 2},$ and $| - \rangle =
\frac{1}{\sqrt{2}}\left (\begin{smallmatrix} 1
  \\ -1 \end{smallmatrix}\right) = \frac{| 0 \rangle - | 1
  \rangle}{\sqrt 2}.$

\paragraph{Bloch sphere.} A geometrical representation of the state
space of a single qubit. States on the surface of the sphere represent
pure states, while those in the interior represent mixed states.

\paragraph{Tensor product notation.}
For brevity, I will usually omit the tensor product symbol from
expressions for states of multi-partite systems; i.e., $| \alpha\beta
\rangle$ and $| \alpha \rangle| \beta \rangle$ should be understood as
shorthand forms of $| \alpha\rangle\otimes| \beta
\rangle$. Additionally, all of the following should be taken to be
equivalent: $$| \alpha \rangle_1 \otimes | \alpha \rangle_2 \otimes
... | \alpha \rangle_n \equiv | \alpha \rangle_1 | \alpha
\rangle_2...| \alpha \rangle_n \equiv | \alpha^n \rangle \equiv |
\alpha \rangle^n \equiv | \alpha \rangle^{\otimes n}.$$

\paragraph{Quantum gates.}
In the network model of quantum computation, logic gates are
implemented as unitary transformations. Some common gates are:
\begin{itemize}
\item the \textbf{H} or Hadamard gate, which takes $| 0 \rangle$ to
  $\frac{| 0 \rangle + | 1 \rangle}{\sqrt 2}$ and $| 1 \rangle$ to
  $\frac{| 0 \rangle - | 1 \rangle}{\sqrt 2}$ and vice-versa;
\item the \textbf{NOT} gate, implemented by the Pauli-\textbf{X}
  transformation, which takes $| 0 \rangle$ to $| 1 \rangle$ and $| 1
  \rangle$ to $| 0 \rangle$;
\item the \textbf{CNOT} or controlled-not gate. This gate takes two
  qubits $| c \rangle| t \rangle$ to $| c \rangle| t \oplus c
  \rangle$, where $| c \rangle$ is the control, $| t \rangle$ the
  target qubit, and $\oplus$ is addition modulo 2 (i.e.,
  `exclusive-or'). Intuitively, the control qubit determines whether
  or not to apply a bit-flip operation (i.e., a NOT operation) to
  the target qubit.
\end{itemize}

\paragraph{Network model of quantum computation.} Also called the
circuit model, this is the standard model of quantum computation, in
which qubits contained in quantum registers are used as inputs to
quantum gates arranged in a network structure (analogous to the
circuit model of classical computation). For instance, the following
is a network specification of the teleportation protocol (cf. Appendix
\ref{ch:tel}):
$$
\Qcircuit @C=0.7em @R=.1em @! {
  \lstick{\ket{\psi}_a} & \ctrl{1} & \gate{H} & \meter &
  \lstick{\raisebox{2.5em}{$M_1$}} \cw & \cw & \control \cw \\
  & \targ & \qw & \meter & \lstick{\raisebox{2.5em}{$M_2$}} \cw &
  \control \cw & \cwx \\
  \lstick{\raisebox{3.25em}{$\ket{\Phi^+}_{ab}$ \huge\{}} & \qw
  & \qw & \qw & \qw & \gate{X^{M_2}} \cwx & \gate{Z^{M_1}} \cwx &
  \rstick{\ket{\psi}_b} \qw
}
$$

\chapter{The Many Worlds Explanation of Quantum
  Computation\footnote{This chapter is a revised version of the
    previously published work, ``Many Worlds, the Cluster-state
    Quantum Computer, and the Problem of the Preferred Basis''
    \citep[]{cuffaro2012}. Full bibliographic details are given at the
    end of this dissertation.}}
\label{ch:mwi}

\settocdepth{subsection}

\section{Introduction}

The source of quantum computational speedup\textemdash the ability of
a quantum computer to achieve, for some problem domains,\footnote{An
  important example is the factoring problem. Factoring is in the
  complexity class \textbf{FNP}; i.e., the class of all function
  problems associated with languages in \textbf{NP}
  (cf. \citealt[\textsection 10.3]{papadim1994}, and also Appendix
  \ref{ch:cc}). It is \emph{also} in the class \textbf{BQP}, the class of
  problems solvable by a quantum computer in polynomial time, as was
  shown by \citet[]{shor1997}. The significance of the latter is that
  the quantum solution to factoring represents an exponential speedup
  over the best known classical factoring algorithm. Shor's algorithm
  has received much attention as a result of its important practical
  implications; it demonstrates, for instance, that quantum computers
  can easily break certain widely used internet encryption schemes. In
  this dissertation we will not directly discuss Shor's algorithm,
  however. For our purposes, no generality is lost, and ease of
  comprehension is gained, by focusing on simpler algorithms such as
  the Deutsch-Jozsa algorithm.} a dramatic reduction in processing
time over any known classical algorithm\textemdash is still a matter
of debate. On one popular view (the `quantum parallelism
thesis'\footnote{I am indebted to \citet[]{duwell2007} for this
  label.}), the speedup is due to a quantum computer's ability to
simultaneously evaluate (using a single circuit) a function for many
different values of its input. Thus one finds, in textbooks on quantum
computation, pronouncements such as the following:

\begin{quote}
[a] qubit can exist in a superposition of states, giving a quantum
computer a hidden realm where exponential computations are possible
... This feature allows a quantum computer to do parallel computations
using a single circuit\textemdash providing a dramatic speedup in many
cases \citep[p. 197]{mcmahon2008}.
\end{quote}

\begin{quote}
Unlike classical parallelism, where multiple circuits each built to
compute $f(x)$ are executed simultaneously, here a \emph{single}
$f(x)$ circuit is employed to evaluate the function for multiple
values of $x$ simultaneously, by exploiting the ability of a quantum
computer to be in superpositions of different states
\citep[p. 31]{nielsenChuang2000}.
\end{quote}

Among textbook writers, N. David Mermin is, perhaps, the most cautious
with respect to the significance of this `quantum parallelism':

\begin{quote}
One cannot say that the result of the calculation \emph{is} 2$^n$
evaluations of $f$, though some practitioners of quantum computation
are rather careless about making such a claim. All one can say is that
those evaluations characterize the \emph{form} of the state that
describes the output of the computation. One knows what the state
\emph{is} only if one already knows the numerical values of all those
2$^n$ evaluations of $f$. Before drawing extravagant practical, or
even only metaphysical, conclusions from quantum parallelism, it is
essential to remember that when you have a collection of Qbits in a
definite but unknown state, \emph{there is no way to find out what
  that state is} \citeyearpar[p. 38]{mermin2007}.
\end{quote}

Mermin's reservations notwithstanding, the quantum parallelism thesis
is frequently associated with (and held to provide evidence for) the
many worlds explanation of quantum computation, which draws its
inspiration from the Everettian interpretation of quantum
mechanics. According to the many worlds explanation of quantum
computing, when a quantum computer effects a transition such as:
\begin{eqnarray}
\label{mwi:eqn:parallel}
\sum_{x=0}^{2^n-1} |
x \rangle | 0 \rangle \rightarrow \sum_{x=0}^{2^n-1} | x \rangle |
f(x) \rangle,
\end{eqnarray}
it literally performs, simultaneously and in different physical
worlds, local function evaluations on all of the possible values of
$x$.

It is all well to say that a quantum computer evaluates a function
simultaneously for many different values of its domain; but one should
also give some physical explanation of \emph{how} this occurs. The
many worlds explanation attempts to do just that; it directly answers
the question of \emph{where} this parallel processing occurs: in
distinct physical universes. For this reason it is also, arguably, the
most intuitive physical explanation of quantum speedup. Indeed, for
some, the many worlds explanation of quantum computing is the only
possible physical explanation of quantum speedup. David Deutsch, for
instance, writes: ``no single-universe theory can explain even the
Einstein-Podolsky-Rosen experiment, let alone, say, quantum
computation. That is because \emph{any} process (hidden variables, or
whatever) that accounts for such phenomena ... contains many
autonomous streams of information, each of which describes something
resembling the universe as described by classical physics''
\citeyearpar[p. 542]{deutsch2010}. Deutsch issues a challenge to those
who would explain quantum speedup without many worlds: ``[t]o those
who still cling to a single-universe world-view, I issue this
challenge: Explain how Shor's algorithm works''
\citeyearpar[p. 217]{deutsch1997}.

Recently, the development of an alternative model of quantum
computation\textemdash the cluster state model\textemdash has cast
some doubt on these claims. The standard network model (which I will
also refer to as the `circuit' model) and the cluster state model are
computationally equivalent in the sense that one can be used to
efficiently simulate the other; however, while an explanation of the
network model in terms of many worlds seems intuitive and plausible,
it has been pointed out by \citet[pp. 474-475]{steane2003}, among
others, that it is by no means natural to describe cluster state
computation in this way.

While Steane is correct, I will argue that the problem that the
cluster state model presents to the many worlds explanation of quantum
computation runs deeper than this. I will argue that the many worlds
explanation of quantum computing is not only unnatural as an
explanation of cluster state quantum computing, but that it is, in
fact, incompatible with it.\footnote{My use of the word `incompatible'
  might strike some readers as a touch strong. I do not mean to convey
  by this any in-principle impossibility, however. Rather, I take it
  that any worthwhile explanation of a process should provide some
  useful insight into its workings, and should be motivated by the
  characteristics of the process, not by predilections for a
  particular type of explanation on the part of the explainer. My
  claim here is that, as I will show below, a many worlds explanation
  of cluster state quantum computing is completely unmotivated and
  useless even as a heuristic device for describing cluster state
  quantum computation, and is in this sense incompatible with it. One
  might call this type of incompatibility `for-all-practical-purposes
  incompatibility'. Since, as we shall see later, the criterion used
  for identifying worlds on the many worlds explanation of quantum
  computation is a for-all-practical-purposes criterion, this is just
  the right sort of incompatibility that must prove problematic for
  the many worlds explanation.} I will show how this incompatibility
is brought to light through a consideration of the familiar preferred
basis problem, for a preferred basis with which to distinguish the
worlds inhabited by the cluster state neither emerges naturally as the
result of a dynamical process, nor can be chosen a priori in any
principled way.

In addition, I will argue that the many worlds explanation of quantum
computing is inadequate as an explanation of even the standard network
model of quantum computation. This is because, first, unlike its close
cousin, the neo-Everettian many worlds interpretation of quantum
\emph{mechanics},\footnote{One should be wary not to treat the
  `Everettian' interpretation of quantum mechanics as if it were a
  unified view. Rather, `Everettian' more properly describes a
  family of views (see \citealt[]{barrett2011} for a list and
  discussion of these), which includes but is not limited to Hugh
  Everett's original formulation \citep[]{everett1957}, `many minds'
  variants \citep[]{albert1988}, and `many worlds'
  variants. Belonging to the last named class are DeWitt's
  \citeyearpar{dewitt1971} original formulation, as well as the, now
  mainstream, `neo-Everettian' interpretation with which we will be
  mostly concerned in this chapter. I follow Hewitt-Horsman (who
  attributes the name to Harvey Brown) in calling `neo-Everettian'
  the amalgam of ideas of
  \citet[]{zurek2003,saunders1995,butterfield2002,vaidman2008}, and
  especially \citet[]{wallace2002,wallace2003,wallace2010}.} where the
\emph{decoherence} criterion is able to fulfil the role assigned to
it, of determining the preferred basis for world decomposition with
respect to macro experience,\footnote{I should not be interpreted here
  as giving an argument \emph{for} the neo-Everettian interpretation
  of quantum mechanics. My views on the correct interpretation of
  quantum mechanics are irrelevant to this discussion. My claim is
  only that the decoherence basis is prima facie well-suited for the
  role it plays in the neo-Everettian interpretation.} the
corresponding criterion for world decomposition in the context of
quantum computing cannot fulfil this role except in an ad hoc
way. Second: alternative explanations of quantum computation exist
which, unlike the many worlds explanation, are compatible with both
the network and cluster state model.

The quantum parallelism thesis, and the many worlds explanation of
quantum computation that is so often associated with it, are
undoubtedly of great heuristic value for the purposes of algorithm
analysis and design, at least with regard to the network model. This
is a fact which I should not be misunderstood as disputing. What I am
disputing is that we should therefore be committed to the claim that
these computational worlds are, in fact, ontologically real, or that
they are indispensable for any explanation of quantum speedup.

The chapter will proceed as follows. I begin, in \textsection
\ref{mwi:s:djalgo}, with an example, often used to motivate the quantum
parallelism thesis and the associated many worlds explanation, of a
simple quantum algorithm specified using the network model of quantum
computation. In \textsection \ref{mwi:s:neo}, I argue that, despite its
intuitive appeal, the many worlds view of quantum computation is not
licensed by, and in fact is conceptually inferior to, the
neo-Everettian version of the many worlds interpretation of quantum
mechanics from which it receives its inspiration. In \textsection
\ref{mwi:s:clu}, I describe the cluster state model of quantum
computation and show how the cluster state model and the many worlds
explanation are incompatible. In \textsection \ref{mwi:s:net} I argue,
based on the conclusions of \textsection \ref{mwi:s:neo} and
\textsection \ref{mwi:s:clu}, that we should reject the many worlds
explanation of quantum computation.

\section{Preliminaries: A simple quantum algorithm}
\label{mwi:s:djalgo}

Deutsch's problem \citep[]{deutsch1985} is the problem to determine
whether a boolean function taking one bit as input and producing one
bit as output (i.e., $f:\{0,1\} \rightarrow \{0,1\},$) is either
constant or balanced. Such a function is constant if it produces the
same output value for each of its possible inputs. For the functions
$f:\{0,1\} \rightarrow \{0,1\}$, the only possible constant functions
are $f(x) = 0$ and $f(x) = 1$. A balanced function, on the other hand,
is one for which the output of one half of the inputs is the opposite
of the output of the other half. For the functions $f:\{0,1\}
\rightarrow \{0,1\}$, the only possible balanced functions are the
identity and bit-flip functions. These are, respectively:

\begin{tabular}{p{6.5cm} p{6.5cm}}
$$
\begin{array}{l}
f(x) = \left\{ 
\begin{array}{ll}
0 & \mbox{if } x = 0 \\
1 & \mbox{if } x = 1, \\
\end{array} \right. \\
\end{array}
$$
& 
$$
\begin{array}{l}
f(x) = \left\{ 
\begin{array}{ll}
1 & \mbox{if } x = 0 \\
0 & \mbox{if } x = 1. \\
\end{array} \right. \\
\end{array}
$$
\\
\end{tabular}

A generalisation of Deutsch's problem, called the Deutsch-Jozsa
problem, enlarges the class of functions under consideration so as to
include all of the functions $f:\{0,1\}^n\to\{0,1\}$. Classically, the
only way to determine whether an arbitrary function from this class is
balanced or constant is to test the function for each of its possible
input values. In a quantum computer, however, we can learn whether
such a function is balanced or constant in (neglecting overhead)
\emph{one} computational step. The quantum solution to the
Deutsch-Jozsa problem is given by the Deutsch-Jozsa algorithm, which I
present here in the improved version due to \citet[]{cleve1998}.

The algorithm begins by initialising the registers of a quantum
computer to $| 0^n \rangle| 1 \rangle$, after which a Hadamard gate is
applied to all $n + 1$ qubits, so that:
\begin{align}
| 0^n \rangle| 1 \rangle & \xrightarrow{H} \left(\frac{1}{2^{n/2}}(| 0
\rangle + | 1 \rangle)^n \right )\left(\frac{| 0 \rangle - | 1
  \rangle}{\sqrt 2}\right) \nonumber \\
\label{mwi:eqn:puredj_pre}
& = \left (\frac{1}{2^{n/2}}\sum_x^{2^n-1}| x \rangle
\right)\left(\frac{| 0 \rangle - | 1 \rangle}{\sqrt 2}\right).
\end{align}
The unitary transformation,
\begin{equation}
\label{mwi:eqn:puredj_uni}
U_f(| x \rangle | y \rangle) \equiv | x \rangle | y \oplus f(x)
\rangle ,
\end{equation}
representative of the function whose character (of being either
constant or balanced) we wish to determine, is then applied, which has
the effect:\footnote{Given the state $| x \rangle(| 0 \rangle - | 1
  \rangle)$ (omitting normalisation factors for simplicity), note
  that when $f(x)=0$, applying $U_f$ yields $| x \rangle(| 0 \oplus
  0 \rangle - | 1 \oplus 0 \rangle) = | x \rangle(| 0 \rangle - | 1
  \rangle)$; and when $f(x) = 1$, applying $U_f$ yields $| x
  \rangle(| 0 \oplus 1 \rangle - | 1 \oplus 1 \rangle) = | x
  \rangle(| 1 \rangle - | 0 \rangle) = -| x \rangle(| 0 \rangle - |
  1 \rangle)$.}
\begin{equation}
\label{mwi:eqn:puredj}
\xrightarrow{U_f} \left (\frac{1}{2^{n/2}}\sum_x^{2^n-1}(-1)^{f(x)}| x
\rangle \right )\left(\frac{| 0 \rangle - | 1 \rangle}{\sqrt
  2}\right).
\end{equation}
Note how the action of the unitary transformation gives the appearance
of evaluating the function over multiple inputs at once.

If $f$ is constant and $= 0$, this, along with a Hadamard
transformation applied to the first $n$ qubits, will result in:
\begin{align*}
f=0: & & \left (\frac{1}{2^{n/2}}\sum_x^{2^n-1}| x \rangle \right )| -
\rangle \xrightarrow{H^n \otimes I} | 0^n \rangle | - \rangle,
\end{align*}
where $| - \rangle \equiv \frac{| 0 \rangle - | 1 \rangle}{\sqrt
  2}$. Otherwise if $f$ is constant and $=1$, then this, along with a
Hadamard transformation applied to the first $n$ qubits, will result
in:
\begin{align*}
f=1: & & -\left (\frac{1}{2^{n/2}}\sum_x^{2^n-1}| x \rangle \right )|
- \rangle \xrightarrow{H^n \otimes I} -| 0^n \rangle | - \rangle.
\end{align*}
In either case, a measurement in the computational basis on the first
$n$ qubits yields the bit string $z = 000 \ldots 0 = 0^n = 0$ with
certainty. If $f$ is balanced, on the other hand, then half of the
terms in the superposition of values of $x$ in \eqref{mwi:eqn:puredj} will
have positive phase, and half negative. After applying the final
Hadamard transform, the amplitude of $| 0^n \rangle$ will be
zero.\footnote{\label{mwi:f:int} To illustrate, consider the case
  where $n=2$. After applying $U_f$, the computer will be in the
  state: $(| 00 \rangle - | 01 \rangle + | 10 \rangle - | 11
  \rangle)| - \rangle.$ Applying a Hadamard transform to the two
  input qubits will yield:
  \begin{eqnarray*}
  & & \Big((| 00 \rangle + | 01 \rangle + | 10 \rangle + | 11 \rangle) -
  (| 00 \rangle - | 01 \rangle + | 10 \rangle - | 11 \rangle) \\
  & + & (| 00 \rangle + | 01 \rangle - | 10 \rangle - | 11 \rangle) -
  (|00 \rangle - | 01 \rangle - | 10 \rangle + | 11 \rangle)\Big)| -
  \rangle \\
  & = & (0| 00 \rangle + \ldots)| - \rangle.
  \end{eqnarray*}
}
Thus a measurement of these qubits \emph{cannot} produce the bit
string $z = 000 \ldots 0 = 0^n = 0.$ In sum, if the function is
constant, then $z = 0$ with certainty, and if the function is
balanced, $z \neq 0$ with certainty. In either case, the probability
of success of the algorithm is 1, using only a \emph{single}
invocation. This is exponentially faster than any known classical
solution.

\section{Neo-Everett and quantum computing}
\label{mwi:s:neo}

Algorithms like the Deutsch-Jozsa algorithm provide strong intuitive
support for the view that quantum speedup is due to a quantum
computer's ability to simultaneously evaluate a function for different
values of its input, and from here it is not a large step to the many
worlds explanation of quantum computation. It is important to note,
however, that one's conception of a world, if one elects to take this
step, cannot be the one that is licensed by the neo-Everettian many
worlds interpretation of quantum mechanics. In superpositions such as
the following, $$\frac{1}{\sqrt 2}(|\alpha\rangle\otimes |\beta\rangle
+ |\gamma\rangle\otimes |\delta\rangle),$$ the neo-Everettian
interpretation will not, in general, license one to identify each term
of this superposition with a distinct world, for such a simplistic
procedure for world-identification will be vulnerable to the so-called
preferred basis objection.

The problem is usually formulated in the context of macro-worlds and
macro-objects; however we can illustrate the basic idea by means of
the following simple example related to quantum computation. The
classical value $\uparrow$ can be represented, in the computational
basis, by a qubit in the state $| 0 \rangle$. We can also represent
the same qubit from the point of view of the $\{| + \rangle, | -
\rangle\}$ basis, however, as\footnote{Since $| + \rangle =
  \frac{1}{\sqrt{2}}\left (\begin{smallmatrix} 1
  \\ 1 \end{smallmatrix}\right ) = \frac{1}{\sqrt 2}(| 0 \rangle +
  | 1 \rangle)$ and $| - \rangle = \frac{1}{\sqrt{2}}\left
  (\begin{smallmatrix} 1 \\ -1 \end{smallmatrix}\right) =
  \frac{1}{\sqrt 2}(| 0 \rangle - | 1 \rangle)$, $\frac{1}{\sqrt
  2}(| + \rangle + | - \rangle) = \frac{1}{2}(| 0 \rangle + | 1
  \rangle + | 0 \rangle - | 1 \rangle) = \frac{1}{2}\cdot 2 | 0
  \rangle = | 0 \rangle$.}

$$\frac{1}{\sqrt{2}} (| + \rangle + | - \rangle).$$

Thus depending on the basis one selects, it will be possible to regard
the qubit as either (if we select the computational basis) in the
definite state $| 0 \rangle$, existing in one world only, or (if we
select the $\{| + \rangle, | - \rangle\}$ basis), as in a
superposition of the two states, $| + \rangle$ and $| - \rangle$, and
thus as existing in two distinct worlds. Yet there seems to be no a
priori reason why we should elect to choose one basis over the other.

Neo-Everettians (see, for instance,
\citealt[]{wallace2002,wallace2003}) attempt to eliminate the
preferred basis problem by appealing to the dynamical process of
decoherence (\emph{cf.} \citealt[]{zurek2003}) as a way of
distinguishing different worlds from one another in the wave
function. Recall that Schr\"odinger's wave equation governs the
evolution of a closed system. In nature, however, there are no closed
systems (aside from the entire universe); all systems interact, to
some extent, with their environment. When this happens, the terms in
the superposition of states representing the system decohere and
branch off from one another. From the point of view of an observer in
a particular world, this gives the appearance of wave-function
collapse\textemdash of definiteness emerging from
indefiniteness\textemdash but unlike actual collapse (i.e., collapse
as per von Neumann's projection postulate), decoherence is an
approximate phenomenon; thus some small amount of residual
interference between worlds always remains. But from the point of view
of our experience of macroscopic objects, this is, for all practical
purposes, enough to give us the appearance of definiteness within our
own world and to distinguish, within the wave-function, macroscopic
worlds that evolve essentially independently and maintain their
identities over time. Thus, a `preferred' basis with which one can
define different worlds emerges \emph{naturally}: ``the basic idea is
that dynamical processes cause a preferred basis to emerge rather than
having to be specified a priori'' \citep[p. 90]{wallace2003}.

On the neo-Everettian view, we identify patterns which are present in
the wave-function and which are more or less stable over time in this
way with macroscopic objects such as measurement pointers, cats, and
experimenters. But note that not every such pattern is granted
ontological status; whether or not we do so depends, not just on the
process of decoherence, but also on the theoretical usefulness of
including that object in our ontology: ``the existence of a pattern as
a real thing depends on the usefulness\textemdash in particular, the
explanatory power and predictive reliability\textemdash of theories
which admit that pattern in their ontology''
\citep[p. 93]{wallace2003}. Thus, while decoherence \emph{is} a
necessary condition for granting ontological status to a pattern, it
is not sufficient; we \emph{also} require that doing so is
theoretically useful and fruitful.

Returning to the quantum computer, it should be clear by now that the
neo-Everettian interpretation, as described above, cannot provide
support for the view that quantum computers simultaneously evaluate
functions for different values of their input \emph{in different
  worlds}, for as we have just seen, \emph{decoherence} determines the
basis according to which we distinguish one world from another on the
neo-Everettian interpretation. The superpositions characteristic of
quantum algorithms, however, are always \emph{coherent}
superpositions. Indeed, the maximum length of a quantum computation is
directly related to the amount of time that the system remains
coherent \citep[p. 278]{nielsenChuang2000}. According to some, in
fact, it is coherence and not parallel processing which is the real
source of quantum speedup \citep[]{fortnow2003}. Decoherence, in the
context of quantum computation, effectively amounts to noise.

It appears, then, that we require a more general criterion for
branching than decoherence if we are to accommodate quantum
computation to a many worlds picture. Thus the many worlds advocate,
\citet[]{hewittHorsman2009}, for instance, rejects the idea that
decoherence is the only possible criterion for distinguishing
worlds. Worlds, for Hewitt-Horsman, are (just as in the neo-Everettian
approach), defined as substructures within the wave-function that `for
all practical purposes' are distinguishable and stable over relevant
time scales. With regards to macro experience these relevant time
scales are long, and the point of using decoherence as an identifying
criterion for distinct worlds, according to Hewitt-Horsman, is that it
is useful for identifying stable macro-patterns over such long time
scales. But the time scales relevant to quantum computation are
generally much shorter: ``they may, indeed, be \emph{de facto}
instantaneous. However, if they are useful then we are entitled to use
them'' \citep[p. 876]{hewittHorsman2009}.

In such a situation we may, according to Hewitt-Horsman, consider
coherent superpositions as representing distinct worlds for the
purposes of characterising quantum computation. ``Defining worlds
within a coherent state in this way is a simple extension of the
FAPP$^[$\footnote{FAPP stands for `for all practical purposes'.}$^]$
principle ... If our practical purposes allow us to deal with rapidly
changing worlds-structures then we may''
\citep[p. 876]{hewittHorsman2009}. As for the preferred basis problem,
it will not arise. Just as with the neo-Everettian interpretation, in
the quantum computer we have a criterion for selecting a basis with
which to decompose the wave function; in this case the basis is that
in which the different evaluations of the function are made manifest,
i.e., the computational basis.

\begin{quote}
This fits in well with intuitions that are often expressed about the
nature of quantum computations ... There are frequently statements to
the effect that it \emph{looks like} there are multiple copies of
classical computations happening within the quantum state. If one
classical state from a decomposition of the (quantum) input state is
chosen as an input, then the computation runs in a certain way. If the
quantum input state is used then it looks as if all the classical
computations are somehow present in the quantum one. ... the
recognition of multiple worlds in a coherent states [\emph{sic.}]
seems both to be a natural notion for a quantum information theorist,
and also a reasonable notion in any situation where `relevant'
time-scales are short \citep[p. 876]{hewittHorsman2009}.
\end{quote}

Certainly it does look as if the computation is composed of many
processes executing in parallel, and plausibly it can be of some
heuristic value to think of these processes as taking place in many
worlds. With this I do not disagree. However, \emph{pace}
Hewitt-Horsman, I do not believe this is enough to justify treating
these worlds as ontologically real, for unlike the criterion of
decoherence with respect to macro experience, Hewitt-Horsman's
criterion for distinguishing worlds in the context of quantum
computation seems quite ad hoc. Declaring that the preferred basis is
the one in which the different function evaluations are made manifest
is like declaring that the preferred basis with respect to macro
experience is the one in which we can distinguish classical states
from one another. But it is, in fact, a rejection of such reasoning
that leads to decoherence as a criterion for world-identification in
the first place. The decoherence basis, on the neo-Everettian view, is
not simply picked from among many possible bases as the one which
serves to capture our experience of definiteness at the
macro-level. To do so would be to commit the same sin (by
neo-Everettian lights) that is committed by other interpretations of
quantum mechanics such as Bohmian mechanics or GRW theory. This is the
sin of adding extra elements to the formalism of quantum theory in
order to preserve classicality at the macroscopic level. For the
neo-Everettian, in contrast, decoherence is appealed to as a known
physical process that \emph{in fact} gives rise to\textemdash and even
then only approximately\textemdash the appearance of distinct
classical worlds \citep[cf.][pp. 55, 63-65]{wallace2010}. The point of
using decoherence as a criterion for distinguishing worlds is not to
save the appearance of classicality, but rather to \emph{explain} why
we experience the world classically, in this case by appealing to a
physical process that gives rise to our experience. The choice of the
computational basis as the basis within which different worlds are to
be distinguished, however, fulfils no such explanatory role. It does
not serve to explain the appearance of parallel classical
computation. It only declares, based on a particular privileged
description of the computation, that parallel computation is occurring
in many worlds.\footnote{I should mention that Wallace, who I am
  taking as representative of the neo-Everettian interpretation of
  quantum mechanics, does seem to cautiously endorse a many worlds
  explanation for \emph{some} quantum algorithms: ``There is no
  particular reason to assume that \emph{all} or even \emph{most}
  interesting quantum algorithms operate by any sort of `quantum
  parallelism' ... But Shor's algorithm, at least, does seem to
  operate in this way'' \citep[p. 70, n. 17]{wallace2010}. Wallace has
  also made similar remarks in informal correspondence. But whatever
  Wallace's views on quantum computation are, they are obviously
  separable from his views on world decomposition for
  macro-phenomena.}

An advocate of the many worlds explanation might make the following
rejoinder: the computational process, considered as a whole, is just
as empirically well-established as the decoherence process is (we know
that a computation has taken place since we have the result). And just
as the decoherence process gives rise to parallel autonomously
evolving decoherent worlds which are (approximately) diagonal in the
decoherence basis, the computational process gives rise to parallel
autonomous computational worlds which are diagonal (at least at the
beginning of the computation) in the computational basis. Thus the
computational process gives rise to and therefore explains the
computational worlds that make up the computation just as well as the
decoherence process explains the decoherent worlds that make up
classical experience.

This response is problematic, however, for it is the computation
itself, in particular what distinguishes it from classical
computation, that we are seeking an explanation for. The many worlds
explanation of quantum computation promises to explain quantum
computation in terms of many worlds, but on this response it appears
that we need to appeal to the computation in order to explain these
many worlds in the first place. This seems circular, and even if the
case can be made that it is not, the response fails to consider that,
as the quote from Mermin with which I began this chapter makes clear,
appearances can be misleading: we must be very cautious when
describing the quantum state characterising a computation. In
particular, we must be cautious when inferring from the form of the
state that describes the computation to the content of that state. For
instance, as Steane \citeyearpar[p. 473]{steane2003} has pointed out,
according to the Gottesman-Knill theorem,\footnote{We will discuss the
Gottesman-Knill theorem in further detail in Chapter \ref{ch:suff}.}
an important class of quantum gates\textemdash the so-called
Clifford-group gates, which include the Hadamard, Pauli, and CNOT
gates\textemdash can be simulated in polynomial time by a classical
probabilistic computer \citep[p. 464]{nielsenChuang2000}. This is
interesting, since several quantum algorithms utilise gates
exclusively from this class. Thus the appearance of quantum
parallelism, in these cases at least, may be deceiving.

Even if true, the quantum parallelism thesis need not entail the
existence of autonomous local parallel computational
processes. \citet[p. 1008]{duwell2007}, for instance, illustrates this
by showing how the phase relations between the terms in a system's
wave function are crucially important for an evaluation of its
computational efficiency. Phase relations between terms in a system's
wave function, however, are global properties of the system. Thus we
cannot view the computation as consisting exclusively of \emph{local}
parallel computations (within multiple worlds or not). But if we
cannot do so, then there is no sense in which quantum parallelism
uniquely supports the many worlds explanation over other
explanations.

Everettian varieties such as the neo-Everettian interpretation of
quantum mechanics and the many worlds explanation of quantum computing
take the branching process \emph{seriously}: they claim ontological
significance for the `worlds' that arise from this process. They are
thus required to confront the preferred basis problem, for they must
determine a criterion for branching. While the neo-Everettian
interpretation of quantum \emph{mechanics} does this admirably well,
the many worlds explanation of quantum \emph{computing}, I have
argued, does not.

Before concluding this section, I should note that not all Everettian
varieties do take branching seriously (in fact, this may have been
true of Everett's own view; see \citealt[]{barrett2011} for a
discussion). Such views are not confronted with the preferred basis
problem and are thus immune to the objections above. However, since
branching is not a real physical process on such views, it is analytic
that they can provide no physical explanation for the quantum
computational process in terms of branching computational worlds. As
an illustration, an Everettian might insist\footnote{I am indebted to
  an anonymous reviewer at the journal \emph{Studies in History and
    Philosophy of Modern Physics} for pointing this out.} that the way
in which one chooses to express the state of a system has no
particular significance. On such an interpretation, one should not
view the universe as having any one particular branching
structure. Rather, the essential point is that in any process such as
quantum computation, the fact is that the Schr\"odinger evolution of
\emph{all} quantum superpositions has been realised. On such a view,
\emph{however} one chooses to decompose the state of a system, the
resulting superposition must be viewed as real. Thus the superposition
of the quantum computational process, as expressed in the
computational basis, is realised, just as is the superposition as
expressed in some other basis.

On this interpretation, however, it cannot be the case that multiple
local parallel computational processes in many worlds are \emph{the
  physical  explanation} for quantum speedup; for any decomposition of
the state of the computer in any given basis can provide the ground
for an equally legitimate `explanation' of the computer's
operation. Rather, we should say that any such decomposition
constitutes, for one who finds Everettian language appealing, a
legitimate \emph{description} of the process. I do not wish to be
misunderstood as attempting to deny to those who find Everettian
language appealing the possibility of, when appropriate, describing
the operation of the quantum computer in this way. But again, this
does not constitute a \emph{physical explanation}.

In any case, the questionable nature of the inference from the
heuristic value of the notion of computational worlds to the
ascription of ontological reality to these worlds is one good reason
to, at the very least, be suspicious of the many worlds explanation of
quantum computing. But let us, for the sake of argument, grant the
inference. Let us focus, instead, on the antecedent clause of the
conditional; i.e., on whether it really is true that the many worlds
description of quantum computation is the most useful one
available. In the next section I will examine the recently developed
cluster state model of quantum computation. I will argue that a
description of the cluster state model in terms of many worlds is, not
only unnatural, but that such a description is incompatible with the
cluster state model. I will then argue that this undermines the
usefulness of the many worlds description of quantum computation, not
just in the cluster state model, but in general.

\section{Cluster state quantum computing}
\label{mwi:s:clu}

On the cluster state model
\citep[]{raussendorf2002,raussendorf2003,nielsen2006} of quantum
computation, computation proceeds by way of a series of single qubit
measurements on a highly entangled multi-qubit state known as the
cluster state.\footnote{For this reason the model has also been given
  the name `measurement based computation'.} The cluster-state quantum
computer ($QC_\mathcal{C}$) is a universal quantum computer; it can
efficiently simulate any algorithm developed within the network
model. In fact it is computationally equivalent to the network model
in the sense that each model may be used to simulate the operation of
the other. Each qubit in the cluster has a reduced density operator of
$\frac{1}{2}\mathbf{I},$ and thus individual qubit measurement
outcomes are completely random. It is nevertheless possible to process
information on the cluster state quantum computer due to the fact that
strict correlations exist between measurement outcomes. These
correlations are progressively destroyed as the computation runs its
course.\footnote{This gives rise to a third name for this model:
  `one-way computation'.}

It is helpful to illustrate the operation of the $QC_\mathcal{C}$ by
exhibiting the way one may use it to simulate a network-based quantum
algorithm. In the network model, single-qubit gates can, in general,
be thought of as rotations of the Bloch sphere (for example, the Pauli
$X$, $Y$, and $Z$ gates can be thought of as rotations of the Bloch
sphere through $\pi$ radians about the $x$, $y$, and $z$ axes,
respectively). It is possible to simulate an \emph{arbitrary} rotation
of the Bloch sphere with the $QC_\mathcal{C}$ by using a chain of 5
qubits as follows (\emph{cf.} \citealt[pp. 446-447]{raussendorf2002},
\citealt[p. 5]{raussendorf2003}). First, we consider the Euler
representation of an arbitrary rotation.\footnote{The Euler
  representation is a way to represent the general rotation of a body
  in three dimensions. The procedure to achieve such a general
  rotation consists of three steps: a rotation of the body about one
  of its coordinate axes, followed by a rotation about a coordinate
  axis different from the first, and then a rotation about a
  coordinate axis different from the second. We represent rotations by
  Rotation operators, and matrix multiplication is used to represent
  combinations of rotations. For example, a rotation of $\alpha$ about
  $\hat{z}$ followed by a rotation of $\beta$ about $\hat{y}$ followed
  by a rotation of $\gamma$ about $\hat{x}$ is represented by
  $R_x(\gamma)R_y(\beta)R_z(\alpha)$. The analogue of the rotation
  operator in a complex state space is the unitary operator.} This is
\begin{eqnarray}
\label{mwi:eqn:euler}
U_{Rot}[\xi,\eta,\zeta] = U_x[\zeta]U_z[\eta]U_x[\xi],
\end{eqnarray}
where the rotations about the $x$ and $z$ axes are given by

\begin{eqnarray}
U_x[\alpha] & = & \mbox{exp}\left (-i\alpha\frac{\sigma_x}{2}\right ),
\\
U_z[\alpha] & = & \mbox{exp}\left (-i\alpha\frac{\sigma_z}{2}\right ).
\end{eqnarray}

The first qubit in the chain is called the input qubit; it will
contain the state that we wish to rotate. It is thus prepared in the
state $|\psi_{in}\rangle,$ while the other four qubits in the chain
are prepared in the $|+\rangle$ state. After applying an
entanglement-generating unitary transformation to the
qubits,\footnote{The procedure for generating entanglement is
  described in \citep[pp. 3-4]{raussendorf2003}.} the first four
qubits are measured one by one in the following way. We begin by
measuring qubit 1 in basis $\mathcal{B}_1(0),$ where 0 is the
measurement angle, $\phi_j$, and the basis is calculated as
\begin{eqnarray}
\label{mwi:eqn:basiseval}
\mathcal{B}_j(\phi_j) =
\left \{\frac{|0\rangle_j +
e^{i\phi_j}|1\rangle_j}{\sqrt{2}},\frac{|0\rangle_j -
e^{i\phi_j}|1\rangle_j}{\sqrt{2}}\right \}.
\end{eqnarray}
The result of this measurement is denoted $s_1,$ where $s_j\in
\{0,1\}$ represents the result of measuring the $j^{th}$ qubit.

We now use $s_1$ to calculate the measurement basis for qubit 2, which
is $\mathcal{B}_2(-\xi(-1)^{s_1})$. Qubit 2 is then measured in this
basis and the result recorded in $s_2$, which is then used to
determine the measurement basis for qubit 3:
$\mathcal{B}_3(-\eta(-1)^{s_2}).$ We then use both $s_1$ and $s_3$ to
determine the basis to use for the measurement of qubit 4:
$\mathcal{B}_4(-\zeta(-1)^{s_1 + s_3}).$ At the end of this process,
the output of the `gate' is contained in qubit 5 (i.e., qubit 5 is in
a state that is equivalent to what would have resulted if we had
applied an actual rotation to $| \psi_{in} \rangle$), which we then
read off in the computational basis.\footnote{I have simplified this
  procedure slightly. The gate simulation actually realises, not
  exactly $U_{Rot}$, but $U'_{Rot}[\xi,\eta,\zeta] =
  U_{\Sigma,Rot}U_{Rot}[\xi,\eta,\zeta]$, where $U_{\Sigma,Rot} =
  \sigma_{x}^{s_2 + s_4}\sigma_z^{s_1 + s_3}$ is called the random
  byproduct operator and is corrected for at the end of the
  computation \citep[p. 5]{raussendorf2003}.}

Similarly, it is possible to implement more specific 1-qubit rotations
such as the Hadamard, $\pi/2$-phase, $X$,$Y$, and $Z$ gates. 2-qubit
gates, such as the CNOT gate, can be implemented using similar
techniques \citep[pp. 4-5]{raussendorf2003} and we can combine all of
these gates together in order to simulate an arbitrary network.

To illustrate the general operation of the cluster state computer,
imagine, once again, that we are simulating a network-based quantum
algorithm. In each individual gate simulation there will be, on the
one hand, those qubits whose measurement depends on the outcomes of
one or more previous measurements for the determination of their
basis, and on the other hand, those that do not. We divide these
qubits into disjoint subsets, $Q_t$, of the cluster $\mathcal{C}$, as
follows. All qubits, regardless of which gate they belong to, which do
not require a previous measurement for the determination of their
basis are added to the class $Q_0$. We then add to $Q_1$ all qubits
which depend solely on the results of measuring qubits in $Q_0$ for
the determination of their basis. $Q_2$ comprises, in turn, all qubits
which depend on the results of measuring qubits in $Q_0 \cup Q_1$ for
the determination of their basis. And so on until we reach
$Q_{t_{max}}$.

We then begin by measuring the qubits in the set $Q_0$. We use the
outcomes of these measurements to determine the measurement bases for
the qubits to be measured in $Q_1$. Once these are measured, the
outcomes of $Q_0$ and $Q_1$ together are used to determine the
measurement bases for $Q_2$. The process continues in this fashion
until all the required qubits have been measured
\citep[p. 19]{raussendorf2003}. Note that the temporal ordering of
measurements on the cluster state will, in general, not depend on what
role\textemdash input, output, etc.\textemdash qubits have with
respect to the network model. In fact, those qubits that play the role
of gates' `output registers' will typically be among the first to be
measured \citep[p. 19]{raussendorf2003}. In general, the temporal
ordering of measurements on a $QC_\mathcal{C}$ that has been designed
to simulate a network does not mirror the temporal ordering the gates
would have had if they had been implemented as a network
\citep[p. 444]{raussendorf2002}.

At this point we must ask ourselves whether it is possible to describe
the cluster state model using a many worlds ontology. At first glance
there does not seem to be anything barring such a description in
principle. We might view each of the qubits as existing simultaneously
in multiple worlds, for example, while the computation is being
performed. But even if this were possible, it is difficult to see what
would be gained by such a description, for this is neither a natural
view of what is happening, nor a particularly useful one: in the
network model it seems natural to conceive of a unitary gate as
effecting a parallel computation by means of a transformation such as
that in equation \eqref{mwi:eqn:parallel}. But such a `step' is missing in
the cluster state model. There is nothing corresponding to such a
unitary transformation. At best we have a simulation of such a gate;
however, it is a simulation that bears no resemblance, in terms of its
physical realisation, to the corresponding network circuit. In
addition, the temporal ordering of computation in the cluster state
has little, if anything, to do with the temporal ordering present in
the simulated network. Thus there is nothing corresponding to
simultaneous function evaluation in the cluster state, for on the
cluster state model gates are only conceptual entities that one may
utilise for algorithm design. When it comes to implementation, the
logical division of the cluster into distinct gates is completely
irrelevant. Indeed, in order to characterise the cluster state model
it is not necessary to begin with the logical layout of the network
model at all, for the cluster state model is, arguably, more
effectively characterised by a graph than by a network
\citep[p. 20]{raussendorf2003}.

Far from being a natural and intuitive picture of cluster state
computation, it seems, rather, that one must work \emph{against} one's
intuition to view the cluster state model as a model of parallel
computation in many worlds, and it is hard to see how such a
description can be useful. Considerations such as these prompt Steane
to write: ``[t]he evolution of the cluster state computer is not
readily or appropriately described as a set of exponentially many
computations going on at once. It is readily described as a sequence
of measurements whose outcomes exhibit correlations generated by
entanglement'' \citeyearpar[p. 474]{steane2003}. I should note that
many worlds advocates such as Hewitt-Horsman, also, reluctantly reject
the view that cluster state computation need involve an appeal to many
worlds \citep[pp. 896-897]{hewittHorsman2009}; though, as we have
seen, she still defends the legitimacy and usefulness of describing
\emph{network} based computation in terms of many worlds and of
treating these worlds as ontologically real
\citep[pp. 890-896]{hewittHorsman2009}.

But the main problem, for one who wishes to defend a many worlds
description of the operation of the cluster state computer, is not
that such a description is neither natural nor useful. The problem is
deeper than this, for it appears that it is for all practical purposes
impossible to specify a preferred basis in which to distinguish the
worlds in which parallel computations take place in the context of the
cluster state computer. Recall that, in general, measurements in the
cluster state model are \emph{adaptive}: the basis for each
measurement will change throughout the computation and will differ
from one qubit to the next. During each time step of the computation,
the (random) results of the measurements performed in that step will
determine the measurement bases used to measure the qubits in
subsequent steps. But this random determination of measurement bases
means that there is no principled way to select a preferred basis a
priori (and even if we did, few qubits would actually be measured in
that basis), and we certainly cannot assert that there is any sense in
which a preferred basis `emerges' from this process. Thus there is no
way in which to characterise the cluster state computer as performing
its computations in many worlds, for there is no way, in the context
of the cluster state computer, to even define these worlds for the
purposes of describing the computation as a whole.

As a possible rejoinder, one might assert that the cluster state model
merely obscures the fact that the computation takes place in many
worlds, and that this would be revealed upon closer analysis by, for
instance, considering how one might go about simulating a
cluster-state computation with circuits. In fact it is possible to
simulate a cluster state using classically controlled
gates. Classically controlled gates are gates whose operation is
dependent on classical bit values (these are typically the results of
measurements). To avoid the problem of the continually changing basis,
one might take the additional step of deferring all measurements to
the end of the process. According to the principle of deferred
measurement \citep[p. 186]{nielsenChuang2000}, this is always
possible.

Such a simulation would require many more qubits and at least one more
two-qubit operation for each single qubit operation in the cluster,
however. In principle, there will be no bound to either the additional
memory or to the number of additional two-qubit gates required to
realise the simulation \citep[p. 2]{deBeaudrap2009}. Practical
methods, therefore, for simulating the cluster state with circuits
allow measurement gates to be a part of the computational process
\citep[]{childs2005,deBeaudrap2009}. They decompose the cluster state
into a series of classically controlled change of basis gates followed
by measurement gates in the standard basis. Thus this will not solve
the problem for the many worlds theorist.

But perhaps some day an ingenious theorist will find a way to simulate
cluster state computation in some other model without the use of
adaptive measurements or classically controlled change of basis
gates. What should we say then? Even in this case I think it would be
misleading to speak of the cluster state model as obscuring the fact
that many worlds are responsible for the speedup it evinces. Recall
that, for those who adhere to the many worlds explanation of quantum
computation, part of the motivation for describing computation as
literally happening in many worlds is that it is useful for algorithm
analysis and design to believe that these worlds are real. This
motivation is absent in the cluster state model irrespective of
whether it can be simulated in some other model. Moreover,
irrespective of whether it can be simulated in some other model, the
cluster state model will, in virtue of its unique characteristics,
surely lead to new ways of thinking about quantum computation that
would not have occurred to a theorist working only with the network
model. To dogmatically hold on to the view that, in actuality, many
worlds are, at root, physically responsible for the speedup evinced in
the cluster state model will at best be useless, for, as we have seen,
it will not help our theorist to design algorithms for the cluster
state. At worst it will be positively detrimental if dogmatically
holding on to this view prevents our theorist from discovering the
possibilities that are inherent in the cluster state model.

\section{The legitimacy of the many worlds explanation for the network
model}
\label{mwi:s:net}

We saw, in \textsection \ref{mwi:s:neo}, that the many worlds
explanation of quantum computing cannot avail itself of many of the
arguments in support of the many worlds interpretation of quantum
mechanics which appeal to decoherence as a criterion for
distinguishing worlds in order to circumvent the preferred basis
objection. Further, we saw that while the decoherence basis is able to
fulfil the role assigned to it, in the many worlds interpretation of
quantum mechanics, of determining the preferred basis for world
decomposition with respect to macro experience, the corresponding
criterion for world decomposition appealed to by those who defend the
many worlds explanation of quantum computing cannot fulfil this role
except in an ad hoc way. Thus we have one reason to reject many worlds
as an explanation of the network model of quantum computation. Let us
put this consideration to one side.

We have just seen, in \textsection \ref{mwi:s:clu}, that the cluster state
model of quantum computation is incompatible with a many worlds
explanation of it. In spite of this, one might still wish to maintain
the view that network-based computation, at least, is computation in
many worlds. There is nothing wrong in principle with such a
stance. What makes this view problematic, however, is the fact that
the cluster-state model is computationally equivalent to the network
model. One must therefore be committed to the view that an algorithm,
when run on quantum circuits, performs its computation in many worlds;
while a simulation of the same algorithm, run on a cluster-state
computer, does not. Moreover, this is in spite of the fact that there
may be no difference in the way in which individual qubits are
physically realised in each computer.

As unfortunate as such a situation would be, it would be forced on us
if there were no other potential unifying explanations of the source
of quantum speedup available. Fortunately, however, there do exist
potential physical explanations for quantum speedup in the network
model which, unlike the many worlds explanation, are compatible with
the cluster state model.

One example of such an explanation is due to Lance
Fortnow. \citet[]{fortnow2003} develops an abstract mathematical
framework for representing the computational complexity classes
associated with classical and quantum computing.\footnote{These are
  \textbf{BPP} (bounded-error probabilistic polynomial time) and
  \textbf{BQP} (bounded error quantum polynomial time). For more on
  computational complexity classes, see Appendix \ref{ch:cc}. For a
  more detailed overview of Fortnow's framework, see Appendix
  \ref{ch:matrix}.} In Fortnow's framework, both classes of
computation are represented by transition matrices which determine the
possible transitions between the configurations of a nondeterministic
Turing machine. This framework shows, according to Fortnow, that the
fundamental difference between quantum and classical computation is
interference: in the quantum case, matrix entries can be negative,
signifying a quantum computer's ability to realise good computational
paths with higher probability by having the bad computational paths
cancel each other out \citep[p. 606]{fortnow2003}.

Another example of a unifying explanation is the physical explanation
for quantum speedup that we will develop in the following chapters.

Unlike the many worlds explanation, these explanations of the source
of quantum speedup do not rely on the particular characteristics of
the network model and seem straightforwardly compatible with cluster
state computation. But the fact that the many worlds explanation of
quantum speedup is not compatible with the cluster state model, while
these other explanations of quantum speedup are, is a reason to
question its usefulness as a description of network-based quantum
computation, and thus one more reason to reject it as an explanation
of quantum speedup \emph{tout court}.

\section{Conclusion}
\label{mwi:s:con}

I hope to have convinced the reader that, whatever the merits of the
neo-Everettian interpretation of quantum mechanics are, the many
worlds explanation of quantum computing is inadequate as an explanation
of either the network or the cluster state model of quantum
computation. We saw above how it depends on a suspect extension of the
the neo-Everettian approach to the interpretation of quantum
mechanics, and we saw how, unlike other explanations of quantum
computing, it is unable to describe the cluster state model of quantum
computation. I hope that the reader agrees that these are convincing
reasons to reject the many worlds explanation of quantum computing.

I do not want to argue that the many worlds explanation of quantum
computation, particularly in regards to the network model, has no
heuristic value. It undoubtedly does, and thinking in this manner has
assuredly led to the development of some important quantum
algorithms. Nevertheless we should take talk of many computational
worlds with a grain of salt. Indeed, taking literally the many worlds
view of quantum computation may be positively detrimental if it
prevents us from constructing models of quantum computation, such as
the cluster state model, in the future.

\section{Next steps}

The many worlds explanation of quantum computation is, arguably, the
best known of the candidate physical explanations for quantum
speedup. It is also, perhaps, the most influential; it has been and
continues to be discussed in the popular, philosophical, and
scientific literature on quantum computation. Given this, thoroughly
considering its merits as an explanation for quantum speedup was both
important and appropriate. Now that we have completed this inquiry,
however, we will take a different approach. In lieu of undertaking a
case by case critical examination of the major candidate explanations
of quantum computation on offer, from here on in we will proceed in a
more constructive manner.

In almost all of the candidate explanations for quantum speedup (e.g.,
\citealt[]{ekert1998, steane2003, duwell2004, bub2006, bub2010}), the
fact that quantum mechanical systems can sometimes exhibit
entanglement plays an important role.\footnote{One important exception
  to this is Fortnow's view (cf. Appendix \ref{ch:matrix}), which
  points to interference, and not entanglement, as the explanation for
  quantum speedup. As I will argue in Chapter \ref{ch:sig}, however,
  interference and entanglement can be seen as, so to speak, two sides
  of the same coin.} On Steane's view, for instance, quantum
entanglement allows one to manipulate the correlations between the
values of a function without manipulating those values themselves. For
proponents of the many worlds explanation, on the other hand, though
they consider computational worlds to be the main component in the
explanation of quantum speedup, they nevertheless view entanglement as
indispensable to its analysis \citep[889]{hewittHorsman2009}. This
circumstance is intriguing, and leads one to wonder whether one
\emph{must} appeal to entanglement in order to explain quantum
speedup; i.e., whether entanglement is a \emph{necessary component} of
any explanation for quantum speedup. This will be the topic of the
next chapter.

Continuing along in this more positive manner, perhaps we will be
fortunate enough to stumble upon some one, or some set, of necessary
and sufficient conditions for the explanation that we seek\textemdash
and in this way assemble together an explanation for quantum speedup,
so to speak, `from the ground up'.

\chapter{Entanglement as a Necessary Component in a Physical
  Explanation for Quantum Computational Speedup}
\label{ch:nec}

\settocdepth{subsection}

\section{Introduction}
\label{nec:s:intro}

The significance of the phenomenon of quantum entanglement\textemdash
wherein the most precise characterisation of a quantum system composed
of previously interacting subsystems does not necessarily include a
precise characterisation of those subsystems\textemdash has been at
the forefront of the debate over the conceptual foundations of quantum
theory, almost since that theory's inception. It is \emph{the}
distinguishing feature of quantum theory, for some
\citep[]{schrodinger1935}.\footnote{For some more recent speculation
  on the the distinguishing feature(s) of quantum mechanics, see,
  for instance, \citet[]{clifton2003,myrvold2010}.} For others, it is
evidence for the incompleteness of that theory
\citep*[]{epr1935}.\footnote{For further discussion, and for
  Einstein's later refinements of the Einstein-Podolsky-Rosen (EPR)
  paper's main argument, see \citet[]{howard1985}.} For yet others,
the possibility of entangled quantum systems implies that physical
reality is essentially non-local \citep[]{stapp1997}.\footnote{For
  responses to Stapp's view and for further discussion, see:
  \citet[]{unruh1999,mermin1998,stapp1999}.} For almost all, it has
been, and continues to be, an enigma requiring a solution.

Logically, entanglement may play the role of either a necessary or a
sufficient condition (or both) in an overall explanation of quantum
speedup. The question of whether entanglement may be said to be a
sufficient condition will be addressed in subsequent chapters. As for
the assertion that entanglement is a necessary component in the
explanation of speedup, this seems, prima facie, to be
supported\footnote{What I take to be supported by
  \citeauthor[]{jozsa2003}'s result is the claim that entanglement
  is required in order \emph{to explain} quantum speedup. As we will
  discuss further in \textsection \ref{nec:s:net}, this is
  distinct from the claim that one requires an entangled quantum
  state in order to achieve quantum speedup.} by a result due to
\citet*[]{jozsa2003}, who prove that for quantum algorithms which
utilise \emph{pure} states, ``the presence of multi-partite
entanglement, with a number of parties that increases unboundedly with
input size, is necessary if the quantum algorithm is to offer an
exponential speed-up over classical computation''
\citeyearpar[p. 2014]{jozsa2003}. When we consider quantum algorithms
which utilise \emph{mixed} states, however, then there appear to be
counterexamples to the assertion that one must appeal to quantum
entanglement in order to explain quantum speedup. In particular,
\citet[]{biham2004} have shown that it is possible to achieve a modest
(sub-exponential) speedup using unentangled mixed states. Further,
\citet[]{datta2005,datta2008} have shown that it is possible to
achieve an exponential speedup using mixed states that contain only a
vanishingly small amount of entanglement. In the latter case, further
investigation has suggested to some that quantum correlations
\emph{other than} entanglement may be playing a more important
role. One quantity in particular, \emph{quantum discord}, appears to
be intimately connected to the speedup that is present in the
algorithm in question. In light of these results, it is tempting to
conclude that it is not necessary to appeal to entanglement at all in
order to explain quantum computational speedup and that the
investigative focus should shift to the physical characteristics of
quantum discord or some other such quantum correlation measure
instead.

In this chapter I will argue that this conclusion is premature and
misguided, for as I will show below, there is an important sense in
which entanglement can indeed be said to be necessary for the
explanation of the quantum speedup obtainable from both of these
mixed-state quantum algorithms.

The chapter will proceed as follows. After introducing the concept of
entanglement and how it is quantified in \textsection
\ref{nec:s:prelim}, I introduce the \emph{necessity of entanglement
  for explanation thesis} in \textsection \ref{nec:s:net}. In
\textsection \ref{nec:s:deq}, I show how what looks like a
counter-example to the necessity of entanglement for explanation
thesis for pure states\textemdash the fact that certain important
quantum algorithms can be expressed so that their states are never
entangled\textemdash is instead evidence for this thesis. Then, in
\textsection\ref{nec:s:mix}, I examine the more serious challenges to
the necessity of entanglement for explanation thesis posed by the
cases of sub-exponential speedup with unentangled mixed states
(\textsection\ref{nec:ss:mixdj}) and exponential speedup with mixed
states containing only a vanishingly small quantity of entanglement
(\textsection\ref{nec:ss:oneq}).

Starting with the first type of counter-example, I begin by arguing
that pure quantum states should be taken to provide a more fundamental
representation of quantum systems than mixed states. I then show that
when one considers the initially mixed state of the quantum computer
as representing the space of its possible pure state preparations, the
speedup obtainable from the computer can be seen as stemming from the
fact that the quantum computer evolves some of these possible pure
state preparations to entangled states\textemdash that the quantum
speedup of the computer can be seen as arising from the fact that it
implements an entangling transformation.

As for the second type of counter-example, where exponential speedup
is achieved with only a vanishingly small amount of entanglement, and
where it is held by some that another type of non-classical
correlation, quantum discord, is responsible for the speedup of the
quantum computer: I argue that, first, it is misleading to
characterise discord as indicative of non-classical correlations. I
then appeal to recent work done by \citet[]{fanchini2011},
\citet[]{brodutch2011}, and \citet[]{cavalcanti2011} who show,
respectively, that when one considers the `purified' state
representation of the quantum computer, there is a conservation
relation between discord and entanglement, and indeed that there is
just as much entanglement in such a representation as there is discord
in the mixed state representation; that entanglement must be shared
between two parties in order to bilocally implement any bipartite
quantum gate; and that entanglement is directly involved in the
operational definition of quantum discord.

Given \citeauthor[]{jozsa2003}'s proof of the necessary presence of
an entangled state for exponential speedup using pure states, and
given the fundamentality of pure states as representations of quantum
systems, the burden of proof is upon those who would deny the
necessity of entanglement for explanation thesis to show either by
means of a counter-example or by some other more principled method
that it is false. Neither of the counter-examples discussed in this
chapter succeeds in doing so. We should conclude, therefore, that the
necessity of entanglement for explanation thesis is true.

\section{Preliminaries}
\label{nec:s:prelim}

\subsection{Quantum entanglement}
\label{nec:ss:ent}

\subsubsection{Pure states}

Consider the following representation of the joint state of two
qubits:
$$| \psi \rangle = | 0 \rangle \otimes | 0 \rangle + | 0 \rangle
\otimes | 1 \rangle + | 1 \rangle \otimes | 0 \rangle + | 1 \rangle
\otimes | 1 \rangle.$$
This expression for the overall state of the system represents the
fact that the two qubits are in an equally weighted superposition of
the four joint states (a)-(d) below:
\vspace{1em}
\begin{center}
\begin{tabular}{lrr}
& $q_1$ & $q_2$ \\
(a) & $| 0 \rangle$ & $| 0 \rangle$ \\
(b) & $| 0 \rangle$ & $| 1 \rangle$ \\
(c) & $| 1 \rangle$ & $| 0 \rangle$ \\
(d) & $| 1 \rangle$ & $| 1 \rangle$. \\
\end{tabular}
\end{center}
\vspace{1em}
This particular state is a \emph{separable} state, for it can,
alternatively, be expressed as a product of the pure states of its
component systems, as follows:
$$| \psi \rangle = (| 0 \rangle + | 1 \rangle) \otimes (| 0 \rangle +
| 1 \rangle).$$

Not all quantum mechanical states can be expressed as product states
of their component systems, and thus not all quantum mechanical states
are separable. Here are four such `entangled' states:\footnote{Note
  that below I have used the shorthand tensor product notation. See
  \textsection \ref{intro:s:term}.}
\begin{align}
\label{nec:eqn:bellstates}
|\Phi^+\rangle & = \frac{| 00 \rangle + | 11 \rangle}{\sqrt{2}}
\nonumber \\
|\Phi^-\rangle & = \frac{| 00 \rangle - | 11 \rangle}{\sqrt{2}}
\nonumber \\
|\Psi^+\rangle & = \frac{| 01 \rangle + | 10 \rangle}{\sqrt{2}}
\nonumber \\
|\Psi^-\rangle & = \frac{| 01 \rangle - | 10 \rangle}{\sqrt{2}}.
\end{align}
The skeptical reader is encouraged to convince himself that it is
impossible to re-express any of these states as a product state of two
qubits. They are called the Bell states, and I will refer to a pair of
qubits jointly in a Bell state as a Bell pair.\footnote{These are
  also sometimes referred to as `EPR pairs'. EPR stands for
  Einstein, Podolsky, and Rosen. In their seminal
  \citeyear[]{epr1935} paper, EPR famously used states analogous to
  the Bell states to argue that quantum mechanics is incomplete.}
Maximally entangled states,\footnote{Note that not all entangled
  states are maximally entangled states. We will discuss this in
  more detail in the next section.} such as these, completely specify
the correlations between outcomes of experiments on their component
qubits without specifying anything regarding the outcome of a single
experiment on one of the qubits. For instance, in the singlet state
(i.e., $|\Psi^-\rangle$), outcomes of experiments on the first and
second qubits are perfectly anti-correlated with one another. If one
performs, say, a $\hat{z}$ experiment on one qubit of such a system,
then if the result is $|0\rangle$, a $\hat{z}$ experiment on the other
qubit will, with certainty, yield an outcome of $|1\rangle$, and vice
versa. In general, the expectation value for joint measurements on the
two qubits is given by $- \hat{m} \cdot \hat{n} = - \cos\theta,$ where
$\hat{m}, \hat{n}$ are unit vectors representing the orientations of
the two experimental devices, and $\theta$ is the difference in these
orientations. Any single $\hat{z}$ experiment on just one of the two
qubits, however, will yield $|0\rangle$ or $|1\rangle$ with equal
probability.

The phenomenon of entanglement has deep implications for our
understanding of the physical world. Consider an alternative theory of
quantum mechanics in which $\lambda$ is an assignment to a set of
hidden variables determining the outcomes of experiments on the two
subsystems of a Bell pair. Suppose $\lambda$ satisfies the condition
that it assigns probabilities to experimental outcomes on the first
subsystem that are independent of experimental outcomes on the second
subsystem (and vice versa); i.e.,
\begin{equation}
\label{nec:eq:oi}
p_\lambda^a(x_a|a,b) = p_\lambda^a(x_a|a,b,x_b).
\end{equation}
This condition has variously been called \emph{completeness}
\citep[]{jarrett1984}, \emph{outcome independence}
\citep[]{shimony1993}, and \emph{separability}
\citep[]{howard1985}. Bell's inequalities imply that any theory
consistent with the predictions of quantum mechanics which satisfies
\eqref{nec:eq:oi} must assign different probabilities to outcomes of
experiments on the first subsystem depending on the \emph{choice of
  test} that is performed on the second subsystem; i.e., it must
violate the condition that
\begin{equation}
\label{nec:eq:pi}
p_\lambda^a(x_a|a,b) = p_\lambda^a(x_a|a,b').
\end{equation}
Jarrett and Howard call this second condition \emph{locality}, while
Shimony calls it \emph{parameter independence}. Together, outcome and
parameter independence yield \emph{factorisability}:

\begin{align}
\label{nec:eq:fact}
p_\lambda^{ab}(x_a, x_b | a, b) = p_\lambda^a(x_a | a) \cdot
p_\lambda^b(x_b | b).
\end{align}

It turns out that Bell's inequalities imply that any theory that is
consistent with the predictions of quantum mechanics must violate
\eqref{nec:eq:fact} and thus violate either \eqref{nec:eq:oi} or
\eqref{nec:eq:pi}. In particular, a fully deterministic hidden
variables theory, which the reader should convince herself must
necessarily satisfy \eqref{nec:eq:oi}, must therefore necessarily
violate \eqref{nec:eq:pi}. On the other hand, standard quantum
mechanics obviously violates \eqref{nec:eq:oi}, but satisfies
\eqref{nec:eq:pi}. It is worthwhile to note that a violation of
\eqref{nec:eq:pi} necessarily brings one into conflict with special
relativity, but that it is at least not obvious that a mere violation
of \eqref{nec:eq:oi} does so; i.e., that `peaceful coexistence'
between the two theories is impossible.\footnote{For further
  discussion of peaceful coexistence, see \citet[]{shimony1993}. For
  a more specific defence of the possibility of peaceful coexistence
  between special relativity and theories which characterise quantum
  mechanical wave function collapse as a real physical process, see
  \citet[]{myrvold2002}.}

We will consider the physical significance of quantum entanglement in
more detail in subsequent chapters, but for the time being we will put
such interpretive questions to one side. For the purposes of this
chapter my intention will be to characterise entanglement as neutrally
and uncontroversially as possible.

\subsubsection{Mixed states}

The concepts of separability and of entanglement are also applicable
to so-called `mixed states'. To illustrate the concept of a mixed
state, imagine that one draws a ball from an urn into which balls of
different types have been placed, and that the probability of drawing
a ball of type $i$ is $p_i$. Corresponding to the outcome $i$, we then
prepare a given system $S$ in the pure state $| \psi_i \rangle$,
representable by the density operator $\rho^S_i = | \psi_i
\rangle\langle \psi_i |$. After preparing $\rho^S_i$, we then discard
our record of the result of the draw. The resulting state of the
overall system will be the mixed state:
\begin{equation}
\label{nec:eqn:mix}
\rho = \sum_i p_i \rho_i^S.
\end{equation}

A \emph{mixed state} is \emph{separable} if it can be expressed as a
mixture of pure product states, and \emph{entangled} otherwise. In
general, determining whether a mixed state of the form
\eqref{nec:eqn:mix} is an entangled state is difficult, because in
general the decomposition of mixtures is non-unique. For instance, the
reader can verify that a mixed state represented by the density
operator $\rho$, prepared as a mixture of pure states in the following
way:
$$\rho = \frac{3}{4}| 0 \rangle\langle 0 | + \frac{1}{4}| 1
\rangle\langle 1 |,$$
can also be equivalently prepared as:
$$\rho = \frac{1}{2}| \psi \rangle\langle \psi | + \frac{1}{2}| \phi
\rangle\langle \phi |,$$
where
$$| \psi \rangle \equiv \sqrt{\frac{3}{4}}| 0 \rangle +
\sqrt{\frac{1}{4}}| 1 \rangle, \quad | \phi \rangle \equiv
\sqrt\frac{3}{4}| 0 \rangle - \sqrt\frac{1}{4}| 1 \rangle.$$
This is so because both state preparations yield an identical density
matrix representation (in the computational basis); i.e.,:
\begin{align*}
\left(
\begin{matrix}
  3/4 & 0 \\
  0 & 1/4
\end{matrix}
\right).
\end{align*}
As we will see in more detail later, a system that is prepared as a
mixture of entangled states will sometimes yield the same density
operator representation as a system prepared as a mixture of pure
product states.

\subsection{Quantifying entanglement}
\label{nec:ss:quantent}

The four Bell states that we encountered in section \ref{nec:ss:ent}
are examples of \emph{maximally entangled} states. Not all entangled
states are maximally entangled states, however. For instance, as will
be clear later, the state
\begin{equation}
\label{nec:eqn:notmaxent}
| \phi \rangle = \sqrt\frac{1}{3}| 01 \rangle + \sqrt\frac{2}{3}| 10
\rangle,
\end{equation}
though entangled, is not maximally entangled.

Entanglement is a potentially useful resource for quantum information
processing. \citet[]{masanes2006} has shown, for instance, that for
any non-separable state $\rho$, some other state $\sigma$ is capable
of having its teleportation fidelity enhanced by $\rho$'s
presence.\footnote{The teleportation fidelity \citep[cf.][\textsection
  9.2.2]{nielsenChuang2000} is a measure of the `closeness' of the
  input and output states in the teleportation protocol
  (cf. Appendix \ref{ch:tel}).} Given this, it will be useful to be
able to quantify the amount of entanglement contained in a given
state. In order to do this, we employ so-called entanglement measures,
the theory of which is outlined below.\footnote{In the exposition
  which follows, I draw substantially from \citet[]{plenio2007}.}

\subsubsection{Local operations and classical communications}

Perhaps the most basic concept of the theory of entanglement measures
is that of \emph{Local Operations and Classical Communications}
(LOCC). Roughly, LOCC refers to local quantum operations (LO) that can
be performed on a quantum system at a given site, and coordinated with
other local operations at other sites using classical communications
links (CC). For instance, in the quantum teleportation protocol
(cf. Appendix \ref{ch:tel}), after the two parties to the
protocol have been spatially separated from one another, all of their
subsequent actions can be classified as LOCC.

LOCC provides the key, within the theory of entanglement measures, for
distinguishing classical from non-classical correlations (which, for
now, we simply identify with entanglement).\footnote{Since
  communication (for instance, of the results of previous
  measurements) between the parties to a quantum informational
  protocol may occur often and at any time during the process, this
  will, in general, introduce highly complex dependencies into our
  description of the process. These make it extraordinarily
  difficult, if not practically impossible, to give a precise
  mathematical characterisation of the set of possible LOCC
  operations. As a workaround, larger more easily characterisable
  classes of operations, which are sufficiently `LOCC-al', are used as
  imperfect proxies for the LOCC class. One such class is the class of
  \emph{separable operations}, described in Appendix \ref{ch:sepop}.}
Recall the discussion which preceded Eq. \eqref{nec:eqn:mix}
above. Now imagine that, upon drawing a ball of type $i$ from the urn,
not only Samantha, but Alice, Bob, and Charles also create their own
individual quantum states, $\rho_i^A, \rho_i^B, \rho_i^C$, based on
the shared information about the outcome $i$. In that case, the
resulting state of the overall system will be the mixed state:
\begin{equation}
\label{nec:eqn:mixmulti}
\rho^{SABC} = \sum_i p_i \rho^S_i \otimes \rho^A_i \otimes
\rho^B_i \otimes \rho^C_i.
\end{equation}
This procedure with urn and balls is an example of a procedure
involving LOCC operations. And since \eqref{nec:eqn:mixmulti} is, in
fact, the general form of a separable state, we may conclude from this
that every separable state is such that it can be created using LOCC
operations alone. Further, since correlations generable using only
LOCC operations can always be described as the result of some common
classical cause (this is built into the very definition of LOCC), it
is reasonable to conclude that \emph{a quantum state $\rho$ can be
  generated perfectly using LOCC alone if and only if it is
  separable.}

Note that one cannot increase the amount of entanglement contained in
a given state using LOCC operations (including combinations of local
unitaries) alone. To see why this is true, note first that, since only
separable states can be created using LOCC operations, it follows that
one cannot generate an entangled state from an unentangled
one. Second, imagine transforming some entangled state $\rho$ into
another state $\rho'$ using LOCC operations. Since $\rho'$ was
obtained using only LOCC operations, anything that can be done with
$\rho'$ + LOCC operations can also be done with $\rho$ + LOCC. Hence,
in terms of the resources made available for information processing,
$\rho'$ is (at best) no more entangled than $\rho$; $\rho'$ and $\rho$,
therefore, will (at best) contain an equal amount of entanglement.

\subsubsection{Maximally entangled states}

Consider bipartite (i.e., two-party) systems of `qudits'; i.e.,
$d$-dimensional quantum systems (a qu\emph{bit}, for instance, is a
qudit for which $d=2$). Any pure state that is local unitarily
equivalent to
\begin{align}
\label{nec:eqn:maxent}
| \Phi_d^+ \rangle = \frac{| 0 \rangle \otimes | 0 \rangle + | 1 \rangle
\otimes | 1 \rangle + \dots + | d-1 \rangle \otimes | d-1
\rangle}{\sqrt d}
\end{align}
is a maximally entangled state. We describe it as such because from
this state + LOCC operations it is possible to prepare (with certainty)
\emph{any} desired two-party $d$-dimensional qudit state. And since
LOCC operations cannot increase the amount of entanglement in a
system, it follows that all states local unitarily equivalent to
\eqref{nec:eqn:maxent} are also maximally entangled. This statement is
absolute; i.e., states of the form \eqref{nec:eqn:maxent} are
maximally entangled irrespective of which entanglement measure (to be
discussed in the next section) is used to impose an ordering on states.

For instance, consider the simple case of two qubits. For $d=2$, a
state of form \eqref{nec:eqn:maxent} is the Bell state $| \Phi^+
\rangle$. We claim that $| \Phi^+ \rangle$ is a maximally entangled
state; i.e., that, with certainty, beginning with $| \Phi^+ \rangle$
(or a state local unitarily equivalent to $| \Phi^+ \rangle$), we can
prepare any arbitrary bipartite state $| \phi \rangle$. To see why
this is so, consider an arbitrary bipartite pure state in Schmidt
decomposed form:\footnote{It is a fact that any bipartite pure state
  can be expressed as:
  $$
  | \psi \rangle = U_a \otimes U_b \sum_{i=1}^N\sqrt{\alpha_i}| i
  \rangle_a| i \rangle_b,
  $$
  where the $\alpha_i$ are positive real numbers called the
  Schmidt-coefficients of $|\psi\rangle$ \citep[]{plenio2007}. Since
  the local unitaries do not affect the entanglement properties of the
  state, we omit them in \eqref{nec:eqn:schm}.}
\begin{equation}
\label{nec:eqn:schm}
| \phi \rangle = \alpha| 00\rangle_{ab}  + \beta| 11\rangle_{ab}.
\end{equation}
(here $a$ and $b$ refer to Alice's and Bob's qubits, respectively). We
will now show how to obtain $| \phi \rangle$ from $| \Phi^+ \rangle$
using exclusively LOCC transformations.

First, to the Bell state, $$|\Phi^+\rangle = \frac{| 00 \rangle_{ab} +
  | 11 \rangle_{ab}}{\sqrt{2}},$$ add an ancilla qubit in state $| 0
\rangle$ at Alice's location: $$\frac{1}{\sqrt 2}(| 00 \rangle_{aa}| 0
\rangle_b + | 01 \rangle_{aa}| 1 \rangle_b).$$ Now perform the unitary
transformation $| 00 \rangle \rightarrow \alpha| 00 \rangle + \beta|
11 \rangle; | 01 \rangle \rightarrow \beta| 01 \rangle + \alpha| 10
\rangle$ on Alice's system. This yields:
\begin{align}
& & \frac{(\alpha| 00 \rangle + \beta| 11 \rangle)_{aa}| 0
  \rangle_b + (\beta| 01 \rangle + \alpha| 10 \rangle)_{aa}| 1
  \rangle_b}{\sqrt 2} \nonumber \\
& = & \frac{\alpha| 00 \rangle_{aa}| 0 \rangle_b + \beta| 11
  \rangle_{aa}| 0 \rangle_b + \beta| 01 \rangle_{aa}| 1 \rangle_b
  + \alpha| 10 \rangle_{aa}| 1 \rangle_b}{\sqrt 2} \nonumber \\
& = & \frac{| 0 \rangle_a(\alpha| 00 \rangle_{ab} + \beta| 11
  \rangle_{ab}) + | 1 \rangle_a(\beta| 10 \rangle_{ab} + \alpha| 01
  \rangle_{ab})}{\sqrt 2}
\end{align}
We now instruct Alice to perform a local measurement on her ancilla
system. If it yields $| 0 \rangle$ then Bob need not do anything,
else if Alice's measurement yields $| 1 \rangle$, Bob applies an
$\mathbf{X}$ (i.e. a ``NOT'') transformation to his qubit. In either
case, the result is $| \phi \rangle,$ as desired. Note that although
we limited ourselves to the pure state case, it is easy to show that
any mixed state, $\rho$, can also be obtained from
\eqref{nec:eqn:maxent}.

\subsubsection{Entanglement measures}

As we have just seen, a bipartite state $\rho$ is said to be maximally
entangled if we can use it and LOCC operations to prepare any
arbitrary bipartite state $\sigma$ with certainty. In the more general
case, where $\rho$ is not necessarily maximally entangled, we will, in
similar fashion, say that the amount of entanglement contained in
$\rho$ is greater than or equal to the amount of entanglement
contained in $\sigma$ if the transformation $\rho \rightarrow \sigma$
can be performed using only LOCC operations.

Because of the limitations inherent in determining an ordering on
entangled states in the single copy setting
\citep[cf.][]{plenio2007}, entanglement measures are sometimes
defined with respect to the asymptotic regime. Here, the basic idea is
that we do not ask whether we may use the single state $\rho$ in order
to exactly prepare the single state $\sigma$. Rather, we ask whether
it is possible to achieve the transformation $\rho^{\otimes n}
\rightarrow \sigma^{\otimes m},$ for large integers $m$ and $n$; and
we use the ratio, $m/n$, as the basis for a measure of the relative
entanglement contained in the two states.\footnote{In fact, even this
  condition is usually relaxed; i.e., rather than ask whether it is
  possible to achieve the transformation $\rho^{\otimes n}
  \rightarrow \sigma^{\otimes m}$, typically we ask only whether it
  is possible to achieve the transformation $\rho^{\otimes n}
  \rightarrow \sigma_m$, where $\sigma_m$ is an approximation of
  $\sigma^{\otimes m}$. In this case, if, for some fixed $r = m/n$, as
  $n \rightarrow \infty$, we can bring the state $\sigma_m$
  arbitrarily close to $\sigma^{\otimes m}$, then we say that the
  rate $r = m/n$ is achievable for this transformation.}

We now consider a few specific entanglement measures.

\paragraph{Entanglement cost and distillable entanglement.} Consider a
bipartite qubit state, $\rho$ and a maximally entangled bipartite
state $\Phi(K) \equiv | \Phi^+_K \rangle\langle \Phi^+_K |$ of
$K$-dimensions. The \emph{entanglement cost}, $E_C(\rho)$, associated
with $\rho$ quantifies the amount of entanglement required in order to
approximate $n$ copies of $\rho$, starting from the maximally
entangled state. More formally, it is defined as the lowest rate $r$
for which the trace norm distance\footnote{We use the trace norm
  distance, $tr|\sigma - \eta|$, for the sake of mathematical
  convenience, as a measure of the distance between quantum
  states. Any suitable measure of distance,  $D(\sigma,\eta)$, could
  have been used instead, however, as the definition of entanglement
  cost is independent of our choice of distance function
  \citep[cf.][]{plenio2007}.} between $\Psi(\Phi(2^{rn}))$ and
$\rho^{\otimes n}$ approaches 0 for large $n$, where $\Psi$ is a trace
preserving (series of) LOCC-al operation(s) performed on
$\Phi(2^{rn})$ with the object of obtaining $\rho^{\otimes n}$; i.e.,
\begin{align}
E_C(\rho) \equiv \inf\left\{r: \lim_{n \to \infty} \left[\inf_\Psi
  \text{tr}|\rho^{\otimes n} - \Psi(\Phi(2^{rn}))|\right] = 0 \right\}.
\end{align}

We would also like to know about the reverse process; i.e., we would
like to know the greatest rate $r$ for which the distance between
$\rho^{\otimes n}$ and $\Phi(2^{rn})$ approaches 0 for large $n$,
where $\Psi$ is now a LOCC operation on $\rho^{\otimes n}$ performed
in order to obtain $\Phi$. This process is known as entanglement
distillation,\footnote{It is also sometimes referred to as
  entanglement concentration, though this name is generally reserved
  for the pure state case. Note that the fact that $\Psi$ produces
  an \emph{approximation} of $\Phi(K)$ is particularly important for
  this case; for, recalling our earlier discussion, no exact
  transformation from $\rho^{\otimes n}$ to even one maximally
  entangled state is in general possible.} and the measure
associated with it is \emph{distillable entanglement}:
\begin{align}
E_D(\rho) \equiv \sup\left\{r: \lim_{n \to \infty}\left[ \inf_\Psi
  \text{tr}|\Psi(\rho^{\otimes n}) - \Phi(2^{rn})|\right] = 0 \right\}.
\end{align}
The distillable entanglement can be thought of as a measure of the
`entanglement potential' of a state; it tells us the maximum possible
rate at which many copies of a `noisy' entangled state may be converted
back into a maximally entangled state using LOCC.

For pure states, these transformations are reversible in the
asymptotic limit; further, for pure states $E_C$ and $E_D$ are both
equal to the \emph{entropy of entanglement} \citep[]{bennett1996},
defined, for a pure state $| \psi \rangle$, as
\begin{align}
  E(| \psi \rangle\langle \psi |) \equiv S(\text{tr}_A| \psi
  \rangle\langle \psi |) = S(\text{tr}_B| \psi \rangle\langle \psi
  |),
\end{align}
where $S$ is the von Neumann entropy: $S(\rho) = -\text{tr}(\rho
\log_2 \rho)$.\footnote{For more on the von Neumann entropy, as well
  as on the other concepts of classical and quantum information
  theory, see Appendix \ref{ch:infth}.} Thus for pure states, there
is a unique total ordering of entangled states in the asymptotic
regime\footnote{It should now be evident why the state
  \eqref{nec:eqn:notmaxent} is not a maximally entangled state
  (indeed, any state of the form $| \phi \rangle = u| 01 \rangle +
  v| 10 \rangle$ is non-maximally entangled for $u,v \neq
  \frac{1}{\sqrt 2}$). The reader who doubts this should compare the
  entropy of entanglement of such a state with the entropy of
  entanglement of any maximally entangled state.} (yielded by the
entropy of entanglement), while the reversibility of $E_C(\rho)$ and
$E_D(\rho)$ allows us to determine the optimal asymptotic rate of
transformation: $E(| \psi_1 \rangle\langle \psi_1 |) / E(| \psi_2
\rangle\langle \psi_2 |)$ between any two pure states $| \psi_1
\rangle$ and $| \psi_2 \rangle$ \citep[cf.][]{plenio2007}.

\paragraph{Entanglement of formation.} Unfortunately, the situation is
more complicated for \emph{mixed states}, where measures of
entanglement are not equivalent to the entropy of entanglement, and
where we do not have reversibility in general \citep[]{vidal2001}. For
mixed states, in fact, the distillable entanglement and entanglement
cost are in general extraordinarily difficult to compute. Very
little progress has been made on solving this problem directly;
however an alternative measure, the \emph{entanglement of formation},
offers some hope in this regard.

Given a mixed state, $\rho$, the entanglement of formation,
$E_F(\rho)$, represents the (lowest possible) average entanglement (as
measured by the entropy of entanglement) for pure state decompositions
of $\rho$; i.e.,
\begin{align}
E_F(\rho) \equiv \inf\left\{\sum_i p_i E(| \psi_i \rangle\langle
\psi_i |) : \rho = \sum_i p_i| \psi_i \rangle\langle \psi_i |\right\}.
\end{align}
Given that $E_F$ is expressed in terms of the entropy of entanglement
associated with the pure states making up particular decompositions of
$\rho$, one should expect $E_F$ to be closely related to $E_C$ and
$E_D$. Indeed, in its asymptotic version,
\begin{align}
E_F^\infty(\rho) \equiv \lim_{n\to\infty}\frac{E_F(\rho^{\otimes n})}{n},
\end{align}
the entanglement of formation has been shown to be equal to the
entanglement cost \citep[]{hayden2001}. While this, by itself, is of
little help in computing entanglement cost (the asymptotic
entanglement of formation is no less difficult to calculate), there
are indications (but no proof as of yet) that the entanglement of
formation is additive; i.e., that $E_F(\rho) =
E_F^\infty(\rho)$. Since the non-asymptotic version of $E_F$ is not
very difficult to calculate, then if the entanglement of formation is
truly additive, it would follow that $E_C$ is easily calculable as
well. Whether or not $E_F$ is additive, therefore, is an important
open question.

\paragraph{Negativity.} The \emph{negativity} \citep[cf.][]{vidal2002}
is an entanglement measure based on the trace norm of the partial
transpose of a bipartite mixed state $\rho^{AB}$. It measures the
degree to which the partial transpose of $\rho^{AB}$: $\rho^{T_A}
\equiv (T \otimes I)\rho_{AB}$ fails to be positive definite;
i.e., the degree to which $\rho^{AB}$ is entangled on Peres's
criterion of separability \citep[]{peres1996}; and it vanishes for
separable states. It is given by
\begin{align}
\label{nec:eqn:neg}
\mathcal{N}(\rho^{AB}) \equiv \frac{\lVert \rho^{T_A} \rVert_1 - 1}{2}.
\end{align}

A variant of the negativity is the \emph{multiplicative negativity}
\citep[]{datta2005}:
\begin{align}
\label{nec:eqn:mulneg}
\mathcal{M}(\rho^{AB}) \equiv 1 + 2 \mathcal{N}(\rho^{AB}).
\end{align}
This quantity is multiplicative in the sense that for a state which is
a product state of pairs of states, $\mathcal{M}$ for the overall
system is equal to the product of the individual values of
$\mathcal{M}$ for each pair.

The negativity is not difficult to calculate, and in its logarithmic
form,
\begin{align}
\label{nce:eqn:logneg}
E_\mathcal{N}(\rho^{AB}) \equiv \log_2 \lVert \rho^{T_A} \rVert_1,
\end{align}
the negativity is additive: $E_\mathcal{N}(\rho_1 \otimes \rho_2) =
E_\mathcal{N}(\rho_1) + E_\mathcal{N}(\rho_2)$ (likewise for the
logarithmic form of the multiplicative negativity). The logarithmic
negativity and logarithmic  multiplicative negativity, unfortunately,
are not monotonic (i.e., they increase under some LOCC
operations).\footnote{A bipartite entanglement measure $E(\rho)$
  mapping density matrices to positive real numbers is
  \emph{monotonic} if (i) $\rho$ is separable whenever $E(\rho) = 0$
  and (ii) $E$ does not increase when LOCC operations are applied to
  $\rho$.}

\subsubsection{Multi-partite entanglement}

The theory of entanglement measures extends beyond bipartite
entanglement to the more general case of multi-partite entanglement
(i.e., entangled systems that are shared between $n$
parties). Unsurprisingly, moving from the bipartite to the
multi-partite setting introduces complications. For instance, in the
multi-partite setting there is no straightforward analogue of a
bipartite maximally entangled state from which all other bipartite
states can be prepared using LOCC operations. In the tripartite
setting, for example, a natural candidate for a maximally entangled
state is the GHZ-state:
\begin{align}
\label{nec:eqn:ghzstate}
| \mbox{GHZ} \rangle = \frac{1}{\sqrt 2}(| 0 \rangle_A| 0 \rangle_B| 0
\rangle_C + | 1 \rangle_A| 1 \rangle_B| 1 \rangle_C).
\end{align}
Unfortunately some states are unobtainable from the GHZ-state using
LOCC alone; one example is the W-state:
\begin{align}
\label{nec:eqn:wstate}
| \mbox{W} \rangle = \frac{1}{\sqrt 3}(| 0 \rangle_A| 0 \rangle_B| 1
\rangle_C + | 0 \rangle_A| 1 \rangle_B| 0 \rangle_C + | 1 \rangle_A| 0
\rangle_B| 0 \rangle_C).
\end{align}

While we will make use of the concept of multi-partite entanglement in
what follows, we will not need to specifically consider multi-partite
entanglement measures. For a more in-depth treatment, see
\citet[]{plenio2007}.

\subsection{Purification}
\label{nec:ss:purify}

Every mixed state can be thought of as the result of taking the
partial trace of a pure state acting on a larger Hilbert space. In
particular, for a mixed state $\rho_A$ acting on a Hilbert space
$\mathcal{H}_A$, with spectral decomposition $\sum_k p_k | k
\rangle\langle k |$ for some orthonormal basis $\{| k \rangle\}$, a
purification (in general non-unique) of $\rho_A$ may be given by $$|
\psi_{AB} \rangle = \sum_k \sqrt{p_k} | k_A \rangle \otimes | k_B \rangle
\in \mathcal{H}_A \otimes \mathcal{H}_B,$$ where $\mathcal{H}_B$ is a
copy of $\mathcal{H}_A$. We then have $\rho_A = \mbox{tr}_B(|
\psi_{AB} \rangle\langle \psi_{AB} |)$, with  $| \psi_{AB} \rangle$ an
entangled state.

\section{The necessity of entanglement for explanation thesis}
\label{nec:s:net}

Recall our discussion of the Deutsch-Jozsa algorithm in \textsection
\ref{mwi:s:djalgo}. In the literature on quantum computation
(cf. \citealt[]{ekert1998, steane2003}) it is often suggested that
entanglement, such as that present in states like
\eqref{mwi:eqn:puredj}, is \emph{required} if a quantum algorithm is
to be capable of achieving a speedup over its classical
alternatives. I will call this the \emph{necessity of an entangled
  state} thesis (NEST). I will call the related claim that
entanglement is a necessary component of any \emph{explanation} for
quantum speedup the \emph{necessity of entanglement for explanation}
thesis (NEXT).\footnote{The attentive reader who has noticed that
  there is actually no entanglement in \eqref{mwi:eqn:puredj} when
  $n = 1$ will be somewhat puzzled by both of these theses. In fact,
  as we will see, entanglement will only appear for $n \geq 3$. In
  what follows I will argue, however, that this turns out to be
  evidence for, not against, the NEXT, and indeed does not contradict
  the NEST. This will be clarified in the next section.}

Note that although the NEXT is related to the NEST, these two claims
are not strictly speaking identical. As we will see in \textsection
\ref{nec:ss:mixdj}, it is possible for the NEXT to be true even if the
NEST is false (in the technical sense of \textsection
\ref{nec:ss:ent}), and it is not incoherent to argue that the NEXT is
false by citing, as a counter-example, a quantum computer whose state
is always entangled, as we shall see in \textsection
\ref{nec:ss:oneq}.

\section{De-quantisation}
\label{nec:s:deq}

At first sight the following consideration seems problematic for both
the NEST and the NEXT. Consider the Deutsch-Jozsa algorithm
(cf. \textsection \ref{mwi:s:djalgo}) for the special case of
$n=1$. This case is essentially a solution for Deutsch's
problem. Deutsch's \citeyearpar[]{deutsch1985} original solution to
this problem is regarded as the first quantum algorithm ever developed
and as the first example of what has since come to be known as quantum
speedup. If one considers the steps of the algorithm as given in
\textsection \ref{mwi:s:djalgo}, however, then the reader can confirm
that, when $n=1$, at no time during the computation are the two qubits
employed actually entangled with one another. The thesis that
entanglement is a necessary condition for quantum speedup thus seems
false. But the situation is not as dark for the NEST and the NEXT as
it appears, since for the case of $n=1$, it is also the case that the
problem can be `de-quantised', i.e., solved just as efficiently using
classical means.

One method for doing this \citep[cf.][]{abbott2010} is with a computer
which utilises the complex numbers $\{1,i\}$ as a computational basis
in lieu of $\{| 0 \rangle, | 1 \rangle\}$. A complex number
$z\in\mathbb{C}$ can be written as $z = a +bi$, where
$a,b\in\mathbb{R}$, and thus can be expressed as a superposition of
the basis elements in much the same way as a qubit.\footnote{Regarding
  the physical realisation of such a computer, note that complex
  numbers can be used, for instance, to describe the impedances of
  electrical circuits and that we can apply the superposition theorem
  to their analysis.} The algorithm proceeds in the following way. We
first note that the action of $U_f$ on the first $n$ qubits in
Eq. \eqref{mwi:eqn:puredj} can, for the case of $n=1$, be expressed
as:\footnote{Note that, since $f(0) = f(0)$, $(-1)^{f(0) \oplus f(0)
  \oplus f(1)} = (-1)^{f(1)}$.}
\begin{align*}
& \frac{1}{\sqrt 2}\Big((-1)^{f(0)}| 0 \rangle + (-1)^{f(1)}| 1
\rangle\Big) \nonumber \\
= & \frac{(-1)^{f(0)}}{\sqrt 2}\Big(| 0 \rangle + (-1)^{f(0) \oplus
  f(1)}| 1 \rangle\Big).
\end{align*}
We now define an operator $C_f$, analogously to $U_f$, that acts on a
complex number as follows:
$$C_f(a+bi) = (-1)^{f(0)}\Big(a + (-1)^{f(0) \oplus f(1)}bi\Big).$$
When $f$ is constant, the reader can verify that $C_f(z) = \pm (a+bi)
= \pm z$. When $f$ is balanced, $C_f(z) = \pm(a-bi) = \pm
z^*$. Multiplying by $z/2$ so as to project our output back on to the
computational basis, we find, for the elementary case of $z = 1 + i$,
that
\begin{eqnarray*}
f \mbox{ constant}: & \frac{1}{2}z\cdot\pm z = \pm i\\
f \mbox{ balanced}: & \frac{1}{2}z\cdot\pm z^* = \pm 1.
\end{eqnarray*}
Thus for any $z$, if the result of applying $C_f$ is imaginary, then
$f$ is constant, else if the result is real, then $f$ is balanced
(indeed, by examining the sign we will even be able to tell
\emph{which} of the two balanced or two constant functions $f$
is). This algorithm is just as efficient as its quantum counterpart.

It can similarly be shown \citep[cf.][]{abbott2010} that no
entanglement is present in \eqref{mwi:eqn:puredj} when $n=2$, and that
for this case also it is possible to solve the problem efficiently
using classical means.
When $n \geq 3$, however, \eqref{mwi:eqn:puredj_uni} is an entangling
evolution and \eqref{mwi:eqn:puredj} is an entangled
state. Unsurprisingly, it is no longer possible to define an operator
$C_f$ analogous to $U_f$ that takes product states to product states,
and it is no longer possible to produce an equally efficient classical
counterpart to the Deutsch-Jozsa algorithm \citep[cf.][]{abbott2010}.

Indeed, for the general case, \citeauthor{abbott2010} has shown that a
quantum algorithm can always be efficiently de-quantised whenever the
algorithm does not entangle the input states. Far from calling into
question the role of entanglement in quantum computational speedup,
the fact that the Deutsch-Jozsa algorithm does not require
entanglement to succeed for certain special cases actually provides
(since in these cases it can be de-quantised) evidence for both the
NEST and the NEXT.

\section{Challenges to the necessity of entanglement for explanation
  thesis}
\label{nec:s:mix}

In their own analysis of de-quantisation, \citet{jozsa2003} similarly
find that, for pure quantum states, ``the presence of multi-partite
entanglement, with a number of parties that increases unboundedly with
input size, is necessary if the quantum algorithm is to offer an
exponential speed-up over classical computation.''\footnote{For some
  earlier results relating to specific classes of algorithms, see
  \citet[]{linden2001,braunstein2002}. For a review, see
  \citet[]{pati2009}.} In the same article, however,
\citeauthor[]{jozsa2003} speculate as to whether it may be possible to
achieve exponential speedup, without entanglement, using \emph{mixed}
states. In fact, as we will now see, it is possible to achieve a
modest (i.e., sub-exponential) speedup using unentangled mixed
states. As I will argue, however, entanglement nevertheless plays an
important role in the computational ability of these states, despite
their being unentangled in the technical sense of \textsection
\ref{nec:ss:ent}. Thus, while such counter-examples demonstrate the
falsity of the NEST, they do not demonstrate the falsity of the NEXT.

\subsection{The mixed-state Deutsch-Jozsa algorithm}
\label{nec:ss:mixdj}

We will call a `pseudo-pure-state' of $n$ qubits any mixed state that
can be written in the form:
\begin{align*}
\rho_{\mbox{\tiny PPS}}^{\{n\}} \equiv \varepsilon| \psi \rangle\langle
\psi | + (1 - \varepsilon)\mathscr{I},
\end{align*}
where $| \psi \rangle$ is a pure state on $n$ qubits, and
$\mathscr{I}$ is defined as the totally mixed state
$(1/2^n)$I$_{2^n}$. It can be shown that such a state is separable
(cf. \textsection\ref{nec:ss:ent}) and remains so under unitary
evolution just so long as $$\varepsilon < \frac{1}{1 + 2^{2n - 1}}.$$

Now consider the Deutsch-Jozsa algorithm once again (cf. \textsection
\ref{mwi:s:djalgo}). This time, however, let us replace the initial
pure state $| 0^n \rangle| 1 \rangle$ with the pseudo-pure state:
\begin{equation}
\label{nec:eqn:djpps}
\rho = \varepsilon| 0^n \rangle| 1 \rangle\langle 0^n |\langle 1 | +
(1 - \varepsilon)\mathscr{I}.
\end{equation}
The algorithm will continue as before, except that this time our
probability of success will not be unity.

To illustrate: assume that the system represented by $\rho$ has been
prepared in the way most naturally suggested by \eqref{nec:eqn:djpps};
i.e., that with probability $\varepsilon$, it is prepared as the pure
state $| 0^n \rangle| 1 \rangle$, and with
probability $1 - \varepsilon$, it is prepared as the completely mixed
state $\mathscr{I}$. Now imagine that we write some of the valid
Boolean functions $f:\{0,1\}^n\to\{0,1\}$ onto balls which we then
place into an urn, and assume that these consist of an equal number of
constant and balanced functions. We select a ball from the urn and
then test the algorithm with this function to see if the algorithm
successfully determines $f$'s type.

Consider the case when $f$ is a constant function. In this case, we
will say the algorithm succeeds whenever it yields the bit string
$z=0$.  We know, from \textsection \ref{mwi:s:djalgo}, that the
algorithm will certainly succeed (i.e., with probability 1) when the
system is actually in the pure state $| 0^n \rangle| 1 \rangle$
initially. Given our particular state preparation procedure,
the system is in this state with probability $\varepsilon$. The rest
of the time (i.e., with probability $1 - \varepsilon$), the system is
in the completely mixed state $\mathscr{I}$. In this latter case,
since there are $2^n$ possible values that can be obtained for $z$,
the probability of successfully obtaining $z=0$ will be $1/2^n$. Thus
the overall probability of success associated with the system when $f$
is constant is:
\begin{equation}
\label{nec:eqn:fconstsucc}
P(z=0|f \mbox{ is constant}) = \varepsilon + (1 - \varepsilon)/2^n.
\end{equation}
The probability of failure is:
\begin{equation}
\label{nec:eqn:fconstfail}
P(z \neq 0| f\mbox{ is constant}) = \frac{2^n - 1}{2^n}\cdot(1 -
\varepsilon).
\end{equation}
In the case where $f$ is balanced, a result of $z \neq 0$ represents
success, and the respective probabilities of success and failure are:
\begin{align}
\label{nec:eqn:fbalsucc}
P(z \neq 0|f \mbox{ is balanced}) &= \varepsilon + \frac{2^n
  -1}{2^n}\cdot(1-\varepsilon), \\
\label{nec:eqn:fbalfail}
P(z = 0|f \mbox{ is balanced}) &= (1-\varepsilon)/2^n.
\end{align}

Note that as I mentioned in \textsection\ref{nec:ss:ent}, mixed states
can in general be prepared in a variety of ways. What I have above
called the `most natural' state preparation procedure associated with
\eqref{nec:eqn:djpps}, in particular, is only one of many possible
state preparations that will yield an identical density matrix
$\rho$. For ease of exposition, and in order to see clearly why
Eqs. (\ref{nec:eqn:fconstsucc}-\ref{nec:eqn:fbalfail}) hold, it was
easiest to assume, as I did above, that the system has been prepared
in the way most naturally suggested by \eqref{nec:eqn:djpps}. But
note that there is no loss of generality here; the identities
(\ref{nec:eqn:fconstsucc}-\ref{nec:eqn:fbalfail}) do not depend on the
fact that we have used this particular preparation procedure.

In any case, consider the alternative to the Deutsch-Jozsa algorithm
of performing \emph{classical} function calls on $f$ with the object
of determining $f$'s type. The reader should convince herself that a
single such call, regardless of the result, will not change the
probability of correctly guessing the type of the function $f$. Thus
the amount of information about $f$'s type that is gained from a
single classical function call is zero.\footnote{This information gain
  is referred to as the \emph{mutual information} between two
  variables (in this case, between the type of the function and the
  result of a function call). For more on the mutual information and
  other information-theoretic concepts, see Appendix
  \ref{ch:infth}.} On the other hand, as we should expect given
(\ref{nec:eqn:fconstsucc}-\ref{nec:eqn:fbalfail}), for the mixed-state
version of the Deutsch-Jozsa algorithm, it can be shown that the
information gained from a single invocation of the algorithm is
greater than zero for all positive $\varepsilon$, and that this is the
case even when $\varepsilon < \frac{1}{1 + 2^{2n - 1}};$ i.e., the
threshold below which $\rho$ no longer qualifies as an entangled
state. Indeed, this is the case even when $\varepsilon$ is arbitrarily
small \citep[cf.][]{biham2004}, although the information gain in this
case is likewise vanishingly small.

\subsubsection{Explaining speedup in the mixed-state Deutsch-Jozsa
  algorithm}

The first question that needs to be answered here is whether the
sub-exponential gain in efficiency that is realised by the mixed-state
Deutsch-Jozsa algorithm should qualify as quantum speedup at all. On
the one hand, from the point of view of computational complexity
theory (cf. Appendix \ref{ch:cc}), the solution to the Deutsch-Jozsa
problem provided by this algorithm is no more efficient than a
classical solution: from a complexity-theoretic point of view, a
solution $S_1$ to a problem $P$ is deemed to be just as efficient as a
solution $S_2$ so long as $S_1$ requires at most a polynomial increase
in the (time or space) resources required to solve $P$ as compared
with $S_2$. From this point of view, only an \emph{exponential}
reduction in time or space resources can qualify as a true increase in
efficiency. Clearly, the mixed-state Deutsch-Jozsa algorithm does not
yield a speedup over classical solutions, in this sense, when
$\varepsilon$ is small. In fact it can be shown
\citep[1148]{vedral2010} that exponential speedup, and hence a true
increase in efficiency from a complexity-theoretic point of view, is
achievable \emph{only} when $\varepsilon$ is large enough for the
state to qualify as an entangled state.

On the other hand, there is a very real difference, in terms of the
amount of information gained, between one invocation of the black box
\eqref{nec:eqn:djpps} and a single classical function call\textemdash
which is all the more striking since the amount of information one can
gain from a single classical function call is actually zero. Further,
one should not lose sight of the fact that the complexity-theoretic
characterisation of efficient algorithms is artificial and, in a
certain sense, arbitrary. For instance, on the complexity-theoretic
characterisation of computational efficiency, a problem, which for
input size $n$, requires $\approx n^{1000}$ steps to solve is
polynomial in terms of time resources in $n$ and thus tractable, while
a problem that requires $\approx 2^{n/1000}$ steps to solve is
exponential in terms of time resources in $n$ and therefore considered
to be intractable. In this case, however, the `intractable' problem
will typically require much less time to compute than the `tractable'
problem, for all but very large $n$.\footnote{For example, for $n =
  1,000,000$, the easy problem requires $(10^6)^{1000} = 10^{6000}$
  steps to complete while the hard problem requires $2^{1000}$
  steps.} Such extraordinary examples aside, for most practical
purposes the complexity-theoretic characterisation of efficiency is a
good one. Nevertheless it is important to keep in mind that this is a
practical definition of efficiency which does not reflect any deep
mathematical truth or make any deep ontological claim about what is
and is not efficient in the common or pre-theoretic sense of that
term.

But let us come back now from this slight digression to our main
discussion, and let us consider the question of whether entanglement
plays a role in the speedup exhibited by this mixed state. The
strongest argument in favour of a negative answer to this question is,
I believe, the following. Recall that what I have called the `most
natural' state preparation procedure associated with
\eqref{nec:eqn:djpps} is only one of many possible ways to prepare the
system represented by $\rho$. It is possible to prepare the system in
an alternate way if we so desire. Likewise, when $\varepsilon$ is
sufficiently small, it is possible to prepare the final state of the
computer, $\rho_{fin}$, as a mixture of product states. This, in fact, is
the significance of asserting that $\rho_{fin}$ is unentangled. Thus
while the state preparation most naturally suggested by
\eqref{nec:eqn:djpps} may well function as a conceptual tool for
\emph{finding} mixed quantum states that display a computational
advantage (i.e., by enabling a facile derivation of the identities
(\ref{nec:eqn:fconstsucc}-\ref{nec:eqn:fbalfail})), \emph{once found},
it seems as though we may do away with this way of thinking of the
system entirely. Hence there seems to be no need to invoke
entanglement in order to explain the speedup obtainable with this
state.

I believe this line of reasoning to be misleading, however, for it
emphasises the abstract density operator representation of the
computational state at the cost of obscuring the nature of the
computational process that is actually occurring in the computer. To
the point: the density operator corresponding to a quantum system
should not be understood as a representation of the actual physical
state of the system. Rather, the density operator representation of a
quantum system should be understood as a representation of our
knowledge of the space of physical states that the system can possibly
be in, and of our ignorance as to which of these physical states the
system is actually in.

From the point of view of quantum mechanics, it is \emph{pure} states
of quantum systems which should be seen as representations of the
`actual' physical states of such systems, for pure states represent
the most specific description of a system that is possible from within
the theory. I have enclosed the word \emph{actual} within inverted
commas in the preceding sentence in order to emphasise the weakness of
the claim I am making. This claim is not intended to rule out that
there may be a deeper physical theory underlying quantum mechanics,
within which quantum mechanical pure states can be seen as merely
derivative representations. Nor is it intended to rule out that
quantum mechanics only incompletely (as a matter of principle)
specifies the nature of the physical world. I am only making what
should be the uncontroversial claim that relative to quantum mechanics
itself, pure states should be interpreted as those which are most
fundamental, in the sense that they represent a maximally specific
description, within the theory, of the systems in question\textemdash
i.e., they represent the best \emph{grasp} available, from within that
theory, of the real physical situation.

Physics is the science of what is real, in the very minimal sense that
physical concepts \emph{purport} to \emph{give us some idea} of what
the world is like. And if pure states represent the best possible,
i.e., the most specific, representation of the physical situation from
the point of view of a theory, then with right should they be treated
as the more fundamental concepts of the theory. Mixed states, on the
other hand, should be seen as derivative in the sense that they are
abstract characterisations of our knowledge of the space \emph{of pure
  states} a system may be possibly in,\footnote{If one prefers, one
  can think of a mixed state as a \emph{statistical} state,
  representing the mean values of a hypothetical ensemble of
  systems. The difference is inessential to this discussion.} and of
our ignorance of precisely which state within this space the system is
actually in.

If the reader accepts this difference in fundamental
status that I have accorded to pure and mixed quantum
states,\footnote{My claim is intended to be weak enough to be
  compatible with interpretations of the quantum state such as
  Spekkens's, in which quantum states are analogous to the state
  descriptions of his toy theory (cf. Appendix \ref{ch:transf}), in
  that they represent \emph{maximal}, though in principle
  incomplete, knowledge \emph{of} the system in question. It is also
  intended to be compatible with Fuchs's statement that ``... the
  quantum state represents a collection of subjective degrees of
  belief about \emph{something} to do with that system ...''
  \citep[989-990]{fuchs2003}. Nevertheless, the compatibility of my
  claim with Fuchs's and Spekkens's views may be doubted by some. This
  is not the place to attempt to give a reading of either Fuchs's or
  Spekkens's opinions on the interpretation of the quantum state
  description, however. While I may be incorrect as regards the
  compatibility of my claim with their views, I hope that most
  readers will, regardless, appreciate the benign nature of and be
  agreeable to the claim that I am making here. In any case I will be
  assuming it in the remainder of this dissertation. (For a more
  in-depth treatment of Fuchs's and Spekkens's interpretation of the
  quantum state description, see: \citealt{tait2012}.)} then she
should agree that if it is an explanation of the physical process
actually occurring in the computer that we desire, then it will not do
to limit ourselves to analysing the characteristics of the computer's
`black box' mixed state; rather, we should attempt to give a more
detailed `white box' characterisation of the operation of the computer
in terms of its underlying pure states. Recall the fact\textemdash
which we noted in our earlier discussion of de-quantisation\textemdash
that the unitary evolution \eqref{mwi:eqn:puredj_uni} is, in general,
an \emph{entangling   evolution}; i.e., it will take pure product
states, such as, for instance, $| 0^n \rangle| 1 \rangle$, to
entangled states. Now imagine that the computer is initially prepared
in the most natural way suggested by the pseudo-pure state
representation \eqref{nec:eqn:djpps}. Call this `most natural' state
preparation: $s_{ini}$. Imagine further that the computer evolves in
accordance with the entangling unitary transformation $U_f$. This will
yield the transformation
$$
| 0^n \rangle| 1 \rangle \xrightarrow{U_f} | \phi
\rangle
$$
with probability $\varepsilon$, and the transformation
$$
\mathscr{I} \xrightarrow{U_f} \mathscr{I}
$$
with probability $1 - \varepsilon$, where $| \phi \rangle$ is an
entangled state. Thus at the end of the computation, the system will
be in the state $| \phi \rangle$ with probability $\varepsilon$ and in
the state $\mathscr{I}$ with probability $1 - \varepsilon$. Call this
combination of possible states for the system $s_{fin}$. Now at the
end of the computation, the state of the computer will be expressible
by means of the density operator
$$
\rho_{fin} = \varepsilon | \phi \rangle\langle \phi | + (1 -
\varepsilon)\mathscr{I}.
$$
The most natural way that suggests itself for preparing the system
represented by $\rho_{fin}$ is $s_{fin}$. However, one may instead
imagine a state preparation procedure $s'_{fin}$ involving only
product states that would result in an equivalent density operator
representation. Because of this, it is concluded by some that
entanglement plays no role in the computational advantage exhibited by
the computer in this case.

The significance of the fact that $U_f$ is an entangling evolution,
however, is that $s_{ini}$, evolved in accordance with $U_f$, will
\emph{not} result in the combination of states $s'_{fin}$\textemdash
rather, it will result in the combination of states $s_{fin}$. Since
both state preparations, $s_{fin}$ and $s'_{fin}$, yield the same
density matrix representation, they are, from this point of view,
equivalent, but one \emph{cannot} directly obtain $s'_{fin}$ from an
application of $U_f$ to $s_{ini}$.\footnote{I am indebted to Wayne
  Myrvold for suggesting this line of thought, and for helping to
  clear up the conceptual confusions regarding this issue that have
  plagued me to date. I am also indebted to the discussion in
  \citet[\textsection 5]{jozsa2003}. I should note, also, that
  \citet[]{long2002} make a similar point to the one made here; but
  in making it they unnecessarily rely on interpreting the density
  matrix of a system as representing the average values of a
  physical ensemble (i.e. of an actual collection of physical
  systems). The objection is equally forceful, however, whether one
  thinks of the mixed state as representing a physical or a
  statistical ensemble, and whether one thinks of the probabilities
  as ignorance probabilities or as representing relative
  frequencies.}

What of the fact, however, that $\varepsilon$ in the state preparation
$s_{fin}$ may be \emph{vanishingly small} in principle and yet
\emph{still} lead to a computational advantage\textemdash does not
this tell against attributing the speedup exhibited by the computer to
entanglement? I do not believe it does. One must not lose sight of the
fact that ``vanishingly small'' $\neq$ 0. If $\varepsilon$ \emph{were}
actually equal to zero, it is evident that there would, in fact, be no
performance advantage.

It is interesting, nevertheless, to consider the question of what can
happen in the quantum computer when $\varepsilon = 0;$ i.e., when the
state of the computer initially just is the totally mixed state
$\mathscr{I}$. Note that this does not signify that it is impossible
for the computer to actually have been prepared in the pure state $|
0^n \rangle| 1 \rangle$ initially. Rather, it represents the
circumstance where we are \emph{completely ignorant} of the initial
state preparation of the quantum computer; for instance, if the
computer has been prepared as an equally weighted mixture of the basis
states:
\begin{align}
\label{eqn:totallymixed}
\rho_{ini} = \mathscr{I} = \frac{1}{2^n}\sum_{x=0}^{2^n - 1}| x
\rangle\langle x |.
\end{align}
Suppose then, that the quantum computer, represented by the density
operator $\rho_{ini} = \mathscr{I}$, actually is in $| 0^n \rangle| 1
\rangle$ at the start of the computation. Is a computational process
occurring which would enable quantum speedup? From one point of view,
the answer is yes, for the entangling unitary evolution $U_f$ evolves
the computer to an entangled state which is then capable of being
utilised in principle in order to solve the problem under
consideration with fewer computational resources than a classical
computer. In fact, it is not even necessary for the computer to
actually be in the state $| 0^n \rangle| 1 \rangle$ initially to
enable a performance advantage. As long as we know, or at least are
not completely ignorant of, the actual initial pure state of the
computer, any of the basis states can, with suitable manipulation, be
used to obtain a performance advantage.

From another point of view, however, the answer is no, for because we
are completely ignorant as to the actual initial state of the
computer, we will be completely ignorant as to which operation to
perform in order to take advantage of this resource. This sounds
paradoxical, but I think it rather illustrates a distinction that
needs to be drawn here which will recur more than once in this
dissertation: between what is actually occurring in a physical system,
on the one hand, and the use which can be made of it by \emph{us}, who
are attempting to achieve some particular end. In the example we are
considering here there assuredly is a process occurring in the
computer that is of the right sort to enable a quantum speedup, but
because we are completely ignorant of the computer's initial
state\textemdash i.e., because there is too much `noise' in the
computer\textemdash we are unable to take advantage of it to achieve
the end of solving the Deutsch-Jozsa problem using fewer computational
resources than a classical computer.

\subsection{The power of one qubit}
\label{nec:ss:oneq}

In the last subsection we saw that it is possible to achieve a
sub-exponential speedup for the Deutsch-Jozsa problem with an
unentangled mixed-state. We concluded that while this does disprove
the NEST, it does not constitute a counter-example to the NEXT, since
the computational algorithm in question is successful only when the
evolution of the state of the computer is an entangling evolution;
therefore the underlying final state of the computer will always
contain some entanglement despite the fact that the density operator
representation of the final state will be unentangled.

We now consider another purported counter-example to the NEXT. This is
the \emph{deterministic quantum computation with one qubit} (DQC1)
model of quantum computation, which utilises a mixed quantum state to
compute the trace of a given unitary operator and displays an
\emph{exponential} speedup over known classical solutions. As we will
see, the claim sometimes made to the effect that the DQC1 achieves
this speedup without the use of entanglement is unsubstantiated. The
NEXT, however, is not the claim that any state that displays quantum
computational speedup must be entangled. That is the NEST. The NEXT
is, rather, the different claim that entanglement must play a role in
any physical \emph{explanation} of quantum speedup. We saw in the last
section how it is possible for the NEST to be false\footnote{I mean
  false in the technical sense explained in \textsection
  \ref{nec:ss:ent}.} and the NEXT to be true. In this section I will
address the objection that the NEXT is false even if it is the case
that the state of the quantum computer is always entangled. Those
defending such a view claim that another measure of quantum
correlations, \emph{quantum discord}, is far better suited for the
explanatory role. In what follows I will argue that this conclusion is
misguided. Quantum discord is indeed an enormously useful theoretical
quantity for characterising mixed-state quantum computation\textemdash
perhaps even more useful than entanglement. Nevertheless, more than
just pragmatic considerations must be appealed to if one is to make
the case that a particular feature of quantum systems explains quantum
speedup. Thus I will argue that when one looks deeper, and considers
the quantum state from the \emph{multi-partite} point of view, one
finds that entanglement is involved in the production, and even in the
very definition, of quantum discord; indeed, there are some
preliminary indications that quantum discord is, in fact, but a
manifestation of and not conceptually distinct from entanglement.

\subsubsection{The DQC1}

\begin{figure}
  \begin{align*}
  \Qcircuit @C=.5em @R=-.5em {
  & \lstick{\frac{1}{2}(I + \alpha Z)} & \gate{H} & \ctrl{1} & \meter &
  \push{\rule{0em}{4em}} \\
  & & \qw & \multigate{4}{U_n} & \qw & \qw \\
  & & \qw & \ghost{U_n} & \qw & \qw \\
  \lstick{\mbox{$I_n/2^n$}} & & \qw & \ghost{U_n} & \qw & \qw \\
  & & \qw & \ghost{U_n} & \qw & \qw \\
  & & \qw & \ghost{U_n} & \qw & \qw \gategroup{2}{2}{6}{2}{.6em}{\{}
  }
  \end{align*}
  \caption[The DQC1 model]{The DQC1 algorithm for computing the
  trace of a unitary operator.}
  \label{nec:fig:dqc1}
\end{figure}
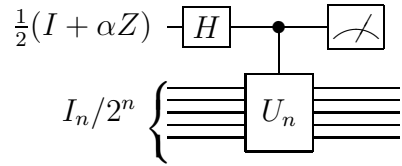

In the DQC1, or as it is sometimes called: `the power of one qubit',
model of quantum computation \cite[cf.][]{knill1998},\footnote{In this
  exposition of the DQC1, I am closely following \citep[]{datta2005}.} a
collection of $n$ `unpolarised' qubits in the completely mixed state
$I_n/2^n$ is coupled to a single `polarised' control qubit, initialised
to $1/2(I + \alpha Z)$. When the polarisation, $\alpha$, is equal to
1, the control qubit is in the pure state $| 0 \rangle \langle 0 | =
1/2(I + Z)$,
otherwise it is in a mixed state. The problem is to compute the trace
of an arbitrary $n$-qubit unitary operator, $\mbox{Tr}(U_n)$. To
accomplish this, we begin by applying a Hadamard gate to the control
qubit,\footnote{This will yield, for instance, when the control qubit
  is pure, $| 0 \rangle\langle 0 | \xrightarrow{H} \frac{1}{2}\big(|
  0 \rangle\langle 0 | + | 0 \rangle\langle 1 | + | 1 \rangle\langle
  0 | + | 1 \rangle\langle 1 |\big).$} which is then forwarded as part
of the input to a controlled unitary gate that acts on the $n$
unpolarised qubits (see Figure \ref{nec:fig:dqc1}). This results in the
following state for all of the $n+1$ qubits:
\begin{align}
\label{nec:eqn:dqc1}
\rho_{n+1} & = \frac{1}{2^{n+1}}\big(| 0 \rangle\langle 0 | \otimes I_n +
| 1 \rangle\langle 1 | \otimes I_n + \alpha | 0 \rangle\langle 1 |
\otimes U_n^\dagger + \alpha | 1 \rangle \langle 0 | \otimes U_n\big)
\nonumber \\
& = \frac{1}{2^{n+1}}
\left(
\begin{matrix}
  I_n & \alpha U_n^\dagger \\
  \alpha U_n & I_n
\end{matrix}
\right).
\end{align}

The reduced state of the control qubit is
$$\rho_c = \left(
\begin{matrix}
  1 & \alpha\mbox{Tr}(U_n)^\dagger \\
  \alpha\mbox{Tr}(U_n) & 1
\end{matrix}
\right),$$
thus the trace of $U_n$ can be retrieved by applying the $X$ and $Y$
Pauli operators to $\rho_c$. In particular, the expectation values of
the $X$ and $Y$ operators will yield the real and imaginary parts of
the trace, $\langle X \rangle = \mbox{Re}[\mbox{Tr}(U_n)]/2^n$ and
$\langle Y \rangle = -\mbox{Im}[\mbox{Tr}(U_n)]/2^n$, respectively; so
in order to determine, for instance, the real part, we run the circuit
repeatedly, measuring $X$ on the control qubit at the end of each run,
while assuming that the results are part of a distribution whose mean
is the real part of the trace.

Classically, the problem of evaluating the trace of a unitary matrix
is believed to be hard, however for the quantum algorithm it can be
shown that the number of runs required does not scale exponentially
with $n$, yielding an exponential advantage for the DQC1 quantum
computer. When $\alpha < 1$, the expectation values, $\langle X
\rangle$ and $\langle Y \rangle$, are reduced by a factor of $\alpha$
and it becomes correspondingly more difficult to estimate the
trace. However as long as the control qubit has non-zero polarisation,
the model still provides an efficient method for estimating the trace
(and thus an exponential speedup over any known classical solution) in
spite of this additional overhead.

We might ask whether, in a way analogous to the mixed-state
Deutsch-Jozsa algorithm, we can make $\alpha$ small enough so that the
overall state of the DQC1 is demonstrably separable. The answer seems
to be no. On the one hand, for any system of $n+1$ qubits there is a
ball of radius $r$ (measured by the Hilbert-Schmidt norm and centred
at the completely mixed state), within which all states are separable
\citep[]{braunstein1999,gurvits2003}. On the other hand, the state of
the DQC1 is at all times at a fixed distance $\alpha 2^{-(n+1)/2}$
from the completely mixed state. Unfortunately the radius of the
separable ball decreases exponentially faster than $2^{-(n+1)/2}$
\citep[2]{datta2005}.

\begin{figure}
\begin{tabular}{lllllll}
\raisebox{2.5em}{(a)} & \includegraphics[scale=0.25]{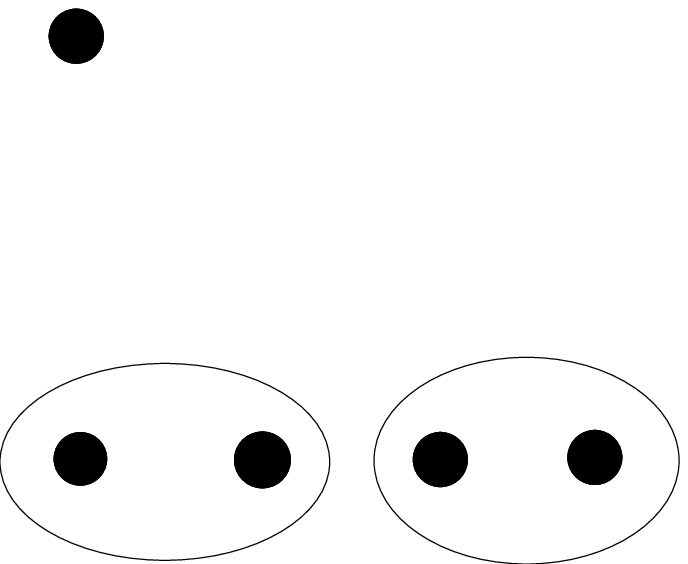} & &
\raisebox{2.5em}{(b)} & \includegraphics[scale=0.25]{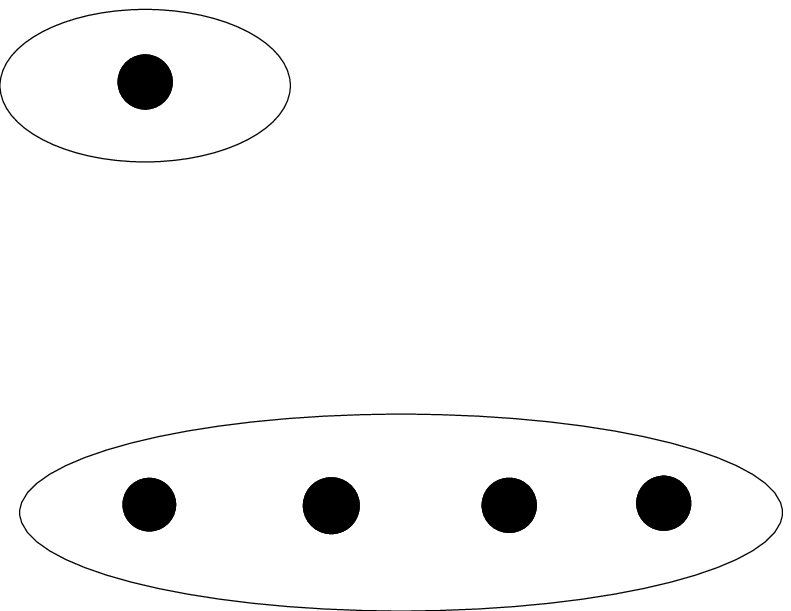}\\
\\
\raisebox{2.5em}{(c)} & \includegraphics[scale=0.25]{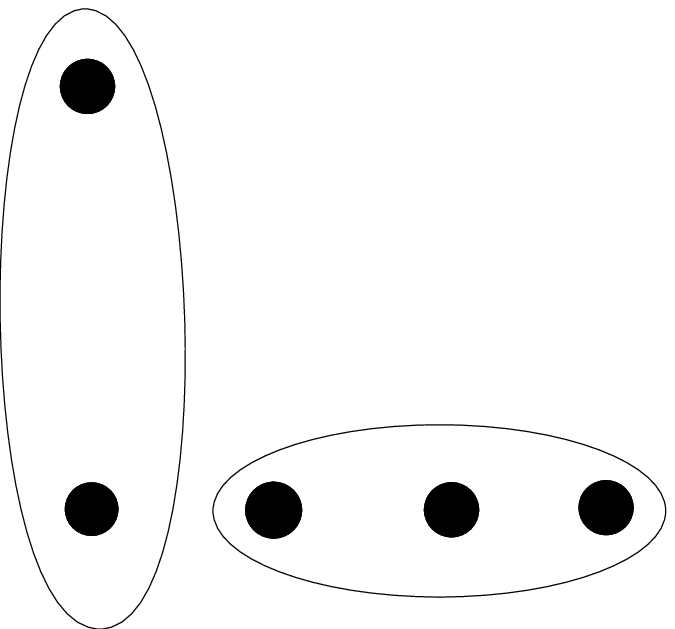} & &
\raisebox{2.5em}{(d)} & \includegraphics[scale=0.25]{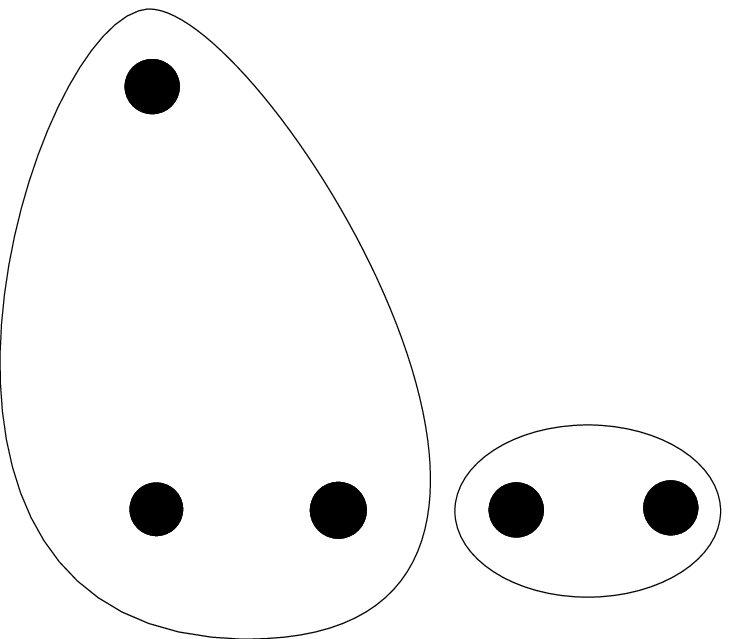} &
\raisebox{2.5em}{(e)} & \includegraphics[scale=0.25]{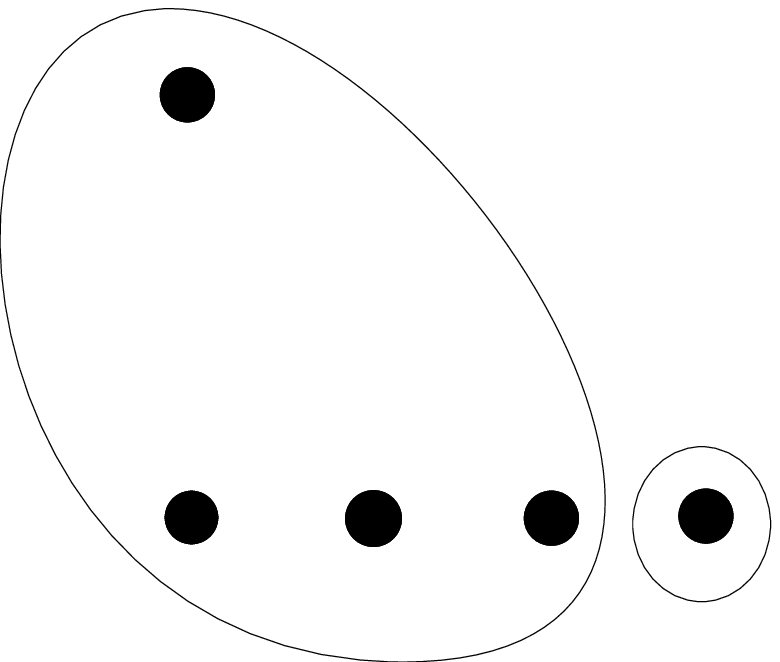}
\end{tabular}
\caption[DQC1 bipartite splits]{Some of the bipartite splits possible
  in the DQC1 for $n=4$. No entanglement can ever occur amongst the
  $n$ unpolarised qubits (a) or between the polarised qubit and the
  rest (b); however, bipartite splits such as (c), (d), and (e) can
  exhibit entanglement \citep[]{datta2005}.}
\label{nec:fig:dqc1_splits}
\end{figure}

Thus, as \citep[2]{datta2005} assert, there appears to be good reason
to suspect that the state \eqref{nec:eqn:dqc1} is an entangled state,
at least for some $U_n$; but it is not obvious where this entanglement
\emph{is}. On the one hand, there is no bipartite entanglement among
the $n$ unpolarised qubits. On the other hand the most natural
bipartite split of the system, with the control qubit playing the role
of the first subsystem and the remaining qubits playing the role of
the second, reveals no entanglement between the two subsystems,
regardless of the choice of $U_n$. When $\alpha > 1/2$, entanglement
can be found when we examine other bipartite divisions amongst the
$n+1$ qubits (see Figure \ref{nec:fig:dqc1_splits}), however, besides
being exceedingly difficult to detect, the amount of entanglement in
the state (as measured by the multiplicative negativity;
cf. \textsection \ref{nec:ss:quantent}) becomes vanishingly small as
$n$ gets large. Commenting on this circumstance, \citet[13]{datta2005}
write ``This hints that the key to computational speedup might be the
global character of the entanglement, rather than the amount of the
entanglement. ... what happier motto can we find for this state of
affairs than \emph{Multam ex Parvo}, or A Lot out of A Little.''

Others have expressed a different viewpoint on the matter. In fact,
both the DQC1 and the mixed-state version of the Deutsch-Jozsa
algorithm have led many (see for instance, \citealt[]{vedral2010}) to
seriously question whether entanglement plays a necessary role in the
explanation of quantum speedup. The result has been a shift in
investigative focus from entanglement to other types of quantum
correlations. One alternative in particular, \emph{quantum discord}
(which I will explain in more detail shortly), has received much
attention in the literature in recent years
\citep[see, e.g.,][]{merali2011}.

On the one hand, the following facts
all seem to run counter to the NEXT: there is no entanglement in the
DQC1 circuit between the polarised and unpolarised qubits\textemdash
the most natural bipartite split that suggests itself\textemdash
during a computation; tests to detect entanglement along other
bipartite splits in the DQC1 when $\alpha \leq 1/2$ have thus far been
unsuccessful;\footnote{The criterion used by \citet[]{datta2005} to
  detect entanglement is the Peres-Horodecki, or Positive Partial
  Transpose (PPT) criterion \citep[]{peres1996,horodecki1996}. The
  partial transpose of a bipartite system, $\sum_{ijkl}p^{ij}_{kl}|
  i \rangle\langle j |\otimes | k \rangle\langle l |$ acting on
  $\mathcal{H}_A \otimes \mathcal{H}_B$ is defined (with respect to
  the system $B$) as: $$\rho^{T_B} \equiv (I \otimes T)\rho =
  \sum_{ijkl}p^{ij}_{kl}| i \rangle\langle j |\otimes (| k
  \rangle\langle l |)^T = \sum_{ijkl}p^{ij}_{kl}| i \rangle\langle j
  |\otimes | l \rangle\langle k |,$$ where $T$ is the transpose map
  on matrices. The PPT criterion states that, if $\rho$ is a
  separable state, then the partial transpose of $\rho$ has
  non-negative eigenvalues. Satisfying the PPT criterion is a
  necessary (but not sufficient) condition for the joint density
  matrix of two systems to be separable. While
  \citeauthor[]{datta2005} were unable to detect entanglement in the
  DQC1 (along any bipartite split) for the case of $\alpha \leq
  1/2$, they nevertheless note that it is very likely that both
  entanglement and bound entanglement are present in the state. A
  state exhibits \emph{bound entanglement} \citep[cf.][]{hyllus2004}
  when, in spite of the fact that it is entangled, no pure entangled
  state can be obtained from it by means of LOCC operations. One
  important characteristic of bound entangled states is that they
  (at least sometimes) satisfy the PPT criterion despite the fact
  that they are entangled.
} and finally, even when $\alpha$ is relatively large, only a
vanishingly small amount of entanglement can be found in the state of
the DQC1 \eqref{nec:eqn:dqc1}. On the other hand, when we consider the
correlations between the polarised and unpolarised qubits from the
point of view of \emph{quantum discord}, it turns out that the discord
at the end of the computation is \emph{always} non-zero along this
bipartite split for \emph{any} $\alpha > 0$
\citep[]{datta2008}. \citet[4]{datta2008} therefore write, \emph{and I
  agree}, that ``for some purposes, quantum discord might be a better
figure of merit for characterizing the quantum resources available to
a quantum information processor.'' All the same, as I will argue
below, it is a mistake to conclude as they and others do that the NEXT
is false; i.e., that entanglement may play no role in the explanation
of the quantum speedup of the DQC1
\citep[]{datta2008,vedral2010,merali2011}; for the NEXT \emph{is
  compatible} with all of these facts.

\subsubsection{Quantum discord}

Quantum discord
\citep[]{zurek2000,henderson2001,ollivier2002}\footnote{Quantum
  discord was introduced independently by both
  \citeauthor[]{henderson2001} and by \citeauthor[]{ollivier2002},
  with slight differences in their respective formulations
  (\citeauthor[]{henderson2001} consider not just projective
  measurements but positive operator valued measures more
  generally). These and other alternative formulations of quantum
  discord do not differ in essentials. The definition of discord I
  introduce here is \citeauthor[]{ollivier2002}'s.} quantifies the
difference between the quantum generalisations of two classically
equivalent measures of mutual information,\footnote{See Appendix
  \ref{ch:infth} for an overview of the basic concepts of classical
  and quantum information theory.}
\begin{align}
\label{nec:eqn:clasmut1}
\mathcal{I}_c(A:B) & = H(A) + H(B) - H(A,B), \\
\label{nec:eqn:clasmut2}
\mathcal{J}_c(A:B) & = H(A) - H(A|B).
\end{align}
These two expressions are not equivalent quantum mechanically, for
while \eqref{nec:eqn:clasmut1} has a straightforward quantum
generalisation in terms of the von Neumann entropy $S$:
\begin{align}
\label{nec:eqn:quantmut1}
\mathcal{I}_q(A:B) & = S(A) + S(B) - S(A,B),
\end{align}
things are more complicated for the quantum generalisation of
\eqref{nec:eqn:clasmut2}. The quantum counterpart, $S(A|B)$, to the
conditional entropy requires a specification of the information
content of $A$ given a determination of the state of $B$. Determining
the state of $B$ requires a measurement, however, which requires the
choice of an observable. But in quantum mechanics observables are, in
general, non-commuting. Thus the conditional entropy will be different
depending on the observable we choose to measure on $B$. If, for
simplicity, we consider only perfect measurements, represented by a
set of one dimensional projection operators, $\{\Pi_j^B\}$, this
yields, for the quantum version of \eqref{nec:eqn:clasmut2}, the
expression:
\begin{align}
\label{nec:eqn:quantmut2}
\mathcal{J}_q(A:B) & = S(A) - S(A|\{\Pi_j^B\}).
\end{align}
We now define discord as the minimum value (taken over $\{\Pi_j^B\}$)
of the difference between \eqref{nec:eqn:quantmut1} and
\eqref{nec:eqn:quantmut2}:
\begin{align}
\label{nec:eqn:discord}
\mathcal{D}(A,B) \equiv \mbox{min}_{\{\Pi_j^B\}}\mathcal{I}_q(A:B) -
\mathcal{J}_q(A:B).
\end{align}
Discord is, in general, non-zero for mixed states, while for pure
states it effectively becomes a measure of entanglement
\citep[3]{datta2008}; i.e., for pure states it is equivalent to the
entropy of entanglement (cf. \textsection \ref{nec:ss:quantent}).

Interestingly, there are some mixed states which, though
\emph{separable}, exhibit non-zero quantum discord. For instance,
consider the following bipartite state:
\begin{align}
\label{nec:eqn:discexamp}
\rho_{\mbox{\tiny disc}} = \frac{1}{2}(| 0 \rangle\langle 0 |_A
\otimes | 0 \rangle\langle 0 |_B) + \frac{1}{2}(| 1 \rangle\langle 1
|_A \otimes | + \rangle\langle + |_B).
\end{align}
This state is obviously separable. Since $| 0 \rangle$ and $| +
\rangle$ are non-orthogonal states, however, $\mathcal{J}_q(A:B)$ will
yield a different value depending on the experiment performed on
system $B$; and thus this state will yield a non-zero quantum
discord. Note that this is impossible for a classical state:
classically, it is \emph{always} possible to prepare a state as a
mixture of \emph{orthogonal} product states.

In most of the literature on this topic, one is introduced to quantum
discord as a quantifier of the non-classical \emph{correlations}
present in a state which are not necessarily identifiable with
entanglement. Such an interpretation of the significance of this
quantity is supported by the fact that, in the classical scenario at
least, the mutual information contained in a system of two random
variables is held to be representative of the extent of the
correlations between them. Since the quantum generalisations of the
two classically equivalent measures of mutual information
$\mathcal{I}_c(A:B)$ and $\mathcal{J}_c(A:B)$ are not equivalent,
then, this is taken to represent the presence of non-classical
correlations over and above the classical ones, some, but not all of
which may be accounted for by entanglement, and some by `quantum
discord'.

Interpreting discord as a type of non-classical correlation is
nevertheless puzzling. Consider, for instance, a classically
correlated state represented by the following probability
distribution:
\begin{align}
\label{nec:eqn:strcl}
\frac{1}{2}([+]_l,[+]_r) + \frac{1}{2}[-]_l[-]_r.
\end{align}
Here, let $[\cdot]_l$ represent the circumstance that Linda (in
Liverpool) finds a letter in her mailbox today containing a piece of
paper on which is inscribed the specified symbol ($+$ or $-$), and let
$[\cdot]_r$ represent the occurrence of a similar circumstance for
Robert (in Ravenna). According to the probability distribution, it is
equally likely that they both receive a letter today inscribed with
$+$ as it is that they both receive one inscribed with $-$, but it
cannot happen that they each today receive letters with non-matching
symbols. These correlations are easily explainable classically, of
course. It so happens that yesterday I flipped a fair coin. I observed
the result of the toss and accordingly jotted down either $+$ or $-$
on a piece of paper, photocopied it, and sent one copy each to Robert
in Ravenna and Linda in Liverpool (by overnight courier, of course).

A quantum analogue for classically correlated states such as
\eqref{nec:eqn:strcl} is a mixed state decomposable into product
states:
\begin{align}
\label{nec:eqn:qucl}
\sum_{ij} p_{ij} | i \rangle\langle i | \otimes | j \rangle\langle j |
\end{align}
such that the $| i \rangle$ and $| j \rangle$ are mutually orthogonal
sub-states of the first and second subsystem, respectively. For such a
state it is easy to provide a `hidden variables' explanation, similar
to the one above, that will account for the observed probabilities of
joint experiments on the two subsystems.

We can equally give such a local hidden variables account of the
discordant state $\rho_{\mbox{\tiny disc}}$: tossing a fair coin, I
prepare the state $| 0 \rangle\langle 0 |_A \otimes | 0 \rangle\langle
0 |_B$ if the coin lands heads, and $| 1 \rangle\langle 1 |_A \otimes
| + \rangle\langle + |_B$ if it lands tails. Let $Pr(X, Y | a, b,
\lambda)$ refer to the probability that Alice's $a$-experiment and
Bob's $b$-experiment determine their qubits to be in states $X$ and
$Y$, respectively, given that the result of the coin toss is
$\lambda$. Then (omitting bras and kets for readability):
\begin{align*}
Pr(0, 0 | \hat{z}, \hat{z}, H) & = Pr(0, \cdot | \hat{z}, \cdot, H)
\times Pr(\cdot, 0 | \cdot, \hat{z}, H) = 1, \\
Pr(1, 1 | \hat{z}, \hat{z}, T) & = Pr(1, \cdot | \hat{z}, \cdot, T)
\times Pr(\cdot, 1 | \cdot, \hat{z}, T) = 1/2, \\
Pr(0, + | \hat{z}, \hat{x}, H) & = Pr(0, \cdot | \hat{z}, \cdot, H)
\times Pr(\cdot, + | \cdot, \hat{x}, H) = 1/2, \\
Pr(1, + | \hat{z}, \hat{x}, T) & = Pr(1, \cdot | \hat{z}, \cdot, T)
\times Pr(\cdot, + | \cdot, \hat{x}, T) = 1,
\end{align*}
and so on. More generally, $Pr(X,Y | a,b, \lambda) = Pr(X, \cdot | a,
\cdot, \lambda) \times Pr(\cdot, Y | \cdot, b, \lambda)$. Thus once we
specify the value of $\lambda$ there are no remaining correlations in
the system and the probabilities for joint experiments are
\emph{factorisable}. This should be unsurprising. Given a
specification of $\lambda$, the state of the system is in a
\emph{product state}, after all, and thus can be prepared (as we saw
earlier) using only LOCC operations.

Contrast this with an entangled quantum system such as, for instance,
the one represented by the pure state $$|\Phi^+\rangle = \frac{| 00
  \rangle + | 11 \rangle}{\sqrt{2}}.$$ Bell's theorem (to be discussed
in more detail in the following chapters) demonstrates that the
correlations between subsystems present in such a state cannot be
reproduced by any local hidden variables theory in the manner
described above. These correlations are non-classical.

There is certainly \emph{something} non-classical about a state such
as $\rho_{\mbox{\tiny disc}}$; viz., a quantum state such as
$\rho_{\mbox{\tiny disc}}$, though separable, cannot be prepared as a
mixture of orthogonal product states. Yet it is always possible to so
prepare classical states. As a result, the information one can gain
about Alice's system through an experiment in the $\{+, -\}$ basis on
Bob's system will be different from the information one can gain about
Alice's system through an experiment in the computational basis on
Bob's system. On the one hand, in the absence of a specification of a
hidden parameter such as $\lambda$, given an experiment on $B$ in the
computational basis which determines $B$ to be in state $| 0 \rangle$,
it is still unclear, because of the way in which system $B$ was
prepared, whether the joint system is in the state $| 0 \rangle
\otimes | 0 \rangle$ or in the state $| 1 \rangle \otimes | +
\rangle$. Given an experiment on $B$ in the $\{+, -\}$ basis which
yields $| + \rangle$, on the other hand, it is perfectly clear which
product state the joint system is in. But these facts by themselves
are certainly not indicative of the presence of non-classical
\emph{correlations} between the two subsystems.

There is one indirect sense, however, in which $\rho_{\mbox{\tiny
  disc}}$ can be said to contain non-classical correlations. Recall
from \textsection \ref{nec:ss:purify} that any mixture can be
considered as the result of taking the partial trace of a pure
entangled state on a larger Hilbert space. Given that, as I argued in
\textsection \ref{nec:ss:mixdj}, the pure state representation of a
quantum system should be taken as fundamental, we can consider the
bipartite state $\rho_{\mbox{\tiny disc}}$ as in reality but a partial
representation of a tripartite entangled quantum system, where the
third party is an external environment with enough degrees of freedom
to purify the overall system. And since entangled systems do not admit
of a description in terms of local hidden variables (or, if one
prefers, in terms of LOCC), it follows that the system partially
represented by $\rho_{\mbox{\tiny disc}}$ can legitimately be said to
contain non-classical correlations.

Even so it is unclear how these non-classical correlations \emph{per
  se} can have anything to do with the quantum discord exhibited by
$\rho_{\mbox{\tiny disc}}$, for it is also the case that a classically
correlated mixture of \emph{orthogonal} product states, i.e. one of
the form \eqref{nec:eqn:qucl}, can be purified in just the same way as
a discordant one and hence also the case that it can be given a
multi-partite representation in which entanglement is present.

As we will now see, however, there is in fact a tight relationship
between the \emph{amount} of discord associated with a bipartite mixed
state and the \emph{amount} of entanglement associated with a
tripartite representation of that state. And, interestingly from our
point of view, what emerges from this is a correspondingly tight
relationship between the quantum speedup exhibited by the DQC1 and the
amount of entanglement associated with its purified tripartite
representation, and thus a confirmation, not a refutation, of the
NEXT.

\subsubsection{Explaining speedup in the DQC1}

Quantum discord was introduced independently by
\citeauthor[]{henderson2001} and by \citeauthor[]{ollivier2002} in
\citeyear[]{henderson2001} and \citeyear[]{ollivier2002},
respectively; however, it was only recently given an operational
interpretation, independently by \citet[]{madhok2011} and by
\citet[]{cavalcanti2011}.\footnote{I present here the definition
  given by \citeauthor[]{cavalcanti2011}, although the conclusion I
  will draw is the same regardless of which definition is used.} On
both characterisations, quantum discord is operationally defined in
terms of the entanglement consumed in an extended version of the
quantum state merging protocol \citep[cf.][]{horodecki2005}.

In the quantum state merging protocol, three parties: Alice, Bob, and
Cassandra, share a state $| \psi_{ABC} \rangle$. Quantum state merging
characterises the process, $$| \psi_{ABC} \rangle \to | \psi_{B'BC}
\rangle,$$ by which Alice effectively transfers her part of the system
to Bob while maintaining its coherence with Cassandra's part. It turns
out that in order to effect this protocol a certain amount of
entanglement must be consumed (quantified on the basis of the quantum
conditional entropy, $S(A|B)$; cf. Appendix \ref{ch:infth}.). When we
add to this the amount of entanglement needed (as quantified by the
entanglement of formation; cf. \textsection \ref{nec:ss:quantent}) to
prepare the state $| \psi_{ABC} \rangle$ to begin with, the result is
a quantity identical to the quantum discord between the subsystems
belonging to Alice and Cassandra at the time the state is prepared.

\begin{figure}
\begin{tabular}{lllllll}
\raisebox{2.5em}{(a)} & \includegraphics[scale=0.3]{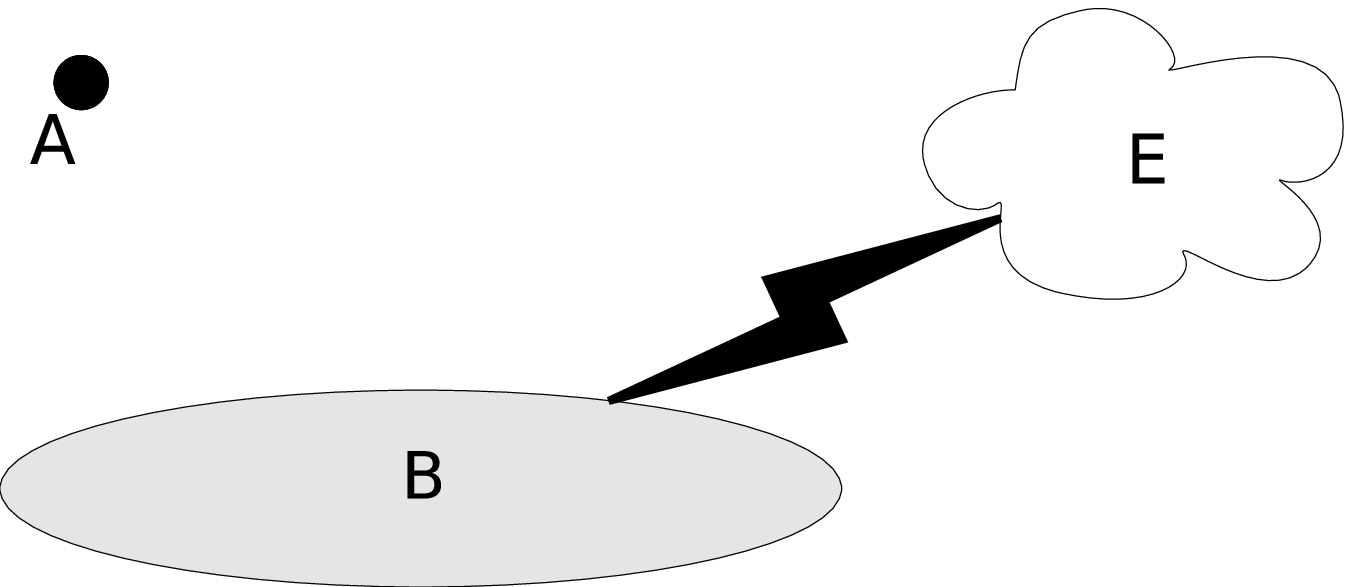} & &
\raisebox{2.5em}{(b)} &
\raisebox{-0.5em}{\includegraphics[scale=0.3]{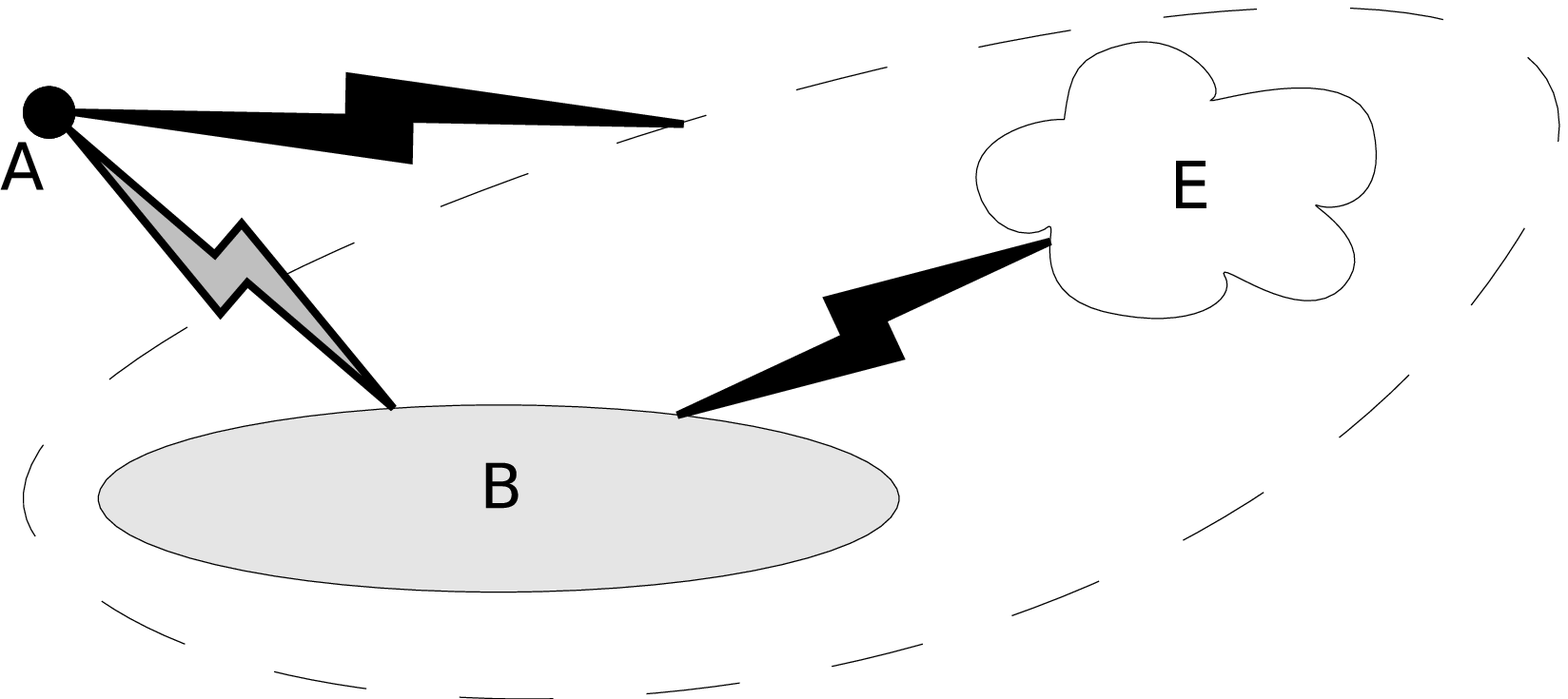}}\\
\end{tabular}
\caption[Tripartite representation of the DQC1]{A (pure) tripartite
  representation of the elements of the DQC1 protocol before (a) and
  after (b) the application of the controlled not gate. Black and
  grey thunderbolts represent entanglement and discord,
  respectively. After the application of the controlled not gate,
  there is an increase in the discord between $A$ and $B$ and a
  corresponding increase in the entanglement between $A$ and the
  combined system $BE$.}
\label{nec:fig:tridqc1}
\end{figure}

The foregoing operational interpretation of discord has an affinity
with an illuminating analysis of the DQC1 circuit due to
\citet[]{fanchini2011}. \citeauthor[]{fanchini2011} show that a
relationship between quantum discord and entanglement emerges when we
consider the DQC1 circuit, not as a bipartite system composed of
polarised and unpolarised qubits respectively, but as a tripartite
system in which the environment plays the role of the third
subsystem. \citeauthor[]{fanchini2011} note that an alternate way of
characterising the completely mixed state of the unpolarised qubits,
$I_n/2^n$, is to view it as part of a bipartite entangled state, with
the second party an external environment having enough degrees of
freedom to purify the overall system. This yields a tripartite
representation for the DQC1 circuit as a whole (see Figure
\ref{nec:fig:tridqc1}).

\citeauthor[]{fanchini2011} show that, for an arbitrary tripartite
pure state, there is a conservation relation between entanglement of
formation and quantum discord. In particular, the sum of the bipartite
entanglement that is shared between a particular subsystem and the
other subsystems of the system cannot be increased without increasing
the sum of the quantum discord between this subsystem and the other
subsystems as well (and vice versa). In the DQC1, after the
application of the controlled not gate (see Figure \ref{nec:fig:dqc1}),
there is an increase in the quantum discord between $B$ and $A$. This
therefore necessarily involves a corresponding increase in the
entanglement between $A$ and the combined system $BE$. All of this
accords with what we would expect given the above operational
interpretation of quantum discord: an increase in quantum discord
requires an increase in the entanglement available for consumption in
a potential quantum state merging process.

Note also that from this tripartite point of view, there is just as
much entanglement in the circuit as there is discord; in particular,
exactly as for quantum discord, there is entanglement in the circuit
whenever it displays a quantum speedup, i.e., for any $\alpha > 0$.

\citeauthor[]{fanchini2011} speculate that it is not the presence of
entanglement or discord (however the latter is interpreted) per se
that is necessary for the quantum speedup of the DQC1, but rather the
ability of the circuit to \emph{redistribute} entanglement and
discord. This thought seems to be confirmed by a theoretical result of
\citet[]{brodutch2011}, who show that shared entanglement is required
in order for two parties to bilocally implement\footnote{Bilocal
  implementation means, in this context, an implementation in which
  Alice and Bob are limited to LOCC operations.} \emph{any} bipartite
quantum gate\textemdash even one that operates on a restricted set
$\mathcal{L}$ of unentangled input states and transforms them into
unentangled output states. This means, in particular, that
entanglement is required in order to implement a gate that changes the
discord of a quantum state.

By themselves, these considerations already amount to confirmations of
the NEXT, for entanglement appears to be involved in the very
definition of discord, and it appears that we require entanglement
even for the production of discord in a quantum circuit. But in
addition, there are indications that quantum discord need not be
appealed to at all to give an account of quantum speedup (though such
a characterisation will of course be less practical, as I have already
mentioned), in light of one other recent theoretical
result. \citeauthor{devi2011} \citeyearpar[]{devi2008,devi2011} have
pointed out that more general measurement schemes than the positive
operator valued measures (POVM) used thus far exist for characterising
the correlations present in bipartite quantum systems.

POVMs are associated with completely positive maps and are well suited
for describing the evolution of a system when we can view the system
as uncorrelated with its external environment. When the system is
initially correlated with the environment, however, the reduced
dynamics of the system may, according to \citeauthor{devi2011}, be
`not completely positive'. But as \citeauthor{devi2011} show, from the
point of view of a measurement scheme that incorporates not completely
positive maps in addition to completely positive maps, all quantum
correlations reduce to entanglement.

In sum, it is, I believe, unsurprising that on the standard analysis
the DQC1 circuit displays strange and anomalous correlations in the
form of quantum discord, for the DQC1 is typically characterised as a
\emph{bipartite} system, and from the point of view of a measurement
framework that incorporates only completely positive maps. As
\citeauthor[]{fanchini2011} have shown, however, the DQC1 circuit is
more properly characterised, not as an isolated system, but as a
system initially correlated with an external environment. The
evolution of such a system is best captured by a measurement framework
incorporating not completely positive maps, and within such a
framework, the anomalous correlations disappear and are subsumed under
entanglement. From this point of view the equivalence of entanglement
and discord for pure bipartite states is also unsurprising, for it is
precisely pure states for which the correlation with the environment
can be ignored and for which a framework incorporating only completely
positive maps is appropriate.

The use of not completely positive maps to characterise the evolution
of open quantum systems is not wholly without its detractors. The
question of whether such not completely positive maps are `unphysical'
is an interesting and important one, though I will not address it
here.\footnote{For a more detailed discussion, and qualified defence
  of the use of not completely positive maps, see
  \citet[]{cuffaro2012a}.} But regardless of the answer to this
question, it should be clear, even without the appeal to this more
general framework, that entanglement has \emph{not} been shown to be
unnecessary for quantum computational speedup. Far from being a
counter-example to the NEXT, the DQC1 model of quantum computation
rather serves to illuminate the crucial role that entanglement plays
in the quantum speedup displayed by this computer.

\section{Conclusion}

Quantum entanglement is considered by many to be a necessary resource
that is used to advantage by a quantum computer in order to achieve a
speedup over classical computation. Given \citeauthor[]{jozsa2003}'s
and \citeauthor[]{abbott2010}'s general results for pure states, and
given that, as I argued in \textsection \ref{nec:ss:mixdj}, a pure
state should be considered as the most fundamental representation of a
quantum system possible in quantum mechanics, the burden is upon those
who deny the NEXT to either produce a counter-example or to show, in
some other more principled way, why the view is false. We examined two
such counter-examples in this chapter. Upon closer examination we
found neither of these, neither the sub-exponential speedup of the
unentangled mixed-state version of the Deutsch-Jozsa algorithm, nor
the exponential speedup of the DQC1 model of quantum computation,
demonstrate that entanglement is unnecessary for quantum speedup; they
rather make clearer than before the role that entanglement \emph{does}
play, and point the way to a fuller understanding of both entanglement
and quantum computation.

\section{Next steps}

We have just concluded that entanglement is a necessary component of
any explanation of quantum speedup\textemdash that the NEXT is
true. The natural next question to ask is whether entanglement is also
sufficient. This question, in turn, can be divided into two
sub-questions. First: is entanglement a sufficient \emph{resource} to
enable quantum speedup? And second: does entanglement suffice to
\emph{explain} quantum speedup. The answer to both of these questions,
I will argue, is yes. We will consider the first question in Chapter
\ref{ch:suff} and the second in Chapter \ref{ch:sig}.

\chapter{Entanglement as a Sufficient Resource to Enable Quantum
  Computational Speedup}
\label{ch:suff}

\settocdepth{subsection}

\section{Introduction}
\label{suff:s:intro}

The answer to the question of whether entanglement is a sufficient
resource to enable quantum computational speedup is commonly held to
be no. To support this conclusion, appeal is usually made to the
\emph{Gottesman-Knill theorem} \citep[\textsection
  10.5.4]{nielsenChuang2000}. According to this theorem, any quantum
algorithm or protocol which exclusively utilises the elements of a
restricted set of quantum operations can be re-expressed using an
alternative formalism which shows us how the algorithm can be
efficiently simulated by classical means. It so happens that among the
quantum computational algorithms and informational protocols which
exclusively utilise operations from this set are some that are
interesting and important\textemdash for instance, the teleportation
and superdense coding protocols. And both of these, and others,
involve the use of \emph{entangled} quantum states.

Reflecting on this circumstance, \citeauthor[]{jozsa2003} write, in
their influential \citeyearpar{jozsa2003} article, in a section
entitled \emph{Is entanglement a key resource for computational
  power?}:

\begin{quote}
Recall that the significance of entanglement for pure-state
computations is derived from the fact that unentangled pure states
... of $n$ qubits have a description involving poly($n$) parameters
(in contrast to $O(2^n)$ parameters for a general pure state). But
this special property of unentangled states (of having a `small'
descriptions [\emph{sic.}]) is contingent on a particular mathematical
description, as amplitudes in the computational basis. If we were to
adopt some other choice of mathematical description for quantum states
(and their evolution), then, although it will be mathematically
equivalent to the amplitude description, there will be a different
class of states which will now have a polynomially sized description;
i.e. two formulations of a theory which are mathematically equivalent
(and hence equally logically valid) need not have their corresponding
mathematical descriptions of elements of the theory being
[\emph{sic.}] interconvertible by a \emph{polynomially bounded}
computation. With this in mind we see that the significance of
entanglement as a resource for quantum computation is not an
\emph{intrinsic} property of quantum physics \emph{itself}, but is
tied to a particular additional (arbitrary) choice of mathematical
formalism for the theory. ... An explicit example of an alternative
formalism and its implications for the power of quantum computation is
provided by the so-called stabilizer formalism and the Gottesman-Knill
theorem ... Thus, in a fundamental sense, the power of quantum
computation over classical computation ought to be derived
simultaneously from \emph{all} possible classical mathematical
formalisms for representing quantum theory, not any single such
formalism and associated quality (such as entanglement),
... \citep[2029-2030]{jozsa2003}.
\end{quote}

Similar considerations, presumably, lead \citeauthor[]{datta2005} to
write: ``the Gottesman-Knill theorem ... demonstrates that global
entanglement is far from sufficient for exponential speedup.''
\citeyearpar[1]{datta2005}. \citeauthor[]{nielsenChuang2000}
\citeyearpar[ibid.]{nielsenChuang2000} writing some years earlier,
are, perhaps, more cautious: ``The Gottesman-Knill theorem highlights
how subtle is the power of quantum computation. It shows that some
quantum computations involving highly entangled states may be
simulated efficiently on classical computers. ... There is much more
to quantum computation than just the power bestowed by quantum
entanglement!'' I say that this statement is more cautious because
while \citeauthor[]{nielsenChuang2000} correctly point out that
an entangled quantum state will not, so to speak, yield a quantum
speedup of its own accord, they (intentionally or not) decline to make
the stronger claim, suggested in my above quote of
\citeauthor[]{jozsa2003}, that further (or perhaps some other)
\emph{physical resources} besides entanglement (which are, according
to \citeauthor[]{jozsa2003}, hidden by the formalism) are required in
order to make quantum speedup \emph{possible}.

Two distinct claims must be distinguished here. The first is this: the
mere presence of an entangled quantum state is sufficient to realise
quantum computational speedup. The Gottesman-Knill theorem shows,
conclusively, that this claim is false. The second, for our purposes
more interesting claim is the following: quantum entanglement is
a resource sufficient \emph{to enable}, or \emph{make possible},
quantum computational speedup; i.e., no other physical resources are
required to make quantum speedup possible if one begins with an
entangled quantum system. This claim, or so I will argue, is true. As
I will explain in the remainder of this chapter, the quantum
operations to which the Gottesman-Knill theorem applies are precisely
those which will never cause a qubit to take on an orientation, with
respect to the other subsystems comprising the total system of which
it is a part, that yields a violation of the Bell inequalities. The
fact that the Gottesman-Knill theorem holds should therefore come as
no surprise. Given this, I will argue that it is misleading to
conclude that more than entanglement is required to enable quantum
computational speedup.

There is, of course, one sense in which more than just entanglement is
required: in order to outperform a classical computer, a quantum
computer realising an entangled quantum state must utilise more
than the relatively small portion of its state space that is
accessible from the Gottesman-Knill group of transformations alone. It
is for this reason that the first thesis which I referred to above is
false. Nevertheless, if one is asked what physical resources are
required in order to make quantum speedup possible, then one can
legitimately answer, or so I will argue, that the answer is no more
than quantum entanglement.

The chapter will proceed as follows. After introducing the
Gottesman-Knill theorem and its implications for the classical
simulability of certain quantum algorithms involving quantum
entanglement, in \textsection \ref{suff:s:gtthm}, I then consider
Bell's theorem, in \textsection \ref{suff:s:bell}, drawing particular
attention to the circumstances in which the Bell inequalities are
satisfied by classical hidden variables theories of the quantum
state. In \textsection \ref{suff:s:suff}, I then argue that the
possibility of an efficient classical simulation of the quantum
algorithms in question is equally evident from a reflection on Bell's
theorem as it is from a reflection on the Gottesman-Knill theorem, and
I discuss the implications of this for our understanding of the
resources involved in quantum speedup, coming to the conclusion that I
have already mentioned.

\section{Preliminaries: The Gottesman-Knill theorem}
\label{suff:s:gtthm}

Call\footnote{The exegesis of the Gottesman-Knill theorem given here
  is indebted to that given in \citet[]{nielsenChuang2000}.} an
operator $A$ a \emph{stabiliser} of the state $| \psi \rangle$ if
\begin{align}
A| \psi \rangle = | \psi \rangle.
\end{align}
For instance, consider the Bell state of two qubits:
\begin{align*}
| \Phi^+ \rangle = \frac{1}{\sqrt 2}(| 0 \rangle| 0 \rangle + | 1
\rangle| 1 \rangle).
\end{align*}
For this state we have
\begin{align*}
(X \otimes X)| \Phi^+ \rangle & = \frac{1}{\sqrt 2}(| 1 \rangle| 1
  \rangle + | 0 \rangle| 0 \rangle) \nonumber \\
  & = \frac{1}{\sqrt 2}(| 0 \rangle| 0 \rangle + | 1 \rangle| 1
  \rangle) = | \Phi^+ \rangle, \\
(Z \otimes Z)| \Phi^+ \rangle & = \frac{1}{\sqrt 2}(| 0 \rangle| 0
  \rangle + (-| 1 \rangle)(-| 1 \rangle)) \nonumber \\
  & = \frac{1}{\sqrt 2}(| 0 \rangle| 0 \rangle + | 1 \rangle| 1
  \rangle) = | \Phi^+ \rangle.
\end{align*}
$X \otimes X$ and $Z \otimes Z$ are thus both stabilisers of the state
$| \Phi^+ \rangle$. Here, $X$ and $Z$ are the Pauli operators:
\begin{align}
  X \equiv \sigma_1 \equiv \sigma_x \equiv
  \left(
  \begin{matrix}
  0 & 1 \\
  1 & 0
  \end{matrix}
  \right), &
  \quad Z \equiv \sigma_3 \equiv \sigma_z \equiv
  \left(
  \begin{matrix}
  1 & 0 \\
  0 & -1
  \end{matrix}
  \right).
\end{align}
The remaining Pauli operators, $I$ (the identity operator) and $Y$,
are defined as:
\begin{align}
  I \equiv \sigma_0 \equiv \sigma_I \equiv
  \left(
  \begin{matrix}
  1 & 0 \\
  0 & 1
  \end{matrix}
  \right), &
  \quad Y \equiv \sigma_2 \equiv \sigma_y \equiv
  \left(
  \begin{matrix}
  0 & -i \\
  i & 0
  \end{matrix}
  \right).
\end{align}
The Pauli group, $P_n$, of $n$-fold tensor products of Pauli operators
(for instance, for $n = 2$, $P_2 \equiv \{I \otimes I, I \otimes X, I
\otimes Y, I \otimes Z, X \otimes I, X \otimes X, X \otimes Y,
... \}$) is an example of a group of operators closed under matrix
multiplication.

Call the set, $V_S$, of states that are stabilised by every element in
$S$, where $S$ is some group of operators closed under matrix
multiplication, the \emph{vector space stabilised by} $S$. Consider a
state $| \psi \rangle \in V_S$. From the definition of a unitary
operator, we have, for any $s \in S$ and any unitary operation $U$,
\begin{align}
U| \psi \rangle = Us| \psi \rangle = UsU^\dagger U| \psi \rangle.
\end{align}
Thus $UsU^\dagger$ stabilises $U| \psi \rangle$ and the vector space
$UV_S$ is stabilised by the group $USU^\dagger \equiv \{UsU^\dagger|s
\in S\}$. Consider, for instance, the state $| 0 \rangle$, stabilised
by the $Z$ operator. To determine the stabiliser of this state after
it has been subjected to the (unitary) Hadamard transformation $H| 0
\rangle = | + \rangle$ we simply compute $HZH^\dagger$. Thus the
stabiliser of $| + \rangle$ is $X$.

Now let $s_1, ..., s_n$ be elements of $S$. $s_1, ..., s_n$ are said
to \emph{generate} the group $S$ if every element of $S$ can be
written as a product of elements from $s_1, ..., s_n$. For instance,
the reader can verify that the subgroup, $A$, of $P_3$, defined by $A
\equiv \{I^{\otimes 3}, Z \otimes Z \otimes I, I \otimes Z \otimes Z, Z
\otimes I \otimes Z\}$ can be generated by the elements $\{Z \otimes Z
\otimes I, I \otimes Z \otimes Z\}$ \citep[\textsection
  10.5.1]{nielsenChuang2000}. We may thus alternately express $A$ in
terms of its generators as follows: $A = \langle Z \otimes Z \otimes
I, I \otimes Z \otimes Z \rangle$.

In order to compute the action of a unitary operator on a group $S$ it
suffices to compute the action of the unitary operator on the
generators of $S$. For instance, $| 0 \rangle^{\otimes n}$ is the
unique state stabilised by $\langle Z_1, Z_2, ..., Z_n\rangle$ (where
the latter expression is a shorthand form of $\langle Z \otimes
I^{\otimes n-1}, I \otimes Z \otimes I^{\otimes n-2}, ...,I^{\otimes
  n-1} \otimes Z\rangle$). Consequently, the stabiliser of the state
$H^{\otimes n}| 0 \rangle^{\otimes n}$ is $\langle X_1, X_2, ...,
X_n\rangle$. Note that this state, expressed in the standard state
vector formalism,
\begin{align*}
H^{\otimes n}| 0 \rangle^{\otimes n} & = \left(\frac{1}{2^{n/2}}(| 0
\rangle + | 1 \rangle)^n \right ) \nonumber \\
& = \left (\frac{1}{2^{n/2}}\sum_x^{2^n-1}| x \rangle \right),
\end{align*}
specifies $2^n$ different amplitudes. Contrast this with the
stabiliser description of the state in terms of its generators
$\langle X_1, X_2, ..., X_n\rangle$, which is linear in $n$ and thus
capable of an efficient classical representation.

Using the stabiliser formalism, it can be shown that all (as well as
all combinations) of the following gates are capable of an efficient
classical representation: \emph{Pauli gates, Hadamard gates, phase
  gates (i.e.,$\pi/2$ rotations of the Bloch sphere for a single qubit
  about the $\hat{z}$-axis), and CNOT gates; as well as state
  preparation in the computational basis and measurements of the Pauli
  observables.} This is the content of the \emph{Gottesman-Knill
  theorem} \citep[\textsection 10.5.4]{nielsenChuang2000}.

What is especially notable about this theorem from the point of view
of our discussion is that some of the states which may be realised
through the operations in this set are actually entangled states. In
particular, by combining a Hadamard and a CNOT gate, one can generate
any one of the Bell states (which one is generated depends on the
value assigned to the input qubits); i.e.,
\begin{align*}
| 0 \rangle| 0 \rangle \xrightarrow{H \otimes I} \frac{| 0 \rangle| 0
\rangle + | 1 \rangle| 0 \rangle}{\sqrt 2} \xrightarrow{CNOT} \frac{|
  0 \rangle| 0 \rangle + | 1 \rangle| 1 \rangle}{\sqrt 2} = | \Phi^+
\rangle, \\
| 0 \rangle| 1 \rangle \xrightarrow{H \otimes I} \frac{| 0 \rangle| 1
\rangle + | 1 \rangle| 1 \rangle}{\sqrt 2} \xrightarrow{CNOT} \frac{|
  0 \rangle| 1 \rangle + | 1 \rangle| 0 \rangle}{\sqrt 2} = | \Psi^+
\rangle, \\
| 1 \rangle| 0 \rangle \xrightarrow{H \otimes I} \frac{| 0 \rangle| 0
\rangle - | 1 \rangle| 0 \rangle}{\sqrt 2} \xrightarrow{CNOT} \frac{|
  0 \rangle| 0 \rangle - | 1 \rangle| 1 \rangle}{\sqrt 2} = | \Phi^-
\rangle, \\
| 1 \rangle| 1 \rangle \xrightarrow{H \otimes I} \frac{| 0 \rangle| 1
\rangle - | 1 \rangle| 1 \rangle}{\sqrt 2} \xrightarrow{CNOT} \frac{|
  0 \rangle| 1 \rangle - | 1 \rangle| 0 \rangle}{\sqrt 2} = | \Psi^-
\rangle.
\end{align*}
In fact many quantum algorithms utilise just such a combination of
gates. One of these, for instance, is the well-known teleportation
algorithm (see Appendix \ref{ch:tel}). If all of the operations from
this set are efficiently classically simulable, however, then it
appears as though entanglement, by itself, cannot be a sufficient
resource for realising quantum speedup, for evidently there are
quantum algorithms utilising entangled states that are efficiently
simulable classically.

In what follows I will argue that this conclusion is not warranted. An
entangled state, I will contend, provides sufficient resources to
enable quantum computational speedup. What the Gottesman-Knill theorem
actually shows, I will argue, is that, in certain special cases, the
resources provided by an entangled state are not utilised to their
full potential. This becomes especially clear when we consider Bell's
theorem, and in particular, the circumstances under which the Bell
inequalities are \emph{satisfied} by classical hidden variables
theories of the quantum state. As we will see, the possibility of an
efficient classical simulation of certain quantum algorithms is
equally evident from a consideration of Bell's theorem as it is from a
consideration of the Gottesman-Knill theorem, and that reflecting on
Bell's theorem helps us to understand better exactly how quantum
entanglement is not being fully exploited in the quantum algorithms we
are considering.

\section{Bell's theorem}
\label{suff:s:bell}

For a system in the singlet state ($| \Psi^- \rangle$), the
expectation value for joint experiments on its subsystems is given by
the following expression:
\begin{align}
\label{suff:eqn:singlet}
\langle \sigma_m \otimes \sigma_n \rangle = - \hat{m} \cdot \hat{n} =
- \cos\theta.
\end{align}
Here $\sigma_m, \sigma_n$ represent spin-$m$ and spin-$n$ experiments
on the first (Alice's) and second (Bob's) subsystem, respectively,
with $\hat{m}, \hat{n}$ the unit vectors representing the orientations
of the two experimental devices, and $\theta$ the difference in these
orientations. Note, in particular, that when $\theta = 0$, $\langle
\sigma_m \otimes \sigma_n \rangle = -1$ (i.e., experimental results
for the two subsystems are perfectly anti-correlated), when $\theta =
\pi$, $\langle \sigma_m \otimes \sigma_n \rangle = 1$ (i.e.,
experimental results for the two subsystems are perfectly correlated),
and when $\theta = \pi/2$, $\langle \sigma_m \otimes \sigma_n \rangle
= 0$ (i.e., experimental results for the two subsystems are not
correlated at all).

Consider the following attempt \citep[]{bell1964} to reproduce the
quantum mechanical predictions for this state by means of a hidden
variables theory. Let the hidden variables of the theory assign, at
state preparation, to each subsystem of a bipartite quantum system, a
unit vector $\hat{\lambda}$ (the same value for $\hat{\lambda}$ is
assigned to each subsystem) which determines the outcomes of
subsequent experiments on the system as follows. Let the functions
$A_\lambda(\hat{m}), B_\lambda(\hat{n})$ represent, respectively, the
outcome of a spin-$m$ and a spin-$n$ experiment on Alice's and Bob's
subsystem. Define these as:
\begin{align}
A_\lambda(\hat{m}) & = \mbox{sign} (\hat{m} \cdot \hat{\lambda}),
\nonumber \\
B_\lambda(\hat{n}) & = - \mbox{sign} (\hat{n} \cdot \hat{\lambda}).
\end{align}
where $\mbox{sign}(x)$ is a function which returns the sign (+, -) of
its argument.

The reader can verify that the probability that both
$A_\lambda(\hat{m})$ and $B_\lambda(\hat{n})$ yield the same value,
and the probability that they yield values that are different
(assuming a uniform probability distribution over $\hat{\lambda}$),
are respectively:
\begin{align}
\mbox{Pr}(+, +) & = \mbox{Pr}(-, -) = \theta/2\pi, \nonumber \\
\mbox{Pr}(+, -) & = \mbox{Pr}(-, +) = \frac{1}{2}\left(1 -
\frac{\theta}{\pi}\right),
\end{align}
with $\theta$ the (positive) angle between $\hat{m}$ and $\hat{n}$. This
yields, for the expectation value of experiments on the combined state:
\begin{align}
\langle \sigma_m \otimes \sigma_n \rangle = \frac{2 \theta}{\pi} - 1.
\end{align}

When $\theta$ is a multiple of $\pi/2$, this expression yields
predictions identical to the quantum mechanical ones: perfect
anti-correlation for $\theta \in \{0, 2\pi, ...\}$, no correlation for
$\theta \in \{\pi/2, 3\pi/2, ...\}$, and perfect correlation for $\theta
\in \{\pi, 3\pi, ...\}$. However, for all other values of $\theta$ there
are divergences from the quantum mechanical predictions.

It turns out that this is not a special characteristic of the simple
hidden variables theory considered above. \emph{No} hidden variables
theory is able to reproduce the predictions of quantum mechanics if it
makes the very reasonable assumption that the probabilities of local
experiments on Alice's subsystem (and likewise Bob's) are completely
determined by Alice's local experimental setup together with a hidden
variable taken on by the subsystem at the time the joint state is
prepared. Consider the following\footnote{In this exposition of the
  CHSH inequality I have followed \citet[]{myrvold2008}.} expression
relating different spin experiments on Alice's and Bob's respective
subsystems for arbitrary directions $\hat{m}, \hat{m}', \hat{n},
\hat{n}'$:
\begin{align}
| \langle \sigma_m \otimes \sigma_n \rangle + \langle \sigma_m \otimes
\sigma_{n'} \rangle | + | \langle \sigma_{m'} \otimes \sigma_n \rangle
- \langle \sigma_{m'} \otimes \sigma_{n'} \rangle |.
\end{align}
As before, let $A_\lambda(\hat{m}) \in \{\pm 1\}, B_\lambda(\hat{n})
\in \{\pm 1\}$ represent the results, given a specification of some
hidden variable $\lambda$, of spin experiments on Alice's and Bob's
subsystems. We make no assumptions about the nature of the `common
cause' $\lambda$ this time\textemdash it may take any form. What we do
assume is that, as I mentioned above, the outcomes of Alice's
experiments depend only on her local setup and on the value of
$\lambda$; i.e., we do not assume any further dependencies between
Alice's and Bob's local experimental
configurations. This `factorisability' (cf. Eq. \ref{nec:eq:fact})
allows us to substitute $\langle A_\lambda(\hat{m}) \cdot
B_\lambda(\hat{n}) \rangle$ for $\langle \sigma_m \otimes \sigma_n
\rangle$, thus yielding:
\begin{align}
\label{suff:eqn:chsh1}
& \big | \big\langle A_\lambda(\hat{m})B_\lambda(\hat{n}) \big\rangle +
  \big\langle A_\lambda(\hat{m})B_\lambda(\hat{n}') \big\rangle \big | +
  \big | \big\langle A_\lambda(\hat{m}')B_\lambda(\hat{n}) \big\rangle -
  \big\langle A_\lambda(\hat{m}')B_\lambda(\hat{n}') \big\rangle \big |
  \nonumber \\
& = \big | \big\langle A_\lambda(\hat{m})\big(B_\lambda(\hat{n}) +
  B_\lambda(\hat{n}')\big)\big\rangle \big | + \big | \big\langle
  A_\lambda(\hat{m}')\big(B_\lambda(\hat{n}) -
  B_\lambda(\hat{n}')\big)\big\rangle \big | \nonumber \\
& \leq \big\langle \big | A_\lambda(\hat{m})\big(B_\lambda(\hat{n}) +
  B_\lambda(\hat{n}')\big)\big | \big\rangle + \big\langle \big |
  A_\lambda(\hat{m}')\big(B_\lambda(\hat{n}) - B_\lambda(\hat{n}')\big)
  \big | \big\rangle,
\end{align}
which, since $|A_\lambda(\cdot)| = 1$, is
\begin{align}
& \leq \big\langle \big | B_\lambda(\hat{n}) + B_\lambda(\hat{n}')\big |
\big\rangle + \big\langle  \big | B_\lambda(\hat{n}) -
B_\lambda(\hat{n}') \big | \big\rangle \nonumber \\
& \leq 2,
\end{align}
where the last inequality follows from the fact that $B_\lambda(\cdot)$
can also only take on values of $\pm 1$. This expression, a variant of
the `Bell inequality' \citeyearpar[]{bell1964}, is known as the
\emph{Clauser-Horne-Shimony-Holt} (CHSH) inequality
\citep[cf.][]{chsh1969,bell1981}.

Quantum mechanics violates the CHSH inequality for some experimental
configurations. For example, let the system be in the singlet state;
i.e., such that its statistics satisfy \eqref{suff:eqn:singlet}; and let
the unit vectors $\hat{m}, \hat{m}', \hat{n}, \hat{n}'$ (taken to lie
in the same plane) have the orientations $0, \pi/2, \pi/4, -\pi/4$
respectively. The differences, $\theta$, between the different
orientations (i.e., $\hat{m} - \hat{n}, \hat{m} - \hat{n}', \hat{m}' -
\hat{n}$, and $\hat{m}' - \hat{n}'$) will all be in multiples of
$\pi/4$ and we will have:
\begin{align}
\langle \sigma_m \otimes \sigma_n \rangle & =
\langle \sigma_m \otimes \sigma_{n'} \rangle =
\langle \sigma_{m'} \otimes \sigma_n \rangle = \sqrt 2/2, \\
\langle \sigma_{m'} \otimes \sigma_{n'} \rangle & = -\sqrt 2/2, \\
| \langle \sigma_m \otimes \sigma_n \rangle & + \langle \sigma_m \otimes
\sigma_{n'} \rangle | + | \langle \sigma_{m'} \otimes \sigma_n \rangle
- \langle \sigma_{m'} \otimes \sigma_{n'} \rangle | = 2\sqrt 2
\not\leq 2.
\end{align}

The predictions of quantum mechanics for arbitrary orientations
$\hat{m}, \hat{m}', \hat{n}, \hat{n}'$ cannot, therefore, be
reproduced by a hidden variables theory in which all correlations
between subsystems are due to a common parameter endowed to them at
state preparation. \emph{They can}, however, be reproduced by such a
hidden variables theory for certain special cases. In particular, the
inequality is \emph{satisfied} (as the reader can verify) when
$\hat{m}$ and $\hat{n}$, $\hat{m}$ and $\hat{n}'$, $\hat{m}'$ and
$\hat{n}$, and $\hat{m}'$ and $\hat{n}'$ are all oriented at angles
with respect to one another that are given in multiples of $\pi/2$.

\section{Entanglement as a sufficient resource}
\label{suff:s:suff}

Recall the content of the Gottesman-Knill theorem: \emph{Pauli gates,
  Hadamard gates, phase gates, and CNOT gates; as well as state
  preparation in the computational basis and measurements of the Pauli
  observables} are efficiently simulable by a classical computer. It
is commonly concluded, from this, that entanglement cannot therefore
be sufficient to enable a quantum algorithm to achieve a speedup over
its classical counterpart. When one notes that all of the operations
which comprise this set involve rotations of the Bloch sphere that are
multiples of $\pi/2$, however, the fact that algorithms restricted to
just these operations are classically simulable should come as no
surprise. In an entangled quantum system, no amount of $k\pi/2$
transformations of one of the constituent systems will cause it to
take on an orientation with respect to the other subsystems that is
not a multiple of $\pi/2$ (unless it was so oriented initially). And
as we have seen above, the statistics of compound states for which the
difference in orientation between subsystems is a multiple of $\pi/2$
are capable in general of being reproduced by a classical hidden
variables theory.

In light of this it is misleading, I believe, to conclude, on the
basis of the Gottesman-Knill theorem, that entanglement is not a
sufficient resource to enable quantum computational speedup. What the
Gottesman-Knill theorem shows us is that simply having an entangled
state is not enough to enable one to outperform a classical computer;
one must also \emph{use} such a state to its full potential; i.e., one
must not limit oneself to transformations which utilise only a small
portion of the system's allowable state space. In \emph{this} sense,
it is indeed correct to say that entanglement is insufficient to
enable quantum speedup. However, if one intends by the claim that
entanglement is insufficient\textemdash something very
different\textemdash that \emph{further physical resources} are
required to enable speedup, then I submit that this claim\textemdash
which is the one most relevant to us\textemdash is incorrect.

Consider the individual state spaces of two quantum mechanical
systems, $\mathcal{H}_1^{d_1}$ and $\mathcal{H}_2^{d_2}$, where $d_1$
and $d_2$ are the dimensionality of the first and second system,
respectively. In quantum mechanics, the overall state space of the
combined system is given by the tensor product of the two systems,
$\mathcal{H}_1^{d_1} \otimes \mathcal{H}_2^{d_2}$, with dimensionality
$d_1 \cdot d_2$. Thus the state space of a combined system of $n$
two-dimensional qubits is $\otimes^n \mathcal{H}^2$, with overall
dimensionality $2^n$. In classical mechanics, on the other hand, the
total state space of two individual subsystems $\omega_1^{d_1}$,
$\omega_2^{d_2}$ is given by the Cartesian product, $\omega_1^{d_1}
\times \omega_2^{d_2}$, with dimensionality $d_1 + d_2$. Thus the
dimensionality of the state space of a classical system of $n$
two-dimensional subsystems is $2n$.

As \citet{ekert1998} note, the possibility of entangled quantum
systems is what is responsible for this difference in the allowable
state space. To illustrate, consider how one would go about
representing a general superposition of $n$ two-dimensional values
classically. It is possible to describe certain classical systems in
terms of superpositions; for instance, the state of motion of a
vibrating string can be characterised as a superposition of its two
lowest energy modes, in the same way that the state of a qubit can be
characterised as a superposition of the states $| 0 \rangle$ and $| 1
\rangle$. The joint state of a system of $n$ strings, however, will
always be a \emph{product} state; \emph{general} superpositions which
include, in particular, values representable by entangled quantum
states, cannot be physically represented using $n$ classical systems
in this way.

It is, of course, possible to classically represent a general
superposition of $n$ two-dimensional values in a more roundabout way;
one may use, for instance, a single classical system which allows for
the discrimination of $2^n$ resource levels within it. The cost of
such a representation scales exponentially with $n$, however, either
(if the spacing between resource levels is kept fixed) in terms of the
total amount of resource required, or (if the total amount of the
resource is kept fixed) in terms of the increasing precision required
to discriminate the different resource levels.

Quantum systems, in contrast, are not subject to this limitation;
because of the possibility of entanglement, a superposition of $n$
$d$-dimensional quantum systems can be used to represent a general
superposition of $n$ $d$-dimensional values \emph{directly}; i.e.,
without incurring the cost associated with the roundabout classical
method.\footnote{\citet[Ch. 8]{duwell2004} calls this
  `well-adaptedness'.} Quantum mechanical systems, therefore, allow us
to efficiently exploit the full representational capacity of Hilbert
space. Classical systems do not; they require exponentially more
resources in order to do so. If we have an $n$-fold entangled quantum
system, therefore, it follows straightforwardly that the possibilities
for representation associated with such a system cannot, \emph{in
  general}, be efficiently simulated classically. (And note that from
this point of view it is quite unsurprising that the Deutsch-Jozsa
algorithm can be classically simulated (\textsection \ref{nec:s:deq})
for $n < 3$: notice that for $n < 3$, $2^n = 2n$.)\footnote{There is
  the caveat, of course, that a quantum computer will never be
  found, when experimented upon, to be in one of these `extra',
  nonseparable, states, and thus the final `readout' of a quantum
  computer will never be one of those states. Any problem,
  therefore, whose solution requires such a representation cannot be
  solved efficiently by a quantum computer. Nevertheless, such
  states represent a wealth of resources that are capable of being
  used as intermediaries in the calculation of a solution which is
  representable as a separable final state.}

Evidently, it is possible to utilise only a small portion of the state
space of a quantum system\textemdash exactly that portion of the state
space which, as the Bell inequalities demonstrate, is accessible
efficiently by an $n$-fold classical system\textemdash but this has no
bearing on the nature of the actual physical resources that are
provided by the quantum system. Analogously, a life vest may be said
to be sufficient to keep me afloat on liquid water. I must actually
wear it if it is to perform this function, of course; but that is not
a fact about this piece of equipment's capabilities, only about my
choice whether to use it or not.

What if the waves are rough? It may be that in this case my life vest
will not be sufficient to save me. Analogously, in the presence of
noise, as noted by \citet[]{linden2001}, entanglement may not be
sufficient to enable one to achieve \emph{exponential} quantum
speedup. Nevertheless, even in rough weather I will at least have a
better chance of surviving with my life vest on than I will without
it. Likewise, as we saw in our discussion of the mixed state
Deutsch-Jozsa algorithm (\textsection \ref{nec:ss:mixdj}), even in the
presence of noise, an entangled quantum state will be sufficient to
enable some (though perhaps only a very small) quantum speedup.

\section{Conclusion}

In this chapter I have argued that there is an important
sense\textemdash the most important sense, for our purposes\textemdash
in which entanglement may be said to provide sufficient physical
resources to enable a quantum computer to achieve quantum
computational speedup. In support of this conclusion, I argued that
claims to the contrary rest on a misunderstanding of the implications
of the Gottesman-Knill theorem\textemdash that indeed, far from being
a problem for the view that entanglement is a sufficient resource, the
Gottesman-Knill theorem serves to \emph{highlight} the role that is
actually played by entanglement in the quantum computer and to clarify
exactly \emph{in what sense} it is sufficient.

As is well known, quantum speedup has not been conclusively proven. It
may be that in every case of purported quantum speedup, there actually
is some hitherto unknown classical algorithm that is capable of
achieving an exponential speedup over its currently known classical
alternatives. From this point of view, therefore, I cannot have
conclusively shown in this chapter that quantum entanglement is
sufficient for quantum speedup; for it may be the case that quantum
speedup is \emph{impossible}.\footnote{It is worthwhile to note,
  however, that even if quantum \emph{speedup} is impossible,
  the\textemdash still interesting\textemdash question as to why it
  is that quantum computers are able to solve certain computational
  problems in polynomial time remains. I am indebted to Filippo Annovi
  for this observation.} I hope, however, that the considerations
that I have brought to the fore in this chapter may serve to do the
following: first, I hope that they will lend weight to the claim that
quantum computers can outperform classical computers, second, I hope
that they will clarify exactly why it is that they should be able to
do so, and finally, I hope that they will point the way to a proof, in
the not too distant future, of this conjecture.

\section{Next steps}

In the last chapter I argued that entanglement is a \emph{necessary}
component of any explanation of quantum speedup, while in this chapter
I argued that, as a physical resource, it is \emph{sufficient to
enable} quantum computational speedup. One is tempted, therefore, to
end our investigation here. Our task is not done yet, however, for
even if one is convinced by all of the arguments I have given thus
far, it will still be possible to object that entanglement is
insufficient in the following sense. What we have been seeking for is
an \emph{explanation} of quantum speedup, and while it may be true
that entanglement is a sufficient \emph{resource} to enable quantum
computational speedup, it does not follow that entanglement is
sufficient \emph{to explain} quantum speedup. The interpretation of
quantum mechanics and of entanglement in particular has long been a
topic of very controversial debate. It may therefore be objected that,
even after determining entanglement to be a necessary and sufficient
resource for enabling quantum speedup, we have not explained quantum
speedup until we have explained quantum entanglement itself. We will
address this issue in the next chapter.

\chapter{Entanglement as the Physical Explanation of Quantum
  Computational Speedup}
\label{ch:sig}

\settocdepth{subsection}

\section{Introduction}

In the previous chapter I argued that, in the sense most relevant to
our investigation, entanglement should be seen as a \emph{sufficient
  resource} for quantum computational speedup; I argued that when the
implications of the Gottesman-Knill theorem are properly understood,
they do not contradict the claim that entanglement is sufficient, but
rather highlight precisely the sense in which this claim is true. In
this chapter I will address the issue of whether entanglement is
\emph{sufficient to explain} quantum speedup as well.

To this purpose I will now proceed in the following way. I will begin,
in \textsection \ref{sig:s:theexp}, by formulating a tentative
explanation for quantum speedup in terms of quantum entanglement,
while at the same time outlining the way in which I take entanglement
to be explanatory; viz., the type of explanation that is being offered
when one appeals to entanglement. I will then consider, beginning in
\textsection \ref{sig:s:physexp}, a possible challenge to the view that
entanglement is the explanation of quantum speedup, to the effect that
one has not explained quantum speedup until one has explained
\emph{why} quantum systems may sometimes become entangled, where one
assumes that the answer to a \emph{why?} question of this kind must
involve a causal-mechanical description.

The envisioned argument begins by considering that, according to John
Stachel, entanglement should not be characterised as essentially
involving physical interactions, but rather as arising from a more
abstract set of requirements known as `Feynman's rules'. But since
entanglement should not, according to this reasoning, be construed as
essentially involving physical interactions, it cannot be explained as
arising from some cause, and therefore, according to this objection,
cannot form the essential part of a physical explanation for quantum
speedup. In \textsection \ref{sig:s:accountfeyn} I argue, in response,
that these abstract requirements themselves can be accounted for in
terms of physical interactions, and that Feynman's rules and quantum
entanglement are, in one sense, but two sides of the same coin.

\section{A physical explanation for quantum speedup: Answering the
  \emph{how-possibly?} question}
\label{sig:s:theexp}

\subsection{Physical explanation}
\label{sig:s:ontexp}

If we consider, in a general way, the `act' of explanation, one way to
characterise it is in terms of the following distinctions. First,
there is that for which an explanation has been requested: some thing
or process which is the \emph{object} of the explanation. Second,
there is the person to whom the explanation is addressed: the
\emph{recipient} of the explanation. Finally, there is the explanatory
text itself. An ideal explanatory text will represent the object for
which an explanation has been requested with perfect accuracy
(relative to a theory of such objects), and at the same time, it will
do so in a way which results in a perfect comprehension of the object
on the part of the recipient of the explanation.\footnote{The notion
  of `perfect comprehension' is, of course, a vague one, but it will
  not be necessary, for our purposes, to elaborate upon it further.}

While an ideal explanation of this sort would be desirable, in
practice (and perhaps even in principle) we must settle for far
less. For on the side of the object, we are not possessed with the
perfect knowledge of its state (its detailed structural features or
detailed initial conditions) which we require in order to produce a
perfect description of it. On the other side, the cognitive
limitations of the recipient must also be taken into account. For even
if a perfect description of the object were available, it would likely
be impossible for a finite agent to comprehend such a description in
its entirety. Further, a perfectly detailed description will
invariably contain information that is irrelevant, for the agent, to
the question being asked; it may thus serve only to distract the
recipient. For \emph{us}, therefore, explanation involves a
choice. Given an explanatory question, one must aim at an explanatory
text which strikes the right balance between these two aspects of the
act of explanation.

In some cases, we will choose to place \emph{most} (but never all) of
the emphasis on describing the object itself. Call this an `ontic'
explanation, where we are to understand by this only that such an
explanatory text attempts to represent the grasp of the objective
features of the object provided to us by our best theory of such
objects. In other cases, for reasons of expediency and ease of
comprehension, we will place \emph{less} emphasis on the more
immediate features of the object in question and situate the
explanatory text at a level removed from the object. In these cases,
it will be understood that such higher-level descriptions are
reducible in principle to lower-level descriptions; thus that these
higher-level descriptions are translatable in principle into `ontic
language'.

Finally, there will be some cases in which high-level descriptions of
a different sort will be employed, either simply for ease of
exposition or, in certain situations, because high-level descriptions
of the reducible sort are not to be had. We may call this last sort of
explanation `analogical', where this is not intended in any particular
technical sense of that term. Here, we can imagine those useful
heuristics which help us to understand certain aspects of
phenomena. And while such analogical descriptions can be explanatory,
in the sense that they help to illuminate certain aspects of the
behaviour of the objects of our investigation, they cannot (and are
not intended to) be construed by the explanatory recipient as
revealing the objective features of these objects. These are not ontic
explanations, in the sense just described.

Now I mentioned that what we are to understand as ontic with respect
to a certain class of objects should be understood to be relative to
whatever \emph{theory} of such objects we take to be true. Our own
investigation concerns the \emph{physical} explanation of quantum
speedup. Such an explanation should therefore describe the features of
quantum systems, as described by \emph{physical} theory, which enable
them to outperform classical systems. A physical explanation for
quantum speedup, therefore, will be an example of ontic explanation,
in the sense in which I have just characterised that mode of
explanation.

There are yet further distinctions among the varieties of explanation,
along a different dimension than the distinctions discussed in the
current section. Here I mean the differences that can be identified
with respect to the characterisation of the explanatory question
itself. We will consider these next.

\subsection{\emph{Why?} questions and \emph{how-possibly?} questions}
\label{sig:s:whyhowp}

Scientific explanations are typically taken to be answers to
\emph{why?} questions. As \citet[135]{hempel1948}, for instance,
write:

\begin{quote}
To explain the phenomena in the world of our experience, to answer the
question ``why?'' rather than only the question ``what?'', is one of
the foremost objectives of all rational inquiry; and especially,
scientific research in its various branches strives to go beyond a
mere description of its subject matter by providing an explanation of
the phenomena it investigates.
\end{quote}

Hempel and Oppenheim's own \emph{Deductive-Nomological} (D-N) model of
scientific explanation, along with Hempel's later
Inductive-Statistical (I-S) model, were, for many years, enormously
influential in the debate over exactly what it means to properly
answer a \emph{why?} question of this sort. Explanations, for Hempel
and Oppenheim, are \emph{arguments}. They involve the subsumption of a
particular set of initial conditions under a law or a set of
laws. Together, the set of initial conditions and laws form the
\emph{explanans} (the premises of the argument). Given the explanans,
the \emph{explanandum} statement (a statement of the fact to be
explained, which is the conclusion of the argument) follows either
deductively (in the case of D-N) or inductively (in the case of
I-S). The explanans is the answer, in just this sense, to the question
of \emph{why} the event expressed by the explanandum statement
occurred.

The counter-examples to Hempel and Oppenheim's characterisation of
scientific explanation which later began to emerge are well-known and
I will not rehearse all of them here (for a survey, see:
\citealt[]{salmon1989}). But let us pause, for a moment, on the
so-called `flagpole' counter-example to the D-N model, where we are
asked to imagine a flagpole standing in a field on a level stretch of
ground under a clear blue sky. It is evident that, from the relevant
set of initial conditions and physical laws, a D-N argument can be
formulated to infer that (and hence explain why) the flagpole casts a
shadow of a particular length. Problematically, however, an equally
good explanation (by D-N lights) of the height of the \emph{flagpole}
that appeals to the length of its shadow can be given. Thus, it was
argued that even if it is admitted that the amenability to D-N form is
necessary for explanation, it does not appear to be sufficient to
capture exactly what we mean when we say that an answer has been given
to (at least some) of the \emph{why?} questions we may want to ask.

With respect to just what those further aspects of explanation might
be, however, there is no consensus. One central debate is over exactly
where the `right balance' between the object and the recipient of
explanation should be struck; particularly, over whether any
scientific explanation worthy of the name \emph{must} be ontic in
nature, in the sense in which I alluded to in the previous
section. Proponents of this view are motivated in part by
considerations such as the flagpole example, which seem to suggest
that what is required in a model of explanation is a way to capture
the asymmetrical cause-effect relationship between the facts cited in
the explanans and the fact cited in the explanandum.

Sylvain Bromberger's \citeyearpar{bromberger1966} and Bas van
Fraassen's \citeyearpar[]{vanFraassen1980} work on \emph{why?}
questions did much to clarify the issues. Not content to focus solely
on the proper characterisation of the \emph{answers} to such
questions, Bromberger and van Fraassen investigated the proper way to
analyse these questions \emph{themselves.} Van Fraassen, in
particular, argued that what in certain contexts may seem like an
inappropriate answer to a request for explanation will, in other
contexts, constitute a perfectly good one. In some contexts, for
instance, the length of a structure's shadow \emph{may} be taken to
explain its height:

\begin{quote}
That tower marks the spot where he killed the maid with whom he had
been in love to the point of madness. And the height of the tower? He
vowed that shadow would cover the terrace where he first proclaimed
his love, with every setting sun\textemdash that is why the tower had
to be so high \citep[133-134]{vanFraassen1980}.
\end{quote}

For van Fraassen, \emph{all} explanations are answers to \emph{why?}
questions, where these are of the form ``Why (is it the case that) $P$
\emph{in contrast to} (other members of) $X$?'', and where the second
half of this schema is taken as implicit in context and typically
left unstated. $X$ is the contrast class: a set of alternatives to
$P$. Thus ``Why did you dye your hair black?'' is, absent an explicit
or implicit contrast class, ambiguous. It can be interpreted, for
example, as either ``Why did you dye your hair black, as opposed to
blond or blue or orange?'', or alternatively, ``Why did you dye your
hair black, as opposed to not dying it at all?''. An \emph{answer} to
a \emph{why?} question will be one that \emph{favours} $P$ over any of
its alternatives in the given contrast class.

I will not go through, in detail, the impressive machinery of van
Fraassen's theory of \emph{why?} questions. It is sufficient to point
out that van Fraassen's theory convincingly shows (at least for this
author) that what we take as the appropriate answer to a particular
\emph{why?} question will depend in large part on the context in which
the question is asked. Thus, while in some contexts it may be that the
appropriate answer to a \emph{why?} question should be ontic in
nature, in other contexts this may not be the case, even when the
context is broadly scientific. In the former cases, we should expect
an answer to appeal to actual causes and causal histories of
phenomena, while in the latter cases we may be satisfied with formal
or informal analogies.

As illuminating as van Fraassen's theory of \emph{why?} questions has
been with respect to these issues, however, if it is taken as a
comprehensive analysis of explanatory questions \emph{as such}, then
it cannot succeed, for there are other questions in addition to
\emph{why?} questions that one may wish to have
answered.\footnote{Note that while \citet[]{vanFraassen1980} takes
  explanatory questions to be exhausted by \emph{why?} questions,
  \citet[90]{bromberger1966} (who nevertheless focuses exclusively on
  \emph{why?} questions in his essay) does not: ````explanation'' may
  refer to the answers of a huge variety of questions besides
  why-questions, the only requirement being that their \emph{oratio
  obliqua} form fit as grammatical object of the verb ``to
  explain'' and its nominalization ``explanation of,'' ...''}

Of these other types of explanatory question, one of these is the
so-called \emph{how-possibly?} question. For instance: ``How can Santa
Claus possibly manage to deliver all of those toys in just one
evening?''. Such a question does not ask for the reason \emph{why}
Santa Claus does this, but for a description of \emph{how he is able
  to} do it. A good answer to this question will consist of an account
of the special characteristics of the sled and of the reindeer (and
especially of Rudolph's nose), it will discuss the circumference of
the earth, the number of deliveries to be made, and the properties of
the chimneys in use in various parts of the globe, among other things.

As Wesley Salmon notes, the answer to a \emph{how-possibly?} question
need not involve a reference to \emph{actual events}:

\begin{quote}
... a DC-9 jet airplane recently crashed upon takeoff at Denver's
Stapleton Airport during a snowstorm. One peculiar feature of this
accident is that the plane flipped over onto its back. There are many
explanations of a crash under the circumstances, but I wondered how it
could have flipped over. Two how-possibly explanations were mentioned
in the news reports. One is that it encountered wing-tip turbulence
from another airplane just after it became airborne. Another was
suggested by the report of a survivor, who claimed that the plane was
de-iced three times during its wait for departure, but that on the
latter two of these occasions one wing, but not the other, was
treated. If one wing had an accumulation of ice on its leading edge
while the other did not, the difference in lift provided by the two
wings might have been sufficient cause for the plane to flip over. As
I write this paragraph I have not yet heard the final determination
regarding the cause of this crash. Both potential explanations I have
mentioned are satisfactory answers to the how-possibly question, but
we do not know the correct answer to the why-question
\citep[137]{salmon1989}.
\end{quote}

Of course, one might always attempt to reframe a \emph{how-possibly?}
question as a \emph{why-possibly?} question: ``Why is it that Santa
Claus can deliver all of those toys in just one night, Mommy?'' is an
example of such an attempted reformulation. This is not the place to
venture into a debate over the proper use of English interrogatives,
and the difference between \emph{how-possibly?} and
\emph{why-possibly?} is less important, for our purposes, than the
difference between \emph{how-possibly?} and \emph{why?}. But that
being said I do not think this reformulation of the Santa Claus
question will quite do. There is clearly a difference in emphasis
between the two questions, for the \emph{why-possibly?} question can
always be answered with: ``because he can afford to buy the proper
equipment,'' while the \emph{how-possibly?} question, in contrast,
seems to demand that we explain exactly \emph{how} it is that his
equipment is `proper' (or, in a different context, exactly \emph{how}
he is able to afford it).

\subsection{The question regarding the source of quantum speedup}

Consider the case in which we would like to explain the fact that a
computer has solved a particular problem. Such an explanation can be
given from either of two points of view: from the `software' point of
view, in which the emphasis is placed on accommodating, what in
\textsection \ref{sig:s:ontexp} we called the \emph{recipient} of
explanation, or from the `hardware' point of view: the point of view
we referred to in \textsection \ref{sig:s:ontexp} as \emph{ontic}, in
which the emphasis is on accurately describing the state of the
\emph{object} (the computer). Thus, imagine sitting at a computer
terminal and being presented with the following prompt:
\begin{quote}
\texttt{Please input a series of integers:}
\end{quote}
Upon entering, for instance, 23, 45, 199, and 17, you receive the
following message:
\begin{quote}
\texttt{Your integers in sorted order are: 17, 23, 45, 199.}
\end{quote}
What is the explanation for the fact that the computer has given the
correct answer? We may, on the one hand, attempt to answer this
question by reverse-engineering some set of high-level instructions
that could have been given to the computer, as in Figure
\ref{sig:fig:prog}. This is not an explanation from the ontic point of
view. Characteristic of the point of view represented by this sort of
explanation is that a solution to a computational problem is described
in terms of a series of high-level black-box (typically function)
evaluations.\footnote{I am using `function' here in a rather loose
  sense. I do not mean to exclude, of course, object-oriented and
  procedural programming models.} No account is taken of the way in
which these instructions are actually implemented in a computer.

\begin{figure}
\begin{lstlisting}[frame=trBL]
void selectionSort(int intsToSort[], int lengthOfList) {
  // Declare list indices:
  int i, j, indexOfLowestNum;
  // For each position in the list,
  for (i = 0; i < lengthOfList - 1; i++) {
    // provisionally assert that it points to the lowest number,
    indexOfLowestNum = i;
    // and then for each of the other list positions,
    for (j = lengthOfList - 1; j > i; j--) {
      // if the number pointed to by it is less than the number
      // pointed to by indexOfLowestNum,
      if (intsToSort[j] < intsToSort[indexOfLowestNum]) {
        // then make this the new provisional miniumum index.
        indexOfLowestNum = j;
      }
    }
    // At the end of the ith iteration, put the number that is in the
    // indexOfLowestNum position into the ith position (and vice versa).
    swap(&intsToSort[i], &intsToSort[indexOfLowestNum]);
  }
}
\end{lstlisting}
\caption{A set of instructions (in C) implementing the `selection
  sort' solution to the problem of sorting a list of given
  integers. The algorithm first puts the lowest integer into position
  0 of the list, then puts the lowest of the remaining integers into
  position 1, and so on.}
\label{sig:fig:prog}
\end{figure}

\begin{figure}
\begin{tikzpicture}[->,>=stealth',shorten >=1pt,auto,node
    distance=2.8cm, semithick]

  \node[initial,state]   (A)                    {$q_0$};
  \node[state]           (B) [right of=A]       {$q_1$};
  \node[accepting,state] (C) [below right of=A] {$a$};

  \path (A) edge  [loop above] node {0} (A)
            edge               node {1} (B)
        (B) edge  [loop above] node {1} (B)
            edge               node {0} (C)
        (C) edge  [bend left]  node {0} (A)
            edge  [bend left]  node {1} (B);
\end{tikzpicture}
\caption[A finite state machine]{A state diagram representation of a
  finite state machine. Binary strings of variable length are input to
  the automaton. They are `accepted' if the machine is found to be
  in the state $a$ after the last character has been read. This
  particular machine will accept any string ending in `10'.}
\label{sig:fig:statediag}
\end{figure}
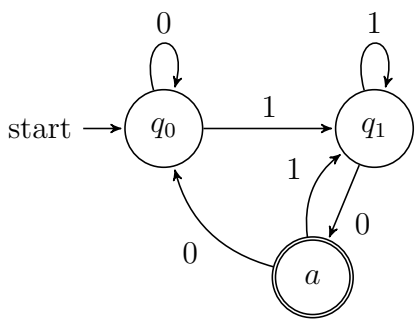

From the ontic point of view, on the other hand, one may attempt to
explain the fact that the computer has solved a computational problem
by imagining a set of possible state transitions of the computer. We
thus imagine a process by which the computer begins in an initial
state $A$, undergoes a series of state transformations, and ends,
finally, in a state $B$, which can then be interpreted as a resolution
to the problem under consideration. Now within the ontic point of
view, there are varying levels of detail which can be employed to
produce such an explanation. We can, for instance, provide a detailed
description of the machine-level instructions required to implement
the algorithm. These instructions will be different, according to the
architecture of the computer on which the algorithm has been
run. Still within the ontic point of view, we can descend some
levels lower, by describing the detailed physical implementation of
the register and memory locations, the bus, etc., of the particular
computer on which the algorithm has been run. We can also ascend
higher in the hierarchy of levels. Perhaps the highest point in this
hierarchy which can still be considered as exemplifying the ontic
point of view is the level of the so-called \emph{state transition
  diagram} (see, e.g., Figure \ref{sig:fig:statediag}). Though
abstract, state transition diagrams can be considered as exemplifying
the ontic point of view in that they purport to describe the
\emph{essential characteristics of the states and state transitions}
associated with a machine capable of implementing the algorithm.

Some explanatory questions effectively admit of only one type of
answer. For instance, if we have been asked to explain the detailed
operation of a modern day computer operating system (i.e., why it is
able to perform the operations that it does), we will typically
employ the software point of view. We will, for instance (though it
may take some time) print out and examine the high-level computer
code; or, alternately, if this is judged to be too cumbersome, we may
employ even higher level descriptions: high-level flowcharts, `use
case' diagrams, and so on. The hardware, or ontic, point of view, on
the other hand, is usually not employed to answer questions of this
type. It is extraordinarily difficult (though not impossible in
principle for an idealised finite being) to explain the detailed
workings of an operating system using a state transition diagram in
which the state of the computer is kept track of at each computational
step. Thus the hardware point of view is more limited in this respect:
above a certain level of complexity it becomes too difficult to give
an explanatory account, from the hardware point of view, of exactly
\emph{why} a computational process has solved (i.e., what steps were
taken by it to solve) a particular instance of a computational
problem.

Yet as we saw in \textsection \ref{sig:s:whyhowp}, \emph{why?}
questions of this sort are not the only types of questions that one
may ask in the computational context. In fact there are other types of
explanatory questions that are also appropriate to ask from the ontic
point of view. Consider, again, the state machine depicted in Figure
\ref{sig:fig:statediag}. This is an example of a deterministic finite
automaton: a state machine implementing a finite set of states and
deterministic transitions between states. Now, there are, of course,
other types of state machine. For instance, there are nondeterministic
finite automata, deterministic and nondeterministic `pushdown'
automata, and deterministic and nondeterministic Turing machines, to
name a few \citep[cf.][]{martin1997}. And these are all described
essentially in terms of the possible states and state transitions
which they are capable of.

One type of question we can ask, from the ontic point of view,
concerns the characteristics of particular classes of automata. We can
ask, for instance, about the class of problems \emph{computable} by
the machines of a particular class. It turns out that finite automata
are severely limited with respect to the class of problems they are
capable of solving, while Turing machines, in contrast, are capable of
solving any effectively calculable function. As another example, we
can ask about the \emph{resources required} to solve certain classes
of computational problems by automata of a particular sort. We can
ask, for instance, about the class of problems solvable by a
deterministic Turing machine in `polynomial time', those solvable by a
nondeterministic Turing machine in `exponential time', and so on
(cf. Appendix \ref{ch:cc}). In order to answer these and other similar
questions, we will appeal to the essential characteristics of the
hardware: to the states and state transitions which can be realised
and which are possible for a particular class of automata, and if we
are asked \emph{how is it possible that} a particular class of
problems is solvable by, for instance, a nondeterministic Turing
machine in polynomial time, we will explain that this is so because of
the state space and state transitions that are possible for the
machine. All of these, and other, questions, are examples of
\emph{how-possibly?} questions.

Let us now come back to the characterisation of \emph{quantum}
computation. The question, `what is the physical source of quantum
speedup?', is a request for ontic explanation that can be framed as
either a \emph{why?} or a \emph{how-possibly?} question. In the former
case we can understand it as asking `why did this particular quantum
computer solve this computational problem in $O(n)$ steps, as opposed
to $O(2^n)$ steps?' Answering this question will involve describing
the actual causal history of the quantum computer\textemdash each
individual transition undertaken by it to solve the computational
problem. While such a causal history may be interesting for some
purposes, it does not strike me as the appropriate answer to give to
the question which is actually being asked; for this question, I
believe, is more appropriately characterised as a \emph{how-possibly?}
question: a request for the structural features of quantum computers
which make it possible for them to outperform classical computers. And
here, just as with the question, `why are Turing machines more
powerful than finite automata?', it is appropriate to answer by
appealing to the state space and state transitions that are possible
for a quantum as opposed to a classical machine.

Now as I explained in detail in \textsection \ref{suff:s:suff},
because of the possibility of quantum entanglement, $n$-fold,
$d$-dimensional quantum systems are capable of efficiently
representing the possibilities associated with a $d^n$-dimensional
Hilbert space, while $n$-fold $d$-dimensional classical systems are
capable of efficiently representing a space of only $d \cdot n$
dimensions. The quantum computer has exponentially more resources at
its disposal than a classical computer, therefore, which it may use in
order to solve a particular computational problem: there are
`shortcuts' through state space, accessible to a quantum computer,
which are inaccessible to classical systems. Thus I submit that it is
the possibility of \emph{entanglement}\textemdash i.e., the fact that
compound states of quantum systems may sometimes transition to
\emph{entangled states}\textemdash which is the explanation for
quantum computational speedup (if quantum speedup is, in fact,
possible) from the physical or ontic point of view. As I argued in
Chapter \ref{ch:nec}, entanglement is necessary for explaining quantum
speedup, and as I argued in Chapter \ref{ch:suff}, it is sufficient as
a resource (if anything is) as well. And as I have just argued, the
states and state transitions made possible by entanglement are
sufficient to explain quantum speedup from the ontic point of
view. In the context of physical theory, ontic explanation just is
physical explanation. Thus I claim that this explanation of quantum
speedup is the physical explanation that we have been seeking.

A higher-level explanation\textemdash one that is closer to the level
of the recipient of explanation but still reducible in principle to the
physical level\textemdash would be desirable and would serve to
illuminate much, for us, about the nature of the physical world. This
is what I take to be the aim of explanations of quantum speedup that
appeal, for instance, to the fact that quantum computers are capable
of massively parallel function evaluation using a single circuit
\citep[]{duwell2004, duwell2007, hewittHorsman2009}, or accounts of
quantum speedup that explain it as arising from the manipulation of
the correlations between these function evaluations instead of the
results of the evaluations themselves \citep[]{steane2003}, or those
which describe quantum computers as computing the global properties of
functions \citep[]{bub2006, bub2010}.

In \textsection \ref{sig:s:ontexp} I made a distinction between i)
properly ontic explanations, ii) higher-level explanations that are
directly translatable (at least in principle) into ontic language, and
finally iii) higher-level explanations that are not so
translatable\textemdash what I there called `analogical'
explanations. The explanations of quantum computation just referred to
seem to fall within the first subdivision, for in these explanations,
quantum algorithms are usually described by something very similar to
what I have, above, characterised as state transition diagrams. And
I previously described these diagrams as belong to the ontic point
of view. When one interprets the action of the unitary gates employed
in quantum algorithms as implementing \emph{function evaluations},
however (or perhaps: operations on the correlations between these
evaluations, or perhaps: global properties of functions), one is,
strictly speaking, employing a concept (`function') that properly
belongs to a higher-level\textemdash the `software'-level\textemdash
of description.

For the case of a \emph{classical} computer, one can typically
translate talk of functions to talk of their low-level implementation
without loss of content. Thus in the classical case, explanations such
as these could still claim to be ontic despite their added emphasis on
the recipient of explanation; i.e., despite being at a level removed
from properly ontic explanation. Thus in the classical case, such
explanations would be classed within the second subdivision. As
was made clear in Chapter \ref{ch:mwi}, however, a description of a
\emph{quantum} state transformation such as
\begin{eqnarray}
\label{sig:eqn:parallel}
\sum_{x=0}^{2^n-1} |
x \rangle | 0 \rangle \rightarrow \sum_{x=0}^{2^n-1} | x \rangle |
f(x) \rangle,
\end{eqnarray}
should not necessarily be taken at face value. Regarding the state
resulting from such a transformation, one cannot say, for instance,
and despite appearances, that 2$^n$ evaluations of the function $f$
are therein represented. Reiterating Mermin: ``One cannot say that the
result of the calculation \emph{is} 2$^n$ evaluations of $f$, ... All
one can say is that those evaluations characterize the \emph{form} of
the state that describes the output of the computation. One knows what
the state \emph{is} only if one already knows the numerical values of
all those 2$^n$ evaluations of $f$.''
\citeyearpar[p. 38]{mermin2007}. This is to say nothing of the
existence of alternative models of quantum computation such as the
cluster state model which, as we have seen in Chapter \ref{ch:mwi},
complicate the situation yet further with respect to the significance
of a state such as \eqref{sig:eqn:parallel}.

The project of providing an answer to the question of the explanation
for quantum speedup from a higher-level, but still reducible, point of
view is both an interesting and important one, and I should not be
here understood as denying that this project may ultimately prove
successful. Nor should I be understood as claiming that all
\emph{existing} attempts at such an explanation must fail. While the
many worlds explanation of quantum speedup, as we saw in Chapter
\ref{ch:mwi}, may be untenable, other high-level explanations of this
sort may still succeed. But I hope it is clear that any explanation
from this point of view which is unable to resolve these
interpretational problems must be seen as, at best, analogical in the
sense in which I defined that term above\textemdash as belonging to
the \emph{third} subdivision. Of course, even here, the label
`analogical' should not be taken in a derogatory sense; explanations
of this sort have been and are \emph{enormously} useful for the
development of our fundamental theories. Even if such explanations are
not ontic, in the sense in which I have defined that term above, they
undoubtedly illuminate a great deal about the objects of our
investigations.

But regardless of whether such a project has any hope of success, an
investigation of the `lower-level' sort\textemdash one undertaken from
a point of view that remains as close to the `hardware' as is both
possible and appropriate\textemdash will be useful, both for its own
sake and also because it may prove informative for the higher-level
project. It is just such an investigation which I have undertaken
here.

\section{Ontic \emph{Why?} questions and causal explanation}
\label{sig:s:physexp}

Yet there will be those who still remain unsatisfied. They will
counter that an explanation for quantum speedup from the physical
point of view has not truly been given, for I have not answered the
question of what entangled quantum states fundamentally represent;
i.e., I have not answered the question of \emph{why} quantum systems
sometimes become entangled\textemdash of what \emph{underlying
  causes} give rise to the observed probabilities for outcomes of
experiments and allowed state transitions associated with entangled
states.

The claim that the only appropriate answer to a \emph{why?} question
in the scientific context is a causal explanation\textemdash that we
can be said to have explained `why $X$?' only when we have answered
that it is be-\emph{cause} of $Y$\textemdash is, I believe, unlikely
to be correct for the general case.\footnote{For a time at least
  \citep[]{salmon1984}, defended such a view, as have
  \citet[]{humphreys1989}, and \citet[]{ruben1990}.} Though it will
not be necessary to defend this claim here, I do believe, for
instance, that mathematics is a science, that there are such things as
mathematical explanations, and that the case cannot be made that
mathematical explanations are causal, unless one means by `causal'
something very far removed from its ordinary signification.

The claim that all answers to \emph{why?} questions must be causal is
more plausible, however, if one restricts one's attention to physics,
or at any rate to \emph{physical processes} (such as the quantum
computational process). Whether or not one agrees with this claim, it
must be admitted that it is at least not absurd to insist that an
explanatory account of a physical process must include an account of
how a particular kind of state of the process comes about or is caused
by the process's immediately prior state. And for those who hold such
a view, a non-causal physical explanation is no physical explanation
at all.

It is common to view quantum entanglement as essentially arising from
the prior physical interaction of two or more quantum systems
\citep[cf.][]{schrodinger1935}. From this point of view, it is
possible to give something like a causal or mechanistic explanation of
the possibility of quantum state transitions to entangled states. Such
an explanation can be construed as causal, at least in the
minimalistic sense that quantum entanglement is explained as having
determinately arisen from the physical interactions of physical
systems.

Such a view has been challenged, however. According to John Stachel,
quantum entanglement should not most generally be understood as the
result of prior physical interactions. Rather, for Stachel, quantum
entanglement should be understood as the manifestation of the effects
consequent upon a set of abstract requirements for determining the
probabilities associated with quantum systems, while these abstract
requirements themselves, according to Stachel, are mechanically
inexplicable and `mysterious'. If this is correct, then it will lead
us to doubt whether a causal characterisation of entanglement (and
hence a physical explanation for quantum speedup) is possible.

In the sequel I will argue that Stachel is perfectly correct to
maintain that the statistics associated with entangled quantum systems
are characterisable in terms of a set of abstract requirements. I will
also argue, however, that it is possible to characterise these
abstract requirements as themselves arising from physical
interactions, and thus that a causal characterisation of entanglement,
at least in this minimalistic sense, can be given. Thus our
explanation of quantum speedup should not be objectionable to those
who insist on the essentially causal nature of physical explanation
(at least as it relates to \emph{why?} questions such as the `why do
quantum systems become entangled?' question).

\section{The mystery of self-interference}
\label{sig:s:mystery}

\subsection{Interfering quantum gates}
\label{sig:ss:intgate}

A classical gate $C$ which flips its input bit with probability $1/2$
(e.g., a very noisy NOT gate) will have the following transition
probabilities: $p^C_{00} = p^C_{01} = p^C_{10} = p^C_{11} = 1/2.$
Since $|\frac{1}{\sqrt{2}}|^2 = |\frac{i}{\sqrt{2}}|^2 = 1/2$, it
follows from the Born rule that a quantum gate $Q$ will yield the same
transition probabilities as $C$ if it is defined to act on a qubit in
the following way:
\begin{align}
| 0 \rangle \xrightarrow{Q} \frac{i}{\sqrt{2}}| 0 \rangle +
\frac{1}{\sqrt{2}}| 1 \rangle, \nonumber \\
| 1 \rangle \xrightarrow{Q} \frac{1}{\sqrt{2}}| 0 \rangle +
\frac{i}{\sqrt{2}}| 1 \rangle.
\end{align}

\begin{figure}
  \begin{center}
  \includegraphics[scale=0.35]{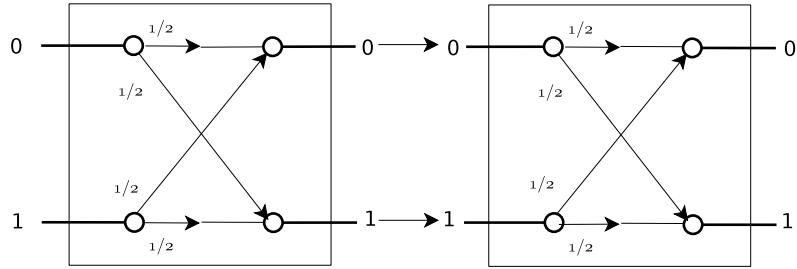}
  \end{center}
  \caption{Conjunction of two $C$ gates. Internal line labels represent
  probabilities for transitions.}
  \label{sig:fig:prob}
\end{figure}

Let us now consider the effect of concatenating two instances of $C$
and two instances of $Q$, respectively. Transition probabilities for
the former (see Figure \ref{sig:fig:prob}) are:
\begin{align}
p^{C_1C_2}_{00} & = p^{C_1}_{00}\times p^{C_2}_{00} + p^{C_1}_{01}
\times p^{C_2}_{10} \nonumber \\
= p^{C_1C_2}_{01} & = p^{C_1}_{01}\times p^{C_2}_{11} + p^{C_1}_{00}
\times p^{C_2}_{01} \nonumber \\
= p^{C_1C_2}_{10} & = p^{C_1}_{10}\times p^{C_2}_{00} + p^{C_1}_{11}
\times p^{C_2}_{10} \nonumber \\
= p^{C_1C_2}_{11} & = p^{C_1}_{10}\times p^{C_2}_{01} + p^{C_1}_{11}
\times p^{C_2}_{11} \nonumber \\
& = 1/4 + 1/4 = 1/2.
\end{align}
Transition probabilities for a concatenation of the two quantum gates,
on the other hand, are:
\begin{align}
p^{Q_1Q_2}_{00} = p^{Q_1Q_2}_{11} = 0, \nonumber \\
p^{Q_1Q_2}_{01} = p^{Q_1Q_2}_{10} = 1.
\end{align}
In other words, these two quantum gates, which by themselves yield
equal probabilities for each of the two possible outcomes, together
yield an outcome that is anti-correlated with the input value with
certainty.\footnote{It is no accident that I have chosen to describe
  the relation between input and output values in terms of
  correlations. I do so in order to highlight the affinities between
  the phenomena of interference and entanglement that are present
  even in simple examples such as this. This will be discussed in
  more depth in the following sections.} That this is so is
evident if we consider, for example, the action of $Q_1$ and $Q_2$ on
a qubit in the initial state $| 0 \rangle$:
\begin{align}
\label{sig:eqn:supprinc}
& | 0 \rangle \xrightarrow{Q_1} \frac{i}{\sqrt{2}}| 0 \rangle +
\frac{1}{\sqrt{2}}| 1 \rangle \nonumber \\
& \xrightarrow{Q_2} \frac{i}{\sqrt 2}\left(\frac{i}{\sqrt 2}| 0
\rangle + \frac{1}{\sqrt 2}| 1 \rangle\right) + \frac{1}{\sqrt
  2}\left(\frac{1}{\sqrt 2}| 0 \rangle + \frac{i}{\sqrt 2}| 1
\rangle\right).
\end{align}
This may be re-expressed as:
\begin{align}
\label{sig:eqn:probamp}
\left(\frac{i}{\sqrt 2} \cdot \frac{i}{\sqrt 2} + \frac{1}{\sqrt
  2} \cdot \frac{1}{\sqrt 2} \right)| 0 \rangle + \left(\frac{i}{\sqrt
  2} \cdot \frac{1}{\sqrt 2} + \frac{i}{\sqrt 2}\cdot\frac{1}{\sqrt
  2} \right)| 1 \rangle.
\end{align}

\begin{figure}
  \begin{center}
  \includegraphics[scale=0.35]{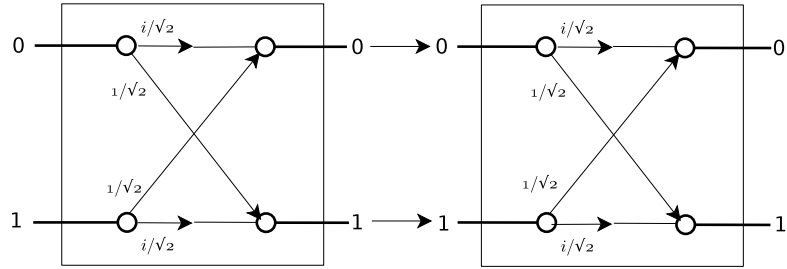}
  \end{center}
  \caption{Conjunction of two $Q$ gates. Internal line labels represent
  probability amplitudes for transitions.}
  \label{sig:fig:probamp}
\end{figure}

Eq. \eqref{sig:eqn:probamp} illustrates the fact, visualised in Figure
\ref{sig:fig:probamp}, that in order to derive the probability for the
outcome of a particular quantum mechanical experiment we must first
calculate the total probability \emph{amplitude} corresponding to that
particular outcome, by summing the probability amplitudes for all of
the possible paths through the state space of the system which yield
that particular result. Some of these paths may `interfere' with one
another. In the current example, the probability amplitudes for the
two possible paths yielding $| 0 \rangle$ are $(i/\sqrt 2)^2 = -1/2$
and $(1/\sqrt 2)^2 = 1/2$. Since these are of opposite sign, they
\emph{destructively} interfere with one another, yielding, in this
case, a total probability amplitude of 0. The probability amplitudes
for the two possible paths which yield $| 1 \rangle$, on the other
hand, \emph{constructively} interfere to yield a total probability
amplitude of $i$. By the Born rule, the probability that the result is
$| 1 \rangle$ is $|i^2| = 1$. Similarly, the reader can verify that
the action of the combined gate on an initial state of $| 1 \rangle$
will yield an outcome of $| 0 \rangle$ with certainty.

It is worthwhile to note, here, a fact which I did not call explicit
attention to in my earlier exposition of the Deutsch-Jozsa
algorithm: recall (cf. \textsection \ref{mwi:s:djalgo},
n. \ref{mwi:f:int}.) that when the function encoded in the unitary
transformation is balanced, the amplitude of $| 0^n \rangle$ in the
superposition \eqref{mwi:eqn:puredj} representing the first $n$
qubits, \emph{owing to destructive interference}, will be zero. Thus a
measurement of these qubits cannot produce the bit string $z = 0$, and
this fact allows us to distinguish constant functions, which always
yield the bit string $z = 0$, from balanced functions, which always
result in a bit string $z \neq 0$. Indeed, based on such
considerations, Lance Fortnow has gone so far as to claim that
interference, in this sense, and not entanglement, is the true source
of quantum speedup. We will not have to consider Fortnow's claim in
detail here (though the interested reader is encouraged to consult
Appendix \ref{ch:matrix}), for as I will argue later, in \textsection
\ref{sig:s:accounteprb} and \textsection \ref{sig:s:accountfeyn},
entanglement and interference can be considered as but two sides of
one and the same coin.

\subsection{The two-slit experiment}
\label{sig:ss:twoslit}

For Richard Feynman, the phenomenon (introduced in the last section)
of `self-interference',\footnote{Though in the context of his
  discussion, Feynman only mentions \emph{electron}
  self-interference, I believe we can charitably take him to be be
  referring to quantum self-interference in general.} which we can
more abstractly characterise as\textemdash a consequence of the
superposition principle\textemdash the need to sum the probability
amplitudes over all of the possible paths through a system's state
space, ``has in it the heart of quantum mechanics. In reality, it
contains the \emph{only} mystery'' \citep[vol. 3,
  1-1]{feynman1964}. What makes this phenomenon so mysterious is the
fact that classically, interference is typically associated
exclusively with wave propagation, but many of the objects which
exhibit interference effects in quantum mechanics also exhibit
characteristically particle-like effects.

\begin{figure}
\begin{tabular}{llllll}
\includegraphics[scale=0.3]{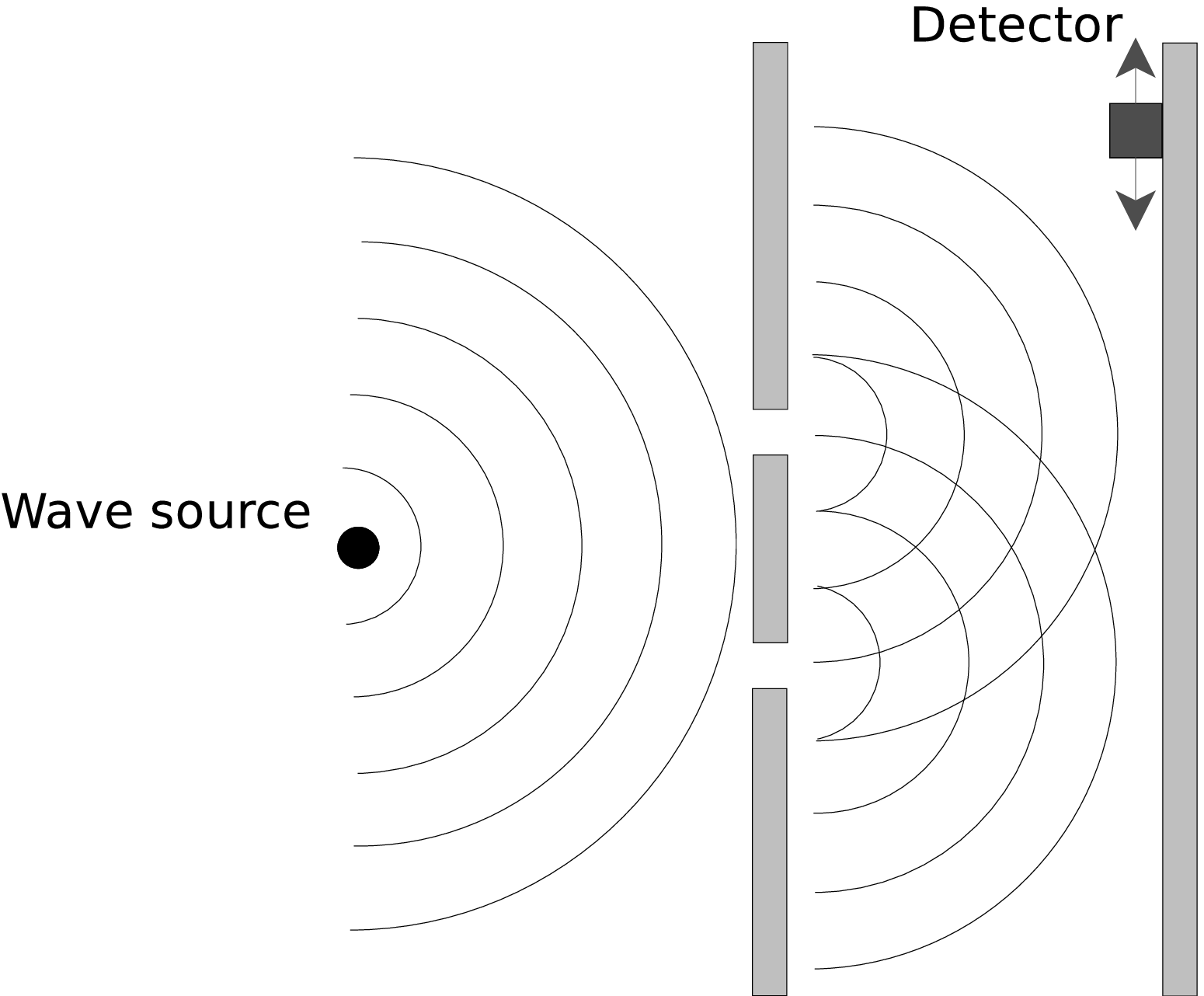} & & & & &
\includegraphics[scale=0.3]{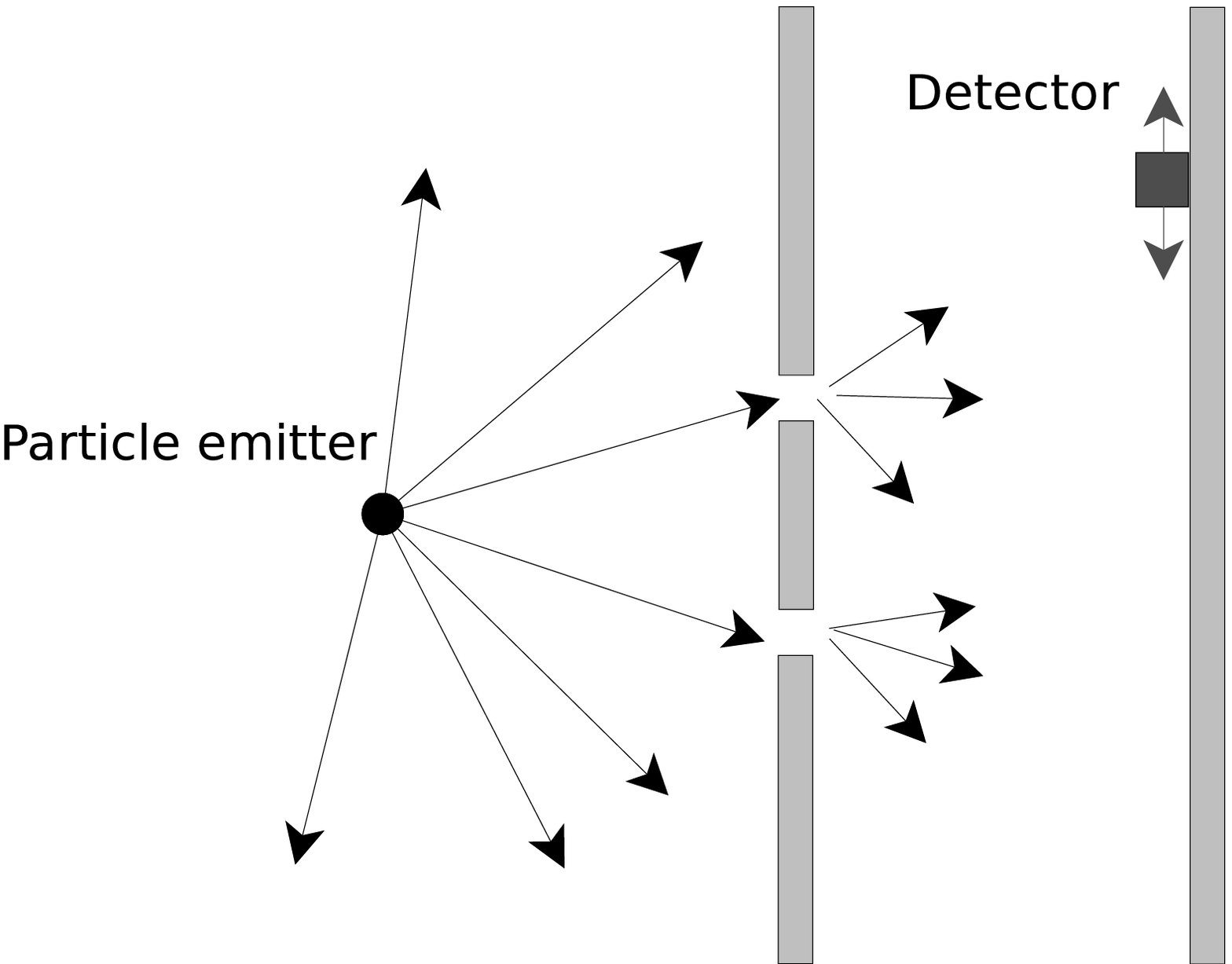}\\
\end{tabular}
\caption{A two-slit experiment with, left: a classical wave
  source, and right: a classical particle emitter.}
  \label{sig:fig:clas}
\end{figure}

Consider, for instance, an experimental setup consisting of a
classical wave source, a diaphragm into which two openings have been
cut, and a movable (in the vertical direction) detector, arranged as
in Figure \ref{sig:fig:clas}. The detector measures the intensity of
the wave motion at that location. We find that in general this
intensity can take on a continuous range of values whose distribution
for different positions of the detector reflects the constructive and
destructive interference of the waves emanating from the
apertures. Consider, on the other hand, a similarly arranged
experimental setup with, in lieu of a classical wave source, a
classical particle emitter (also depicted in Figure
\ref{sig:fig:clas}), which emits, one at a time, particles of
identical shape and size in random directions. Since the particles are
fired from the gun one at a time we will of course find no
interference effects. As for the detector, it will either detect a
particle or it will not, thus the distribution of intensity values
will decidedly not be continuous.

When we come to perform similarly arranged experiments with quantum
objects, however, things begin to get more puzzling. For instance,
suppose that, analogously to our experiment with the classical
particle emitter, we set up an experimental apparatus consisting of a
diaphragm with two apertures, a movable detector, and an
\emph{electron} gun. In this case we find that, on the one hand, as we
would expect on the assumption that electrons are particles, they
arrive at the detector one at a time and are registered with equal
intensity. On the other hand, the probability that an electron will
arrive at any given position on the back wall is distributed
analogously to the intensity distribution for a classical
wave\textemdash i.e., the probability distribution displays
interference effects, as we saw in our comparison of classical and
quantum computer gates above. Quantum objects like electrons,
therefore, manifest both particle and wave effects.

This is extremely counter-intuitive. It is difficult if not impossible
to imagine how particles, shot one at a time through a slitted
diaphragm can interfere with one another. For Feynman, this phenomenon
is simply a brute fact\textemdash one which ``is impossible,
\emph{absolutely} impossible, to explain in any classical way''
\citeyearpar[vol. 3, 1-1]{feynman1964}.

It is possible, of course, to account for these statistics by
appealing to the formal requirement of which
Eq. \eqref{sig:eqn:probamp} is an example; however, Feynman does not
consider such an account to be explanatory; for him it is \emph{only}
an account: ``We cannot make the mystery [of self-interference] go
away by `explaining' how it works. We will just \emph{tell} you how it
works'' \citeyearpar[vol. 3, 1-1]{feynman1964}. For Feynman, what is
missing from such an account is precisely a causal or mechanistic
description of the process by which self-interference phenomena
arise. Feynman is of the opinion that explanations of physical
phenomena should account for the mechanisms that give rise to
them\textemdash something which he adamantly believes cannot be done
in the case of self-interference phenomena:\footnote{I should note
  that implicit in this is a denial, by Feynman, of the possibility
  that a classical description can associate a wave with a single
  particle. This presupposition is denied by proponents of the de
  Broglie-Bohm interpretation.}

\begin{quote}
One might still like to ask: ``How does it work? What is the
machinery behind the law?'' No one has found any machinery behind the
law. No one can ``explain'' any more than we have just ``explained.''
No one will give you any deeper representation of the situation. We
have no ideas about a more basic mechanism from which these results
can be deduced \citeyearpar[vol. 3, 1-10]{feynman1964}.
\end{quote}

\section{Accounting for correlations in EPRB composite
  systems\footnote{An EPRB system is a system analogous to that
  utilised in the \emph{gedankenexperiment} of
  \citeauthor*[]{epr1935} \citeyearpar[]{epr1935}, which was
  designed to demonstrate the incompleteness of the standard
  quantum mechanical state description. The `B' is for Bohm,
  whose conceptually streamlined version
  \citeyearpar[]{bohm1951} of the \emph{gedankenexperiment} will
  be the one referred to in the remainder of this chapter.}}
\label{sig:s:accounteprb}

Consider two fermions (spin-1/2 systems) initially brought into
interaction with one another to form a composite system with zero
total spin in every direction. The system is said to be in the
`singlet state'.\footnote{This is the Bell state $|\Psi^-\rangle$ from
  Eq. \eqref{nec:eqn:bellstates}.} Since fermions may only take on
spin values of $\pm 1/2$, this requires that the spins of the
individual subsystems be oppositely correlated with one another; i.e.,
the only possibilities for their respective spins are: (a) $1/2,
-1/2$; (b) $-1/2, 1/2$. Conservation of angular momentum dictates that
as long as there are no further interactions, the subsystems must
maintain their correlation with one another even if, after some
elapsed time, they become spatially separated. In particular, if we
perform, for instance, a $\sigma_z$ experiment on one subsystem and
receive a positive result, then a $\sigma_z$ experiment on the second
subsystem must yield a negative result with certainty, and vice
versa. This will be the case regardless of the orientation of the
experimental device; i.e., we have the following relation for the
expectation value of experiments on the joint system for any direction
$\hat{m}$:
\begin{align}
\langle \sigma_m \otimes \sigma_m \rangle = -1.
\end{align}
On the other hand, if we perform a $\sigma_z$ experiment on the first
system and a $\sigma_x$ experiment on the second we will find no
correlation between the respective results. In general, for unit
vectors $\hat{m},\hat{n}$:
\begin{align}
\langle \sigma_m \otimes \sigma_n \rangle = -\hat{m}\cdot\hat{n},
\end{align}
where the scalar product $\hat{m}\cdot\hat{n} \equiv \lVert m \rVert
\lVert n \rVert cos\theta = cos\theta$ for unit vectors
$\hat{m},\hat{n}$. States such as the singlet state are examples of
entangled states. As we have already discussed (\textsection
\ref{nec:ss:ent}, \textsection \ref{suff:s:bell}), there is no local
hidden variables theory which can reproduce all of the predictions of
quantum mechanics for states such as these.

Physically, entanglement is usually given a characterisation
essentially similar to the one I have just given; i.e., when two (or
more) quantum systems, existing independently of one another in
different parts of space, are brought into temporary physical
interaction to form a composite system, then if after a time the
subsystems become spatially separated once again, it may happen that
as a result of their interaction, probabilities for outcomes of
experiments on the individual subsystems are no longer independent of
one another.\footnote{Compare: ``When two systems, of which we know
  the states by their respective representatives, enter into
  temporary physical interaction due to known forces between them,
  and when after a time of mutual influence the systems separate
  again, then they can no longer be described in the same way as
  before, viz. by endowing each of them with a representative of its
  own'' \citep[555]{schrodinger1935}.} Once entered into, this
situation will persist indefinitely and will only cease when the
subsystems undergo further interactions with other (external)
systems.

Key in the foregoing account are the ideas of spatially distinct
quantum systems and of the physical interactions between them. Yet
spatially separated ensembles are not the only quantum systems to
display statistical dependence. The two-slit experiment with
electrons, which we considered in \textsection \ref{sig:ss:twoslit}, for
instance, can be thought of as a series of experiments on a collection
of identically prepared quantum systems which together comprise a
\emph{temporally} separated ensemble.

As we saw, the results of experiments on such systems may display
statistical correlations with one another, which we normally conceive
of as arising from self-interference. Self-interference, meanwhile, is
typically considered to be an aspect of quantum mechanics that is
fundamentally distinct from whatever gives rise to the statistical
correlations observed in EPRB experiments (these, as we have just
seen, are usually conceived of as being due to the physical
interaction between spatially distinct subsystems).

According to John Stachel, however, quantum entanglement \emph{just
  is} a species of statistical dependence, and is exhibited by
\emph{both} of these phenomena. Stachel attributes no special
significance, in particular, to \emph{physical} interactions between
quantum systems: ``Rather than a physical interaction, it is precisely
the \emph{quantum entanglement} of their members\textemdash
non-interacting or interacting\textemdash that distinguishes quantum
from classical ensembles'' \citeyearpar[246]{stachel1997}.

Ultimately, for Stachel, the statistical dependence observed in both
cases is due to the requirement, illustrated by
Eq. \eqref{sig:eqn:probamp}, that probability amplitudes for all of
the possible paths through a system's state space be summed in order
to derive the probabilities for outcomes of experiments on that
system. Let us call the collection of rules that encapsulate this
requirement the `Feynman rules' for short.\footnote{Specifically, they
  are: the Born rule, the quantum law of superposition of
  amplitudes, the classical law for addition of probabilities, the
  quantum law of multiplication of amplitudes, and the classical law
  of multiplication of probabilities \citep[][\textsection
  5.5]{stachel1986}.} We saw an example of how to apply these rules to
single systems in \textsection \ref{sig:ss:intgate}. As for EPRB (and
similar) systems composed of more than one subsystem, Stachel argues
that one can think of experiments on such systems as composed of two
steps. The first step, consisting of an experiment on the first
subsystem, yields a non-maximal experimental outcome for the system as
a whole. It is followed by an experiment on the second subsystem,
which together with the first experiment can be considered as yielding
a maximal experimental outcome for the total system. Given such a
description of the experiment, Feynman's rules can be shown to
correctly account for the observed statistics.\footnote{L\"uders'
  rule \cite[cf.][]{bub1977}, $$\rho \to \rho' = \frac{P_{a_i}\rho
  P_{a_i}}{\mbox{tr}(P_{a_i}\rho P_{a_i})},$$ an alternative form of
  the von Neumann projection postulate applicable to non-maximal
  experiments, gives us the updated state of a system consequent
  upon a possibly non-maximal projective measurement of some
  observable $A$ yielding the experimental outcome $a_i$. We can use
  L\"uders rule to obtain the correct probabilities for maximal
  experimental outcomes conditional upon non-maximal experimental
  outcomes, and L\"uders rule can be shown to follow from Feynman's
  rules \citep[331-333]{stachel1986}. Note that L\"uders rule is a
  special case (for projection operators) of the more general
  measurement rule $$\rho \to \rho' = \frac{M_{\alpha_i}\rho
  M_{\alpha_i}^\dagger}{\mbox{tr}(M_{\alpha_i}\rho
  M_{\alpha_i}^\dagger)},$$ where $M$ is in general not a projection
  operator \citep[cf.][\textsection 2.4.2]{nielsenChuang2000}.}

\section{Accounting for the Feynman rules}
\label{sig:s:accountfeyn}

\subsection{Physical interactions}
\label{sig:ss:phys}

In the last section we saw that it is possible to characterise the
statistics manifested by EPRB-type composite systems as stemming from
the need to sum the probability amplitudes for all of the paths a
system may take through its state space (i.e., from Feynman's
rules). From this, Stachel has concluded that EPRB-type effects,
usually taken as a paradigm example of the effects consequent upon
physically interacting quantum systems, are in reality just
consequences of Feynman's rules. The requirement expressed by
Feynman's rules, in fact, for Stachel (just as for Feynman himself) is
the true and only quantum mystery.

This would seem to undercut my claim that the explanation of quantum
speedup I have given above can be construed in a causal way. But
before we accept this conclusion, let us see if something rather more
subtle may be at work. In particular, let us determine whether it is
possible to characterise the requirements expressed by Feynman's rules
as themselves stemming from physical interactions of some sort. If we
could show this, we might then conclude that characterising quantum
systems in terms of the requirements imposed by Feynman's rules, on
the one hand, and in terms of physical interactions, on the other, are
merely two different ways of regarding one and the same quantum
mystery. Those with a predilection for causal-mechanical descriptions,
of course, will prefer the latter.

In fact, we \emph{can} provide such a characterisation if we focus on
the system's \emph{interaction with the state preparation device}. In
particular, the common source of fermions in the EPRB experiment may
be taken to represent a \emph{physical source} for the entanglement
present in a system in (for instance) one of the Bell
states\textemdash a physical source, moreover, that is in the common
causal past of both subsystems.

That is all well and good for an experimental setup such as the EPRB.
But, one might object, if we are to answer Stachel's challenge we must
provide a physical source for the entanglement present in
temporal ensembles as well as in spatial ensembles, for the
entanglement present in \emph{single} particle experiments as well as
in the EPRB-type experiments.

\begin{figure}
\begin{center}
  \includegraphics[scale=0.25]{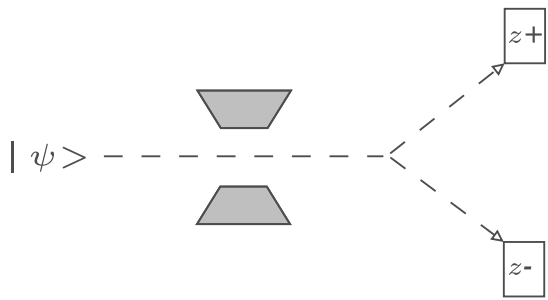}
\end{center}
\caption{A Stern-Gerlach-like experimental setup (implementing a
  $\sigma_z$ experiment) in which a two-dimensional quantum system
  is sent through a state preparation device, after which it impinges
  on one of two detectors.}
\label{sig:fig:sgexp}
\end{figure}

The following consideration should allay this concern. Imagine a
spin-$\nicefrac{1}{2}$ particle that has been sent through a
Stern-Gerlach apparatus oriented in the $\hat{z}$ direction (see
Figure \ref{sig:fig:sgexp}). Once it has passed through the apparatus,
there is, henceforward, an important sense in which only \emph{joint}
experiments on the system are possible, for now an experiment to
determine whether the particle occupies a particular spatial region is
\emph{implicitly also} an experiment to determine whether the particle
is in a complementary spatial region. For supposing that the effect of
the magnet is that the particle is now in a superposition of being in
the spatial regions occupied by the $z+$ and $z-$ detectors. Then in
that case the combined state of the two spatial regions will be
expressible in the occupation number formalism
\citep[cf.][Ch. 7]{mattuck1976} as follows:
\begin{align}
\label{sig:eqn:altqbz}
| \psi \rangle = a| 1 \rangle_{z+}| 0 \rangle_{z-} + b| 0 \rangle_{z+}|
1 \rangle_{z-}.
\end{align}
Here, $| 1 \rangle_\alpha$ signifies that one particle occupies the
spatial region inhabited by the $\alpha$ detector, while $| 0
\rangle_\alpha$ signifies that no particles occupy the spatial region
inhabited by the $\alpha$ detector. Eq. \eqref{sig:eqn:altqbz}
expresses the fact that if we perform a $\sigma_z$ experiment on
$\psi$ and detect a particle at the $z+$ detector, then we cannot also
detect a particle at the $z-$ detector, and vice versa. Thus we can
think of the statistics associated with a \emph{single} particle as,
from another point of view, the statistics associated with an
entangled state of two spatial regions,\footnote{It is worth noting
  that the situation described here is essentially similar to the
  situation described by Einstein in the argument for the
  incompleteness of quantum mechanics which he gave at the Solvay
  Congress of 1927 \citep[cf.][115-121]{jammer1974}, and also to the
  situation described in his letter to Schr\"odinger of 19 June,
  1935. \citet[]{norsen2005} has argued that this argument is in
  fact a conceptually simpler and superior version of the more
  well-known EPR argument. The EPR argument figures prominently, of
  course, in almost all discussions of entanglement. For more on
  this topic, see \citet[]{shimony2005, norton2011}.} where this
entangled state has been brought about via the influence of the
Stern-Gerlach apparatus, which we can think of as representing ``an
influence on the very conditions which define the possible types of
predictions regarding the future behaviour of the system''
\cite[700]{bohr1935}.\footnote{This entire passage is emphasised in
  the original, thus there is no harm in not reproducing the
  emphasis, as I have done here.} In this vein one recalls Shimony:
``It must be emphasized that the concept of entanglement is
inseparable from the role of potentiality in quantum mechanics''
\citep[142-143]{shimony1993}. In an entangled state such as
\eqref{sig:eqn:altqbz}, Shimony writes, the two observables involved
are ``... merely potential, but in an interlocked manner'' (ibid.).

Considering the state of a system from varying points of view is all
very well, one might interject at this point, but the proof is in the
pudding: do such states manifest detectable correlations between their
subsystems? The, perhaps surprising, answer seems to be yes; the state
of a single system, such as a photon, can give rise to EPRB-type
correlations that are detectable in principle by experiment. This was,
in fact, illustrated with a \emph{gedankenexperiment}, some time ago,
by Lucien \citet[]{hardy1994}. We will consider this
\emph{gedankenexperiment} in the next section.

\subsection{Entanglement of a single photon}
\label{sig:ss:hardyexp}

\begin{figure}
  \begin{center}
  \includegraphics[scale=0.25]{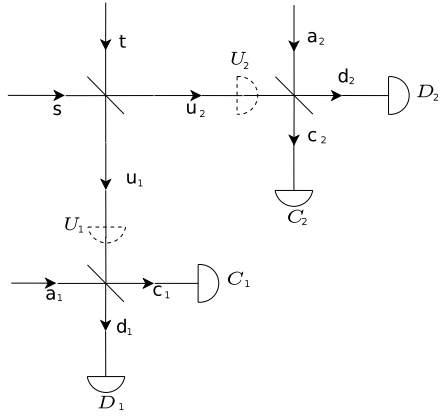}
  \end{center}
  \caption{The experimental setup for Hardy's
  \emph{gedankenexperiment}. A photon in the state $q| 0 \rangle +
  r| 1 \rangle$ is incident on the $s$ mode; the vacuum state is
  incident on the $t$ mode; $| \alpha_1 \rangle$ and $| \alpha_2
  \rangle$ are incident on the $a_1$ and $a_2$ modes,
  respectively. Source: \citet[]{hardy1994}.}
  \label{sig:fig:hardy}
\end{figure}

Hardy's thought experiment consists of three 50:50 beam
splitters\footnote{The experiment is conducted with photons, which
  unlike the spin-1/2 qubits we considered in the previous section,
  are spin-1 systems. The difference is inessential.} each
implementing the following state transformations, expressed in the
occupation number formalism as:
\begin{align}
\label{sig:eqn:splittersa}
| 0 \rangle_a| 0 \rangle_b & \to | 0 \rangle_c| 0 \rangle_d, \\
\label{sig:eqn:splittersb}
| 0 \rangle_a| 1 \rangle_b & \to \frac{1}{\sqrt 2}(| 0 \rangle_c| 1
| \rangle_d + i| 1 \rangle_c| 0 \rangle_d), \\
\label{sig:eqn:splittersc}
| 1 \rangle_a| 0 \rangle_b & \to \frac{1}{\sqrt 2}(i| 0 \rangle_c| 1
| \rangle_d + | 1 \rangle_c| 0 \rangle_d).
\end{align}
Here $a,b$ are the input and $c,d$ are the output modes, and it is
assumed for simplicity that the $a$ mode is transmitted into the $c$
mode and likewise for $b$ and $d$. A photon, prepared in the state $q|
0 \rangle_s + r| 1 \rangle_s$, is directed at the $s$ input of one of
the beam splitters (see Figure \ref{sig:fig:hardy}), while the input to
$t$ is the vacuum state $| 0 \rangle_t$. The outputs of this splitter,
$u_1$ and $u_2$, are fed as inputs to two further beam splitters,
where they are each mixed with the coherent states $| \alpha_1
\rangle_{a_1}$ and $| \alpha_2 \rangle_{a_2}$. The outputs of these
beam splitters, $c_1, d_1, c_2, d_2$ are then fed to photon number
detectors, $C_1, D_1, C_2, D_2$. Additionally, two more detectors,
$U_1, U_2$, may be optionally inserted into the paths $u_1, u_2$,
respectively.

It turns out that \citep[cf.][]{hardy1994} when neither $U_1$ nor
$U_2$ are removed from paths $u_1$ and $u_2$, it is impossible for
both $U_1$ and $U_2$ to register a photon; writing $X_i = n$ to
indicate that $n$ photons were detected at detector $X_i$, we have:
\begin{align}
\label{sig:eqn:exp1}
U_1 = 1 \mbox{ and } U_2 = 1 \mbox{ never happens}.
\end{align}
When $U_1$ is removed, we have the result that:
\begin{align}
\label{sig:eqn:exp2}
\mbox{if } F_1 = 1 \mbox{ then } U_2 = 1,
\end{align}
where we have written $F_1 = 1$ as shorthand for $C_1 = 0, D_1 =
1$. Similarly, when $U_2$ is removed:
\begin{align}
\label{sig:eqn:exp3}
\mbox{if } F_2 = 1 \mbox{ then } U_1 = 1,
\end{align}
where $F_2 = 1$ is shorthand for $C_2 = 0, D_2 = 1$.
When $U_1$ and $U_2$ are both removed, we find that
\begin{align}
\label{sig:eqn:exp4}
F_1 = 1 \mbox{ and } F_2 = 1 \mbox{ happens sometimes.}
\end{align}

To appreciate the significance of these results, imagine that Alice
and Bob are at ends 1 and 2 respectively, and suppose that Alice
chooses to perform experiment $F_1$ and Bob chooses to perform
experiment $F_2$. Further suppose that these yield $F_1 = 1, F_2 =
1$. Alice can deduce from $F_1 = 1$ and \eqref{sig:eqn:exp2} that the
photon from the source would have been detected in $u_2$ if Bob had
placed $U_2$ there. From $F_2 = 1$ and \eqref{sig:eqn:exp3}, Bob can
deduce that the photon from the source would have been detected in
$u_1$ if the detector $U_1$ had been placed there by Alice. They
cannot both be correct, however, for only one photon has been emitted
from the source; i.e., \eqref{sig:eqn:exp1} will then be violated.

According to Hardy, it is possible to avoid this contradiction if we
are willing to drop the assumption of
\emph{locality}:\footnote{\label{sig:foot:newhardy} Hardy's
  interpretation of these results, when first published, were quite
  controversial (cf. \citealt[]{vaidman1995}; \citealt*[]{ghz1995};
  \citealt[]{hardy1995}). Since then, an improved, and less
  controversial, version of Hardy's experiment has been proposed which
  is both feasible and capable of demonstrating
  Bell-inequality-violations (cf. \citealt[]{dunningham2007};
  \citealt[]{terraCunha2007}). I have limited myself here to an
  exposition of Hardy's original scheme as it is conceptually
  simpler, while the differences between the various versions are
  inessential to the point I am making.}

\begin{quote}
... there is an implicit assumption of locality in this reasoning, and
... without this assumption there is no contradiction. Alice obtains
$F_1 = 1$. Bob is actually measuring $F_2$. Alice might deduce from
her result and the prediction [\eqref{sig:eqn:exp2}] that had Bob
measured $U_2$ instead he would have gotten $U_2 = 1$. However,
without assuming locality, this deduction is wrong, because if Bob had
decided to measure $U_2$ instead, there might then have been a
nonlocal influence from Bob's end to Alice's end
\citep[2281-2282]{hardy1994}.
\end{quote}

Let us forego evaluating Hardy's interpretation of the significance of
this experiment (specifically, his attribution of \emph{nonlocality}
to the effects manifested by the experiment). It is enough to note
that, irrespective of whether we interpret these effects as nonlocal,
the Hardy \emph{gedankenexperiment} manifests effects which we would
normally associate with physically interacting multi-particle
systems. The experiment thus illustrates, in a concrete manner, that
the quantum superposition of a single system can also be thought of in
terms of the correlations consequent upon its physical interactions
with an experimental device.\footnote{As was explained previously, one
  can think of the entanglement effects that are manifested by the
  subsystems of single particle systems as arising from the
  interaction of the system in question with a state preparation
  device in the common causal past of its subsystems. In the Hardy
  experiment, the vacuum state incident on the $t$ mode of the first
  beam splitter comprises part of the experimental setup, yet it
  seems strange to interpret the vacuum as part of the \emph{cause} of
  the correlations subsequently manifested by the experiment. This
  should not be controversial, however, as long as we remember that we
  are not dealing, in this experiment, with a naturally occurring
  vacuum, but rather with a vacuum that has been specifically
  prepared by a laboratory technician (by, for instance, physically
  placing a screen in front of this part of the apparatus). While
  vacuums do occur (rarely) in nature we do not, at least at this
  point, have any capability to use them for the purposes of
  experiment. Vacuums must be created in order to be used in this
  way, thus this state preparation may be interpreted as a part of
  the cause of the correlations manifested in the Hardy thought
  experiment, in the manner outlined in the previous section. I
  thank Wayne Myrvold for this point.} In this vein,
\citet[1]{dunningham2007} conclude:

\begin{quote}
Feynman once famously claimed that superposition is the only mystery
in quantum mechanics. Others would add nonlocality to the list. If,
however, single particles can exhibit nonlocality, then these two
mysteries become one and the same.
\end{quote}

Entanglement represents a real physical feature of quantum systems,
whether or not we maintain that we should require of such physical
features that they be explicable in terms of the physical interactions
of systems. An explanation of quantum speedup, therefore, according to
which it is entanglement which makes it possible for quantum systems
to outperform their classical counterparts, is a physical explanation
of quantum speedup on a reasonable interpretation of what it means for
an explanation to be physical.

\chapter{Summary and Conclusion}
\label{ch:conc}

\settocdepth{section}

In Chapter \ref{ch:mwi} I began this dissertation by considering the
most popular of the candidate physical explanations for quantum
speedup: the so-called \emph{many worlds explanation of quantum
  computation}. I argued that, although it is inspired by the
neo-Everettian interpretation of quantum mechanics, unlike the latter
it does not have the conceptual resources required to overcome the
preferred basis objection. I also argued that the many worlds
explanation, at best, can serve as a good description of the physical
process which takes place in network-based computation, but that it is
incompatible in an important sense with other models of computation
such as cluster state quantum computing. I next considered, in Chapter
\ref{ch:nec}, a common component of most other candidate explanations
of quantum speedup: \emph{quantum entanglement}. I investigated
whether entanglement can be said to be a necessary component of any
explanation for quantum speedup, and I considered two major purported
counter-examples to this claim. I argued that neither of these, in
fact, show that entanglement is unnecessary for speedup, and that, on
the contrary, we should conclude that it is. In Chapters \ref{ch:suff}
and \ref{ch:sig} I then asked whether entanglement can be said to be
sufficient as well. In Chapter \ref{ch:suff} I argued that despite a
result that seems to indicate the contrary, entanglement, considered
as a resource, can be seen as sufficient to enable quantum
speedup. Finally, in Chapter \ref{ch:sig} I argued that entanglement
is sufficient \emph{to explain} quantum speedup.

In this dissertation I have neither proved any original theorems, nor
provided any new experimental results. Rather, my conclusions are the
result of an investigation and analysis of the valuable scientific
contributions which have already been made. As compared to these, my
own contribution is slight. But I hope that the reader agrees that it
is not unimportant\textemdash that even if, for all of my efforts, my
conclusions are in fact incorrect, that there has been some
clarification of the underlying issues, with the hoped-for result that
there will, in the not too distant future, be new theorems and new
results in the directions pointed to by this dissertation.

In his closing remarks to the \emph{Logical Syntax of
  Language}, \nocite{carnap1937} Carnap wrote of a vision of ``fruitful
co-operative work on the part of the various investigators working on
the same problems\textemdash work fruitful for the individual
questions of the logic of science, for the scientific domain which is
being investigated, and for science as a whole.'' For `logic of
science' I would put, in its place, `philosophy of science'. But I
wholeheartedly agree with the spirit of these words. And it is my
sincere hope that this dissertation makes some approximation to
this so eloquently expressed ideal.

\appendix
\appendixpage

\chapter{Computational complexity theory}
\label{ch:cc}

\settocdepth{chapter}

Alan Turing \citeyearpar[]{turing1937, turing1938} inaugurated the
field of Computer Science by arguing persuasively for the equivalence
of the class of computable, or `effectively calculable' functions with
the class of problems computable by a Turing machine. The statement of
this equivalence is known as the Church-Turing thesis, and it is the
fundamental principle of computer science. In its early period,
research in computer science was focused primarily on the question of
computability; i.e., on the question of whether a given problem is or
is not computable by Turing machine. More recently, another focus of
research has emerged: the field known as Computational Complexity
Theory. This field is dedicated to the more practical question
concerning the \emph{cost} of solving a given computational problem.

A basic distinction, in Complexity Theory, is between those
computational problems that are amenable to an \emph{efficient}
solution in terms of time and/or space resources, and those that are
not. Easy (or `tractable', `feasible', `efficiently solvable', etc.)
problems are those for which solutions exist which involve resources
bounded by a polynomial in the input size, $n$. Hard problems are
those which are not easy, i.e., they are those whose solution requires
resources that are `exponential' in $n$, i.e., that grow faster than
any polynomial in $n$ \citep[p. 139]{nielsenChuang2000}.\footnote{The
  term `exponential' is being used rather loosely here. Functions
  such as $n^{\log n}$ are called `exponential' but do not grow as
  fast as a true exponential such as $2^{n}$.}

For example, a problem, which for input size $n$, requires $\approx
n^c$ steps to solve (where $c$ is some constant) is polynomial in
terms of time resources in $n$ and thus tractable according to our
definition. A problem that requires $\approx c^n$ steps to solve, on
the other hand, is exponential in terms of time resources in $n$ and
is therefore intractable according to our definition. The definition
is a coarse one, and its usefulness will depend, in a given case, on
the values of $c$ and $n$. Nevertheless it is adequate for most cases
of practical interest.

An important theoretical reason for adopting the definition is the
following principle, usually referred to as the `Strong' Church-Turing
thesis in the literature. In order not to confuse this thesis with the
more fundamental (`weak') Church-Turing thesis that I mentioned
earlier, I will refer to it as the \emph{Computational Efficiency
  Thesis} (CET), which states that:

\begin{quote}
\emph{Any model of computation can be simulated on a probabilistic
  Turing machine with at most a polynomial increase in the number of
  elementary operations required}
\citep[p. 140]{nielsenChuang2000}.\footnote{A probabilistic Turing
  machine is one for which transitions between states are chosen
  from a set according to some probability distribution, rather than
  assigned deterministically.}
\end{quote}

If, now, we identify easy problems with those having polynomial
resource solutions, then CET tells us that in our analysis of
computational complexity, we can restrict our attention to the
probabilistic Turing machine model of computation
\citep[p. 140]{nielsenChuang2000}. Our definition of an easy problem,
coupled with the CET, thus provides us with an elegant,
model-independent theory of computational complexity. But note that
while CET is what gives computational complexity theory its elegant
model-independent character, and that without it ``computational
concepts and even computational kinds such as `an efficient algorithm'
or `the class \textbf{NP}' will become machine-dependent, and recourse
to `hardware' will become inevitable in any analysis of the notion of
computational complexity'' \citep[p. 245]{hagar2007b}, it is not a
foundational principle to the field of computational complexity theory
in the same way that the Church-Turing thesis is to computer
science.\footnote{See, for example, Lance Fortnow's blog entry
  \citeyearpar{fortnow2006}: ``By no means does computational
  complexity ``rest upon'' a strong Church-Turing thesis. The goals
  of computational complexity is [\emph{sic.}] to consider different
  notions of efficient computation and compare the relative
  strengths of these models. Quantum computing does not break the
  computational complexity paradigm but rather fits nicely within
  it.''}

Problems that are decidable in polynomial time by a deterministic
Turing machine are said to be in the complexity class \textbf{PTIME},
usually referred to simply as \textbf{P}. Problems for which a
deterministic Turing machine can verify whether a given solution is,
in fact, a solution are said to be in the complexity class
\textbf{NP}.\footnote{This stands for ``nondeterministic polynomial
  time,'' as it can be equivalently defined as the class of problems
  solvable in \emph{polynomial} time by a \emph{nondeterministic}
  Turing machine.} For example, consider a language $L = \{10, 11,
101, 111, 1011, ...\}$ over the alphabet $\Sigma = \{0,1\}$, the set
of binary digits. $L$ is the set of binary representations of prime
numbers. If it is possible for a deterministic Turing machine to
decide, using only polynomial time resources, whether an arbitrary
binary input string is in the language (i.e., whether it is a prime
number), we say the problem is in \textbf{P}. If, on the other hand,
one is given a string in the language at the outset, then if it is
possible for a deterministic Turing machine to \emph{verify}, in
polynomial time, that the string is, in fact, in the language, then
the problem is said to be in \textbf{NP}.

A long-standing question in complexity theory is the nature of the
relationship between \textbf{P} and \textbf{NP}. \textbf{P} is clearly
a subset of \textbf{NP}. If it can be decided in polynomial time
whether an arbitrary string is a member of $L$, then, trivially, it
can be decided in polynomial time whether a member of $L$ is a member
of $L$. It is strongly suspected that \textbf{P} is a \emph{proper}
subset of \textbf{NP}, however this has not yet been proven. This is
known as the \textbf{P} $\neq$ \textbf{NP} problem in complexity
theory.

An important notion in complexity theory is
\emph{reducibility}. Intuitively, problem $B$ is reducible to problem
$A$ if, with no more than polynomial overhead, we can convert an
algorithm for deciding $B$ into an algorithm for deciding $A$. In
other words, $B$ is reducible to $A$ if a solution for $A$ can be used
to solve $B$. Reducibility leads us to our next important complexity
class, \textbf{NP}-complete. A problem, $C$, is called
\textbf{NP}-complete if $C$ is in \textbf{NP} and every other problem
in \textbf{NP} is reducible (in polynomial time) to $C$. The concept
of an \textbf{NP}-complete problem is important for the resolution of
the \textbf{P} $\neq$ \textbf{NP} problem, for if it can be shown that
an \textbf{NP}-complete problem is solvable in polynomial time by a
deterministic Turing machine (i.e., if it can be shown that it is in
\textbf{P}), then it follows that all other problems in \textbf{NP}
are also in \textbf{P}, and hence that \textbf{P} = \textbf{NP}.

Besides \textbf{P} and \textbf{NP}, the two most relevant complexity
classes with respect to quantum computation are \textbf{BPP} and
\textbf{BQP}. \textbf{BPP} stands for bounded-error probabilistic
time. A problem, $A$, is in \textbf{BPP} if there is a (classical)
probabilistic Turing machine that will accept a string $x$ with
probability $1/2 \leq k \leq 1$ if $x \in L$ (the language
representing $A$) and reject it with probability $1/2 \leq k \leq 1$
if $x \notin L$ \citep[p. 152-153]{nielsenChuang2000}. The quantum
analogue of \textbf{BPP} is \textbf{BQP} (bounded error quantum
polynomial time), the set of problems such that a \emph{quantum
  computer} will accept $x$ with probability $1/2 \leq k \leq 1$ if $x
\in L$ and reject $x$ with probability $1/2 \leq k \leq 1$ if $x
\notin L$ \citep[pp. 200-202]{nielsenChuang2000}. \textbf{BPP}
$\subseteq$ \textbf{BQP} since a quantum computer can efficiently
simulate a classical probabilistic Turing machine
(\citealt[p. 30]{nielsenChuang2000};
\citealt[p. 240]{hagar2007b}). However, it is not clear whether
\textbf{BQP} $\neq$ \textbf{BPP}.

It is important to note that proving \textbf{BQP} $\neq$ \textbf{BPP}
amounts to proving that quantum computers are more powerful than
classical computers; but while it is strongly suspected that
\textbf{BQP} $\neq$ \textbf{BPP}, this question has not yet been
resolved. Factoring, the most famous problem for which a quantum
algorithm has been developed, has not been proven to be outside
\textbf{P}. Thus solving the factoring problem does not show us that
\textbf{P} $\neq$ \textbf{BQP}, let alone that \textbf{BPP} $\neq$
\textbf{BQP}. Note also that as of yet no quantum algorithm has been
developed which can efficiently solve a problem inside the class
\textbf{NP}-complete, and the relation between \textbf{NP} and
\textbf{BQP} is still unknown.

For more on computational complexity, see: \citet[]{papadim1994},
\citet[]{nielsenChuang2000}, \citet[]{aaronson2012}.

\chapter{Information Theory}
\label{ch:infth}

\section{Shannon entropy}

The birth and development of classical information theory is due, in
large part, to the pioneering work of Claude Shannon. In his seminal
article, ``A Mathematical Theory of Communication''
\citeyearpar[]{shannon1948}, Shannon introduced the scientific
community to the fundamental information-theoretic concept of
\emph{entropy}. Entropy is a measure of the information one gains when
one comes to know the value of a random variable. Equivalently, it can
be thought of as the uncertainty associated with a random variable;
e.g., a message produced by an information source. We define the
\emph{Shannon entropy}, $H$, with respect to the random variable
$x$, as:
\begin{align}
H(x) = -K\sum_{i=1}^{n}p(x_i)\log p(x_i),
\end{align}
where $K$ is a positive constant (amounting to a choice of unit
measure) normally chosen to be 1 \citep[p. 11]{shannon1948}, $0\log 0$
is conventionally defined to be 0, and $p(x_i)$ refers to the
probability of receiving message $i$, given a set of $n$ possible
messages. The $\log$ is typically taken to base 2. For example,
suppose an information source transmits sequences of binary digits
with the probabilities of the next digit in the sequence being a 0 or
a 1, $1/3$ and $2/3$, respectively. In this case our uncertainty with
respect to the next bit, or our entropy, is $-(1/3 \times \log 1/3 +
2/3 \times \log 2/3) = 0.92.$

Considered as a measure of our uncertainty with respect to the
messages produced by an information source, we should expect $H$ to be
0 if we are certain of the result, i.e., if one of the $p(x_i) =
1$. It is easily verified that this is the case. We should also expect
$H$ to be at a maximum when the bits are received with equal
probability, for this is the situation in which we are most uncertain
of the result. This is also easily verified
\citep[cf.][p. 11]{shannon1948}.

The \emph{joint entropy} of two random events, $x$ and $y$ is the
total uncertainty associated with $x$ and $y$. To determine it one
must take into account the probabilities of all possible combinations
of values for $x$ and $y$. Thus,
\begin{align}
H(x,y) = -\sum_{i,j}p(x_i,y_j)\log p(x_i,y_j).
\end{align}
Here, $p(x_i,y_j)$ refers to the probability that message $x_i$ and
$y_j$ occur together.

Note that it can be shown that $H(x,y) \leq H(x) + H(y)$, with
equality only if the events $x_i$ and $y_j$ are independent. This is
called subadditivity (intuitively, the total uncertainty associated
with $x$ and $y$ is equal to the sum of the uncertainties of $x$ and
$y$ unless they share information in common). Strong subadditivity,
$H(x,y,z) \leq H(x,y) + H(y,z) - H(y),$ also holds for the Shannon
entropy.\footnote{We subtract $H(y)$ from the RHS since the
  uncertainty associated with $y$ is common to $H(x,y)$ and $H(y,z).$}

The \emph{conditional entropy} of $x$ with respect to $y$,
\begin{align}
H(x|y) = H(x,y) - H(y),
\end{align}
is the total uncertainty associated with $x$ and $y$ minus the
uncertainty that disappears once we come to know $y$.

The information shared in common between $x$ and $y$, or \emph{mutual
  information} of $x$ and $y$ is defined as
\begin{align}
H(x:y) = H(x) + H(y) - H(x,y).
\end{align}
This definition is easily grasped if one expresses the equation in
terms of the joint information, i.e., $H(x,y) = H(x) + H(y) - H(x:y),$
which is the total information gain associated with $x$ and $y$ minus
the information shared in common (to avoid double counting)
\citep[p. 506]{nielsenChuang2000}.

\section{Von Neumann entropy}

The von Neumann entropy plays the same role in quantum information
theory as the Shannon entropy plays in classical information
theory. It is defined as
\begin{align}
S(\rho) = -\mbox{tr}(\rho\log\rho),
\end{align}
for a quantum system represented by the density matrix, $\rho$. As
before, by convention, $0\log 0 \equiv 0$. Since the trace of a matrix
$A$ is equal to the sum of its eigenvalues; i.e., since $\mbox{tr}(A)
= \sum\lambda_i$; the von Neumann entropy can be more usefully
expressed as
\begin{align}
S(\rho) = -\sum_x\lambda_x\log\lambda_x
\end{align}
where $\lambda_x$ are the eigenvalues of $\rho$.

The \emph{joint entropy} of a state with two components $A$ and $B$ is
defined as
\begin{align}
S(A,B) = -\mbox{tr}(\rho^{AB}\log (\rho^{AB})),
\end{align}
where $\rho^{AB}$ is the density matrix of the composite system $AB$
\citep[p. 514]{nielsenChuang2000}.

Conditional entropy and mutual information are defined analogously to
their classical counterparts. The \emph{conditional entropy} is given
by
\begin{align}
S(A|B) = S(A, B) - S(B).
\end{align}
The \emph{mutual information} is given by
\begin{align}
S(A:B) = S(A) + S(B) - S(A, B).
\end{align}

There are interesting disanalogies between the von Neumann and the
Shannon entropy. For instance, the inequality $S(A) \leq S(A,B)$ does
not hold in quantum information theory, as it does for the classical
case. In the classical case it is intuitively obvious that the
uncertainty associated with the state of one random variable cannot be
more than the uncertainty associated with the joint state of two. But
in the quantum case, this relation will fail to hold, for instance, in
the case where we have a maximally entangled state of two
subsystems. In this case, the joint state of the two systems is pure,
and hence $S(A,B) = 0$, but the marginals are completely mixed and
thus $S(A) = S(B) = 1.$ One other disanalogy, between the classical
and quantum versions of mutual information, is discussed in greater
detail in Chapter \ref{ch:nec}.

\chapter{Quantum teleportation}
\label{ch:tel}

One of the most well-known applications of
entanglement in quantum information processing is as a resource in the
so-called teleportation protocol
\citep[cf.][]{nielsenChuang2000,mermin2007}.\footnote{The quantum
  teleportation protocol originally appeared in
  \citet[]{bennett1993}. The name `teleportation' is something of a
  misnomer. To a layperson, teleportation usually brings to mind the
  idea of physically transporting objects around, possibly
  instantaneously. Quantum teleportation, however, is a protocol for
  transferring information, not physical objects, and the speed at
  which information is transferred, since it involves the exchange
  of a classical signal, is limited by the speed of light.} Consider
Alice and Bob, two spatially separated experimenters, who have the
ability to send classical information to one another (e.g., Alice may
call Bob on the telephone, send him an email, and so on). Imagine that
Alice would like to send the state (which she does not know) of some
arbitrary qubit, $| \psi \rangle = \alpha| 0 \rangle + \beta| 1
\rangle$, to Bob. Classically, this seems like a very difficult task,
for even if Alice knows the state of the qubit, she seems, in
principle, to require an infinite amount of classical information to
describe it \emph{precisely}, for the state of a qubit will in
general take on a continuum of values.\footnote{It turns out that this
  claim is actually false. Surprisingly, the teleportation protocol
  has been shown to be efficiently simulable classically. This is
  explained in Chapter \ref{ch:suff}.}

But suppose that Alice and Bob are given one extra resource: suppose
that the Bell state $| \Phi^+\rangle$ is generated, and that one half
of the Bell pair is given to Bob and the other half to Alice. Alice
may now proceed as follows. First, she interacts the qubit represented
by $| \psi \rangle$, whose state she wishes to send to Bob, with the
Bell pair; i.e., $$| \psi \rangle_a| \Phi^+ \rangle_{ab} =
\frac{1}{\sqrt 2}[\alpha| 0 \rangle_a(| 00\rangle + |
  11\rangle)_{ab} + \beta| 1 \rangle_a(| 00 \rangle + |
  11\rangle)_{ab}],$$ where $a$ and $b$ indicate whether the qubits
are in Alice's or Bob's possession. Alice then applies a
controlled-not (CNOT) operation to the qubits in her possession, using
the qubit represented by $| \psi \rangle$ as the control and her
member of the Bell pair as the target qubit. This results
in: $$\frac{1}{\sqrt 2}[\alpha| 0 \rangle_a(| 00\rangle + |
  11\rangle)_{ab} + \beta | 1\rangle_a (| 10\rangle + |
  01\rangle)_{ab}].$$ Now Alice sends the qubit represented by $| \psi
\rangle$ through a Hadamard gate,\footnote{Alice's purpose in
  performing the CNOT and Hadamard transformations is to implement,
  in a roundabout way, a measurement in the Bell-basis (which we
  assume she does not have the technology to perform directly).}
which results in:

\begin{align*}
  & \frac{1}{2}[\alpha(| 0\rangle + | 1\rangle )_a(|
  00\rangle + | 11\rangle)_{ab} + \beta (| 0\rangle - |
  1\rangle)_a(| 10\rangle + | 01\rangle)_{ab}] \nonumber \\
  & = \frac{1}{2}[| 00\rangle_{aa}(\alpha| 0\rangle + \beta|
  1\rangle)_b + | 01\rangle_{aa}(\alpha| 1\rangle + \beta|
  0\rangle)_b \nonumber \\
  & + | 10\rangle_{aa}(\alpha| 0\rangle - \beta| 1\rangle)_b  + |
  11\rangle_{aa}(\alpha| 1\rangle - \beta| 0\rangle)_b].
\end{align*}

In the next step, Alice measures her two qubits. This will yield one
of four possible measurement results (00, 01, 10, 11), and Bob's qubit
will correspondingly be in one of the following four states:
\begin{align*}
00:\quad & | \psi \rangle_b \equiv (\alpha| 0 \rangle + \beta| 1
\rangle)_b \nonumber \\
01:\quad & | \psi' \rangle_b \equiv (\alpha| 1 \rangle + \beta| 0
\rangle)_b \nonumber \\
10:\quad & | \psi''\rangle_b \equiv (\alpha| 0 \rangle - \beta| 1
\rangle)_b \nonumber \\
11:\quad & | \psi'''\rangle_b \equiv (\alpha| 1 \rangle - \beta| 0
\rangle)_b
\end{align*}

Alice now communicates her result to Bob using a classical
communications link (e.g. a telephone line). If Alice's result is 00,
then Bob's state is $| \psi \rangle_b = \alpha| 0 \rangle + \beta| 1
\rangle = | \psi \rangle_a$, i.e., the state that Alice had originally
intended to transfer. Otherwise, Bob can apply a unitary
transformation to his qubit which will transform it into the state $|
\psi \rangle_b$. For instance, if Alice's result is 01, Bob will apply
the Pauli \textbf{X} transformation. Recalling that $| 0 \rangle
\equiv \left (
\begin{smallmatrix}
1 \\
0
\end{smallmatrix}
\right )$ and $| 1 \rangle \equiv
\left (
\begin{smallmatrix}
0 \\
1
\end{smallmatrix}
\right )
$, we see that
$$\mathbf{X}| \psi' \rangle_b = \left(
\begin{matrix} 0 & 1 \\ 1 & 0 \end{matrix}
\right )
\left [ \beta\left (
\begin{matrix}1 \\ 0 \end{matrix}
\right ) + \alpha\left (
\begin{matrix} 0 \\ 1 \end{matrix}
\right )\right ] = \left(
\begin{matrix} 0 & 1 \\ 1 & 0 \end{matrix}
\right )\left(
\begin{matrix} \beta \\ \alpha \end{matrix}
\right ) = \left(
\begin{matrix} \alpha \\ \beta \end{matrix}
\right ) = | \psi \rangle_b.$$

If Alice's result is 10, then Bob applies a Pauli \textbf{Z}
transformation:
$$\mathbf{Z}| \psi'' \rangle_b = \left (
\begin{matrix} 1 & 0 \\ 0 & -1 \end{matrix}
\right )\left [ \alpha\left (
\begin{matrix} 1 \\ 0 \end{matrix}
\right ) - \beta\left (
\begin{matrix} 0 \\ 1 \end{matrix}
\right )\right ] = \left (
\begin{matrix} 1 & 0 \\ 0 & -1 \end{matrix}
\right )\left (
\begin{matrix}
\alpha \\ -\beta \end{matrix}
\right ) = \left (
\begin{matrix} \alpha \\ \beta \end{matrix}
\right ) = | \psi \rangle_b.$$

Finally, if Alice's result is 11, then the reader can verify that Bob
should apply the combined transformation \textbf{ZX}.

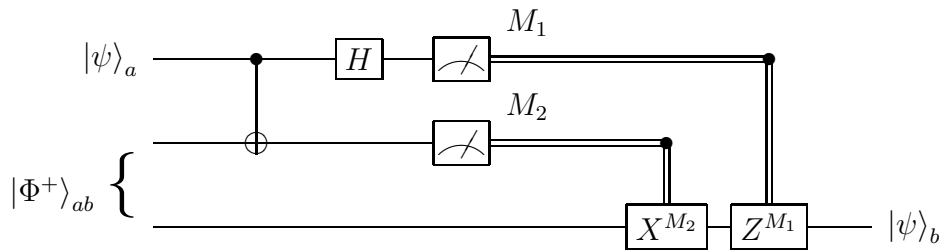
\begin{figure}
\label{fig:teleport_circuit}
$$
\Qcircuit @C=0.7em @R=.1em @! {
  \lstick{\ket{\psi}_a} & \ctrl{1} & \gate{H} & \meter &
  \lstick{\raisebox{2.5em}{$M_1$}} \cw & \cw & \control \cw \\
  & \targ & \qw & \meter & \lstick{\raisebox{2.5em}{$M_2$}} \cw &
  \control \cw & \cwx \\
  \lstick{\raisebox{3.25em}{$\ket{\Phi^+}_{ab}$ \huge\{}} & \qw
  & \qw & \qw & \qw & \gate{X^{M_2}} \cwx & \gate{Z^{M_1}} \cwx &
  \rstick{\ket{\psi}_b} \qw
}
$$
\caption[Teleportation circuit]{A quantum circuit for the
  teleportation of a qubit. The top two lines represent the parts of
  the system accessible to Alice; the bottom line is the part of the
  system accessible to Bob.}
\end{figure}

\chapter{Separable operations}
\label{ch:sepop}

\settocdepth{chapter}

An open quantum system (e.g., a noisy quantum circuit) is one in which
the state changes of the system of interest, $S$, are due both to its
own internal dynamics and to its interaction with an external
environment or `reservoir', $R$. Let the initial state of the overall
system be the product state,\footnote{This is typically a safe
  assumption to make as the process of state preparation will destroy
  any correlations between the system and the environment. See
  \citet[]{cuffaro2012a} for a discussion of the case where this
  assumption does not hold.} $\rho \otimes \omega_R$, with $\rho \in
\mathcal{H}_S$, $\omega_R \in \mathcal{H}_R$ where $\mathcal{H}_S,
\mathcal{H}_R$ are the Hilbert spaces associated with $S$ and
$R$. Then a state change of $S$ can be expressed as:
\begin{equation}
\label{eqn:dynev_a}
\rho \mapsto \Lambda\rho = \mbox{tr}_R(U\rho \otimes \omega_R
U^\dagger),
\end{equation}
where $\Lambda$ is the dynamical transformation map for $S$ which maps
density operators to density operators, $U \equiv
e^{-iH_{S+R}t/\hbar}$ is the time evolution operator for the combined
system, and $\mbox{tr}_R$ is the partial trace over $R$.


For many purposes it is more convenient to express $\Lambda$
exclusively in terms of $S$. Take $| f_\nu \rangle$ to be an
orthonormal basis for the state space of the reservoir (which we
assume, without loss of generality, to be pure\footnote{If the
  reservoir begins in a mixed state, it is always possible to purify
  it by means of an extra system. Cf. \textsection
  \ref{nec:ss:purify}}), with $\omega_R = | f_0 \rangle\langle f_0 |$
the reservoir's initial state. Since the partial trace, over $R$, of
$\rho \otimes \omega_R$ is given by $\mbox{tr}_R(\rho \otimes
\omega_R) = \langle f_\nu | \rho \otimes \omega_R | f_\nu \rangle$, we
can rewrite \eqref{eqn:dynev_a} as:
\begin{align}
\label{eqn:dynev_b}
\rho \to \Lambda\rho & = \sum_\nu\langle f_\nu |U\big[\rho \otimes |
  f_0 \rangle \langle f_0 |\big]U^\dagger | f_\nu \rangle \nonumber \\
& = \sum_\alpha E_\alpha \rho E_\alpha^\dagger,
\end{align}
where $E_\alpha \equiv \langle f_\nu |U| f_0 \rangle$ is an operator
on the state space of $S$, and the $E_\alpha$, known as \emph{Kraus
  operators} or \emph{operation elements}, satisfy the completeness
relation:
\begin{align}
\sum_\alpha E_\alpha^\dagger E_\alpha = I.
\end{align}

\emph{Separable operations} are those operations that can be
decomposed as a product of Kraus operators as follows:
\begin{align}
\label{nec:eqn:sepop}
\Lambda\rho = \sum_k A_k\otimes B_k \rho A^\dagger_k\otimes
  B^\dagger_k
\end{align}
such that $\sum_kA^\dagger_kA_k\otimes B^\dagger_kB_k =
\mathbf{1}\otimes\mathbf{1}$.

If Alice and Bob perform only LOCC (`local operations plus classical
communications') operations on a shared system $\rho$, then their
individual Kraus operators may be joined together into product Kraus
operators; i.e., into the form of a separable operation. The converse
is false \citep[]{bennett1999}. Separable operations are nevertheless
a convenient proxy for LOCC operations, as the optimal implementation,
via separable operations, of a given task provides strong bounds for
what can be achieved using LOCC (see, for instance,
\citealt[]{rains2001, virmani2003}).

\chapter{Lance Fortnow's Matrix Framework}
\label{ch:matrix}

Let us first consider a classical nondeterministic Turing machine. We
begin by defining the transition function, $\delta$, of the machine in
terms of a transition matrix, such that there is an entry in the
matrix corresponding to every possible transition of the
machine. We allow matrix entries to contain arbitrary nonnegative
rational numbers. We then define the matrix entry, $T(c_a,c_b),$ as
the probability that the computer goes to configuration $c_b$ from
configuration $c_a$ in one computational step. $T^r(c_a,c_b),$
correspondingly, is the probability of getting to $c_b$ from $c_a$ in
$r$ steps; it is the sum of the probabilities of each computational
path of length $r$ leading from $c_a$ to $c_b$, with the restriction
that the sum of all possible computational paths of length $r$
beginning from $c_a = 1$.

\begin{table}
$$
\begin{matrix}
      & c_a & c_b & c_c & c_d & c_e & c_f & c_g & c_h \\
  c_a & 0.0 & 0.2 & 0.3 & 0.5 & 0.0 & 0.0 & 0.0 & 0.0 \\
  c_b & 0.0 & 0.0 & 0.0 & 0.5 & 0.0 & 0.0 & 0.0 & 0.5 \\
  c_c & 0.0 & 0.0 & 0.0 & 0.5 & 0.0 & 0.0 & 0.5 & 0.0 \\
  c_d & 0.0 & 0.0 & 0.0 & 0.0 & 0.6 & 0.4 & 0.0 & 0.0 \\
  c_e & 0.0 & 0.0 & 0.0 & 0.0 & 1.0 & 0.0 & 0.0 & 0.0 \\
  c_f & 0.0 & 0.0 & 0.0 & 0.0 & 0.0 & 1.0 & 0.0 & 0.0 \\
  c_g & 0.0 & 0.0 & 0.0 & 0.0 & 0.0 & 0.0 & 1.0 & 0.0 \\
  c_h & 0.0 & 0.0 & 0.0 & 0.0 & 0.0 & 0.0 & 0.0 & 1.0
\end{matrix}
$$
\caption[State transition matrix]{A sample state transition
  matrix. Entries represent probabilities of transition between
  states.}
\label{tab:statemat}
\end{table}

\begin{figure}
\begin{tikzpicture}[->,>=stealth',shorten >=1pt,auto,node
    distance=2.8cm, semithick]

  \node[initial,state]   (A)                    {$c_a$};
  \node[state]           (B) [above right of=A] {$c_b$};
  \node[state]           (C) [below right of=A] {$c_c$};
  \node[state]           (G) [below right of=C] {$c_g$};
  \node[state]           (H) [above right of=B] {$c_h$};
  \node[state]           (D) [below right of=B] {$c_d$};
  \node[state]           (E) [below right of=D] {$c_e$};
  \node[accepting,state] (F) [above right of=D] {$c_f$};

  \path (A) edge              node {0.2} (B)
            edge              node {0.3} (C)
            edge              node {0.5} (D)
        (B) edge              node {0.5} (D)
            edge              node {0.5} (H)
        (C) edge              node {0.5} (D)
            edge              node {0.5} (G)
        (D) edge              node {0.4} (F)
            edge              node {0.6} (E)
        (E) edge [loop below] node {1}   (E)
        (G) edge [loop right] node {1}   (G)
        (H) edge [loop right] node {1}   (H);
\end{tikzpicture}
\caption[State diagram representation of a transition matrix]{State
  diagram representation of the transition matrix in Table
  \ref{tab:statemat}. The edge labels represent the probability of a
  transition between the states connected by that edge.}
\label{fig:statediag}
\end{figure}
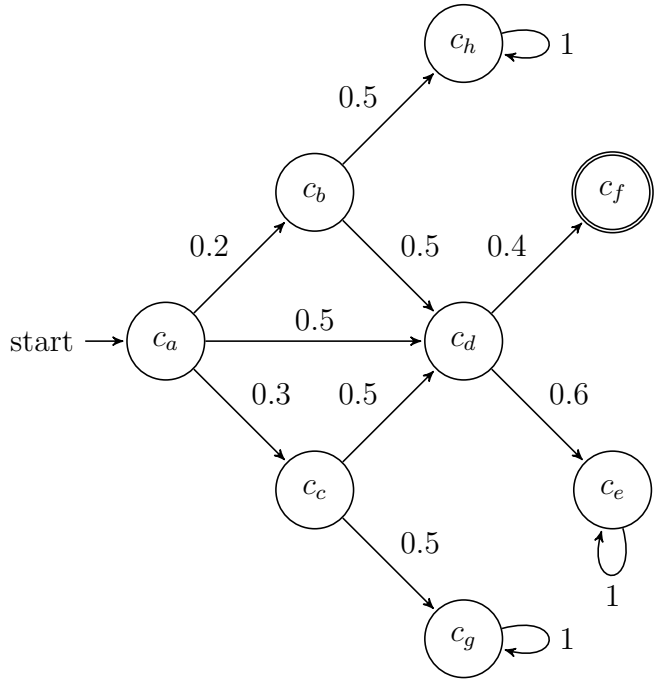

For instance, given the matrix in Table \ref{tab:statemat}, we can
determine that

\begin{eqnarray*}
  T^3(c_a,c_e) & = & [T(c_a,c_b) \times T(c_b,c_d) \times
    T(c_d,c_e)] + \\
  & & [T(c_a,c_c) \times T(c_c,c_d) \times T(c_d,c_e)] \\
  & = & (0.2 \times 0.5 \times 0.6) + (0.3 \times 0.5 \times 0.6) =
  0.15.
\end{eqnarray*}

We can now define $T^t(c_a,c_f)$ as the probability of success for our
nondeterministic Turing machine ($c_a$ and $c_f$ are the initial and
accepting states, respectively) in $t$ time steps. It can be shown
that a language $L$ is in the computational complexity class
associated with classical probabilistic
computation\footnote{Cf. Appendix \ref{ch:cc}.} (\textbf{BPP}) if
there is a probabilistic matrix $T$ such that, for $x \in L$ and $1/2
\leq k \leq 1$ (typically taken to be 2/3), $$T^t(c_a,c_f) \geq
k,$$ and for $x \notin L$, $$T^t(c_a,c_f) \leq k,$$ for polynomial $t$.

To capture the case of the quantum nondeterministic Turing machine, we
omit the restriction that the matrix entries be nonnegative, and we
redefine the probability of acceptance as $(T^t(c_1,c_A))^2.$ It can
be shown that a language $L$ is in the computational complexity class
associated with quantum computation (\textbf{BQP}) if there is a
matrix $T$, as just defined, such that, for $x \in L$,
$$(T^t(c_a,c_f))^2 \geq k,$$ and for $x \notin L$, $$(T^t(c_a,c_f))^2
\leq k,$$ for polynomial $t$.

According to Fortnow, the fundamental difference between quantum and
classical computing is interference. The matrix framework,
according to Fortnow, shows us that, in a quantum computer, `bad'
computational paths are associated with negative matrix entries,
allowing other computational paths to occur with higher
probability. Fortnow writes: ``The strength of quantum computing lies
in the ability to have bad computation paths eliminate each other thus
causing some good paths to occur with larger probability''
\citep[pp. 605-606]{fortnow2003}.

\chapter{Spekkens vs. Quantum Transformations}
\label{ch:transf}

In Robert Spekkens' toy theory \citeyearpar[]{spekkens2007}, a system
consists of a ball that can be in one of four boxes. A state, in the
theory, is an expression of our knowledge of the location of the
ball. For instance, if we know that the ball is in either the first or
the second box, we write $1 \lor 2$. Knowledge is restricted in the
Spekkens theory. Aside from the `completely mixed state', $1 \lor 2
\lor 3 \lor 4$, the only other allowable states are the following six
states of maximal knowledge:

\begin{align*}
\includegraphics[width=15mm]{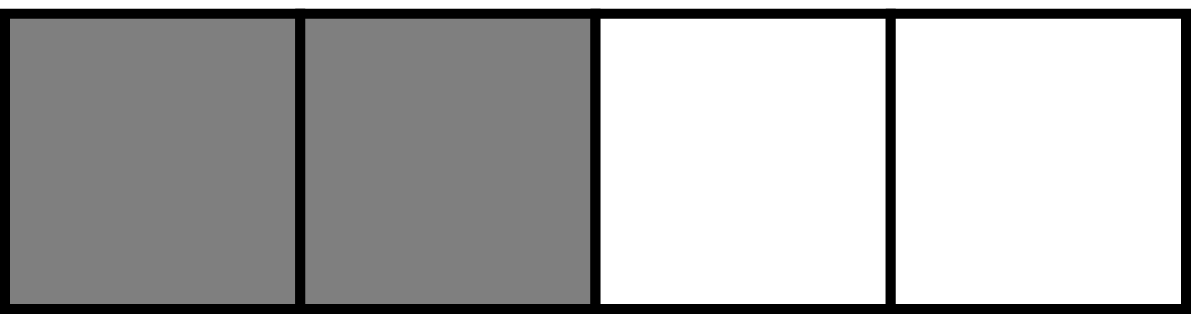} & \equiv 1 \lor 2 \equiv | 0
\rangle \tag{$z+$} \\
\includegraphics[width=15mm]{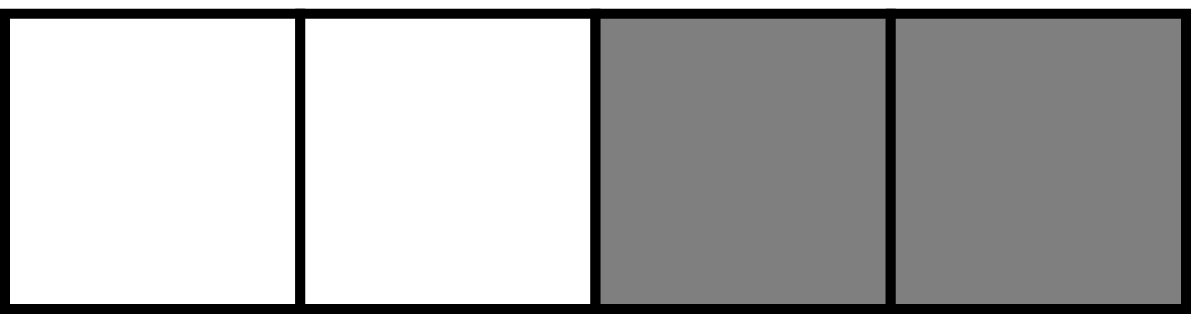} & \equiv 3 \lor 4 \equiv | 1
\rangle \tag{$z-$} \\
\includegraphics[width=15mm]{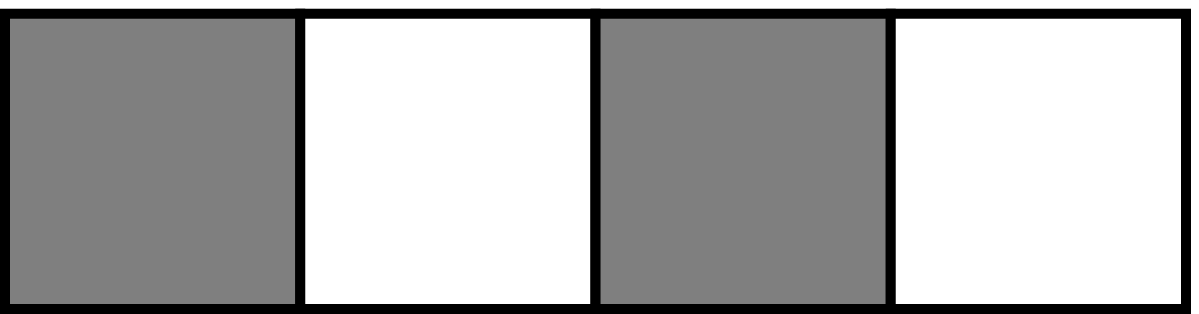} & \equiv 1 \lor 3 \equiv | +
\rangle \tag{$x+$} \\
\includegraphics[width=15mm]{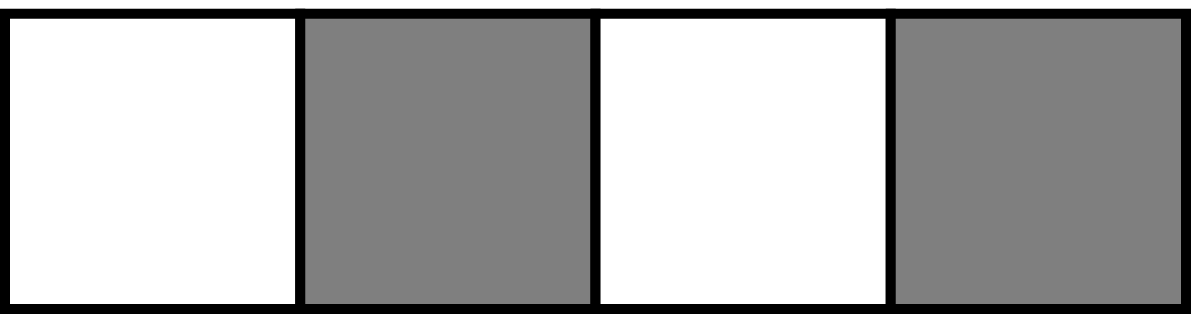} & \equiv 2 \lor 4 \equiv | -
\rangle \tag{$x-$} \\
\includegraphics[width=15mm]{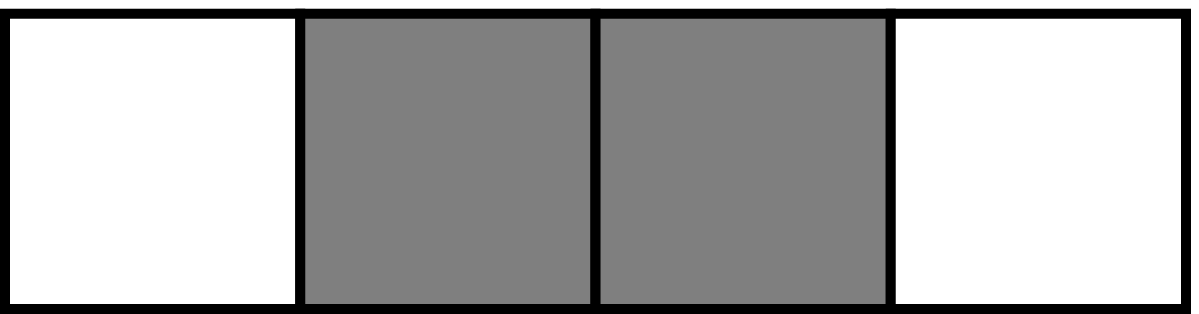} & \equiv 2 \lor 3 \equiv | +i
\rangle \tag{$y+$} \\
\includegraphics[width=15mm]{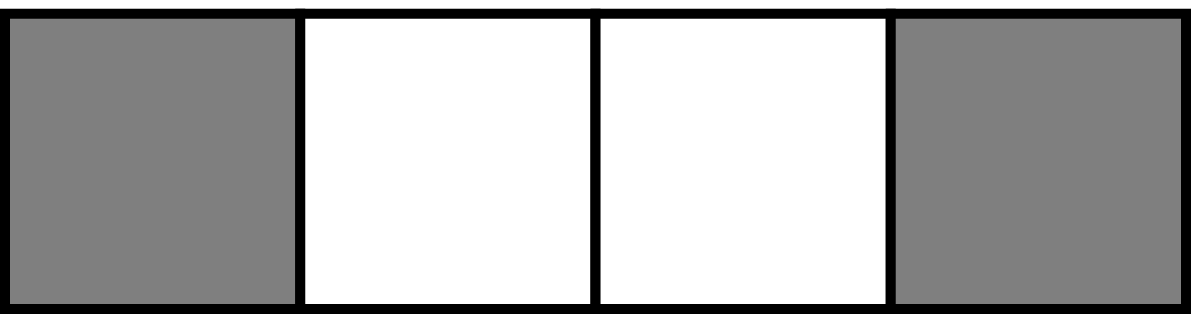} & \equiv 1 \lor 4 \equiv | -i
\rangle \tag{$y-$}
\end{align*}

Transformations of the Spekkens states are just permutations of the
boxes. For instance, if we subject the state $1 \lor 2$ to the
permutation $\langle 1 \rightarrow 2 \rightarrow 3 \rightarrow 1\rangle,$
the resulting state will be $2 \lor 3$. Subjecting $3 \lor 4$ to this
permutation will result in $1 \lor 4$.

We can associate some of the permutations of boxes in Spekkens' toy
theory with rotations of the Bloch sphere in quantum theory
\citep[]{myrvold2010}. In quantum theory, a $2\pi/3$ rotation of the
Bloch sphere about the direction of $\hat{x} + \hat{y} + \hat{z}$
takes $x+ \rightarrow y+$, $y+ \rightarrow z+$, and $z+ \rightarrow
x+.$ In the Spekkens toy theory, this corresponds to the permutation
$\langle 1 \rightarrow 3 \rightarrow 2 \rightarrow 1 \rangle$. Let us
call this transformation $T_1$. A $\pi/2$ rotation of the Bloch sphere
about the $z$-axis, in quantum theory, leaves $z+$ invariant but takes
$x+$ to $y+$ and $y+$ to $x-$. Let us call this transformation
$T_2^q$. It cannot be achieved in the Spekkens theory. An alternative,
however, is a $\pi/2$ rotation about the $z$-axis followed by a
reflection in the $xy$ plane, which corresponds, in the Spekkens
theory to $\langle 1 \rightarrow 3 \rightarrow 2 \rightarrow 4
\rightarrow 1 \rangle$. Call this transformation $T_2^S$. Note that
while $T_2^q$ leaves $z+$ invariant, $T_2^S$ takes $z+$ to $z-$. It
can be shown that the sets $\{T_1,T_2^q\}$ and $\{T_1,T_2^S\}$ are
sufficient to generate the Spekkens and quantum groups of
transformations, respectively.

Spekkens's toy theory contains entangled states, but because of the
differences in the allowable transformations between the toy theory
and quantum theory, the set of entangled states that the toy theory
contains is not identical to the set of entangled states contained in
quantum theory; specifically, none of the entangled states in
Spekkens's toy theory yield correlations between outcomes of
experiments that violate the Bell inequalities.

\newpage

\singlespacing

\bibliographystyle{apa-good}
\addcontentsline{toc}{chapter}{Bibliography}
\bibliography{Bibliography}{}

\newpage

\addcontentsline{toc}{chapter}{Vita}
\begin{center}\textsc{Vita}\end{center}

\begin{table}[ht]
\begin{tabular}{ll}
\textbf{Name:} & Michael E. Cuffaro \\\\
\textbf{Post-Secondary} & Concordia University\\
\textbf{Education and}& Montr\'eal, Qu\'ebec, Canada\\
\textbf{Degrees:}& 1993--1994, 1997--2000 B.CompSc.\\\\
& Concordia University\\
& Montr\'eal, Qu\'ebec, Canada\\
& 2007--2008 M.A.\\\\
& The University of Western Ontario\\
& London, Ontario, Canada\\
& 2008--2013 Ph.D.\\\\
\textbf{Honours and}& Province of Ontario Graduate Scholarship\\
\textbf{Awards:}& 2010--2011, 2011--2012, 2012--2013\\\\
& University of Western Ontario Graduate Teaching\\
& Assistantship Research Grant \\
& Fall 2009\\\\
\textbf{Related Work}& Instructor\\
\textbf{Experience:}& The University of Western Ontario\\
& Winter 2011, Winter 2013\\\\
& Teaching Assistant\\
& The University of Western Ontario\\
& 2008--2010\\\\
& Teaching Assistant\\
& Concordia University\\
& 2007--2008
\end{tabular}
\end{table}

\subsubsection*{Publications:}

\mbox{} \\

\setlength\leftskip{0.1in}
\setlength\parindent{-0.1in}

Cuffaro, M. (2012a). Many Worlds, the Cluster-state Quantum Computer,
and the Problem of the Preferred Basis. \emph{Studies in History and
  Philosophy of Modern Physics, 43}, 35-42. \\
\mbox{} \\
Cuffaro, M. (2012b). Kant and Frege on Existence and the Ontological
Argument. \emph{History of Philosophy Quarterly, 29}, 337-354. \\
\mbox{} \\
Cuffaro, M. (2012c). Kant's Views on Non-Euclidean Geometry. To
appear in the \emph{Proceedings of the Canadian Society for History
  and Philosophy of Mathematics}. \\
\mbox{} \\
Cuffaro, M. \& Myrvold, W. (2012). On the Debate Concerning the Proper
Characterisation of Quantum Dynamical Evolution. Accepted for
publication in \emph{Philosophy of Science}. \\
\mbox{} \\
Muldoon, R., Borgida, M., \& Cuffaro, M. (2012) The Conditions of
Tolerance. \emph{Politics, Philosophy and Economics, 11},
322-344. \\
\mbox{} \\
Cuffaro, M. (2011). On Thomas Hobbes's Fallible Natural Law
Theory. \emph{History of Philosophy Quarterly, 28}, 175-190. \\
\mbox{} \\
Cuffaro, M. (2010a). The Kantian Framework of
Complementarity. \emph{Studies in History and Philosophy of Modern
  Physics, 41}, 309-317. \\
\mbox{} \\
Cuffaro, M. (2010b). Wittgenstein on Prior
Probabilities. \emph{Proceedings of the Canadian Society for History
  and Philosophy of Mathematics, 23}, 85-98. \\
\mbox{} \\
Cuffaro, M. (2008). Nativist Models of the Mind. \emph{GNOSIS,
  9.3}. \\
\mbox{} \\
Cuffaro, M. (2007). Which Rights are Basic Rights?
\emph{GNOSIS, 9.1}. \\

\end{document}